\documentclass[11pt]{article}

\usepackage[T1]{fontenc}
\usepackage{lmodern}
\usepackage{microtype}
\usepackage[margin=1.05in]{geometry}
\usepackage{amsmath,amssymb,amsthm,mathtools,bm,mathrsfs}
\usepackage{booktabs,array,longtable}
\usepackage{enumitem}
\usepackage{float}
\usepackage[hidelinks]{hyperref}

\hypersetup{
  pdftitle={The Infrared Universe},
  pdfauthor={Jonathan Holland},
  pdfsubject={Stationary infrared transport, galactic response, and radiative states},
  pdfkeywords={open systems, stochastic pumps, Carnot--Caratheodory geometry,
  nilpotent groups, stationary cosmology, galactic rotation curves,
  baryonic Tully--Fisher relation, cosmic microwave background}
}

\title{The Infrared Universe}
\author{Jonathan Holland}
\date{August 2026}

\newcommand{\dd}{\mathrm d}
\newcommand{\ii}{\mathrm i}
\newcommand{\e}{\mathrm e}
\newcommand{\Tr}{\operatorname{Tr}}
\newcommand{\STF}{\operatorname{STF}}
\newcommand{\im}{\operatorname{im}}

\newcommand{\Var}{\operatorname{Var}}
\newcommand{\E}{\mathbb E}
\newcommand{\R}{\mathbb R}

\newcommand{\cU}{\mathcal U}
\newcommand{\cX}{\mathcal X}
\newcommand{\cL}{\mathcal L}
\newcommand{\cK}{\mathcal K}

\newcommand{\cW}{\mathcal W}
\newcommand{\cE}{\mathcal E}
\newcommand{\bsxi}{\boldsymbol\xi}
\newcommand{\bsomega}{\boldsymbol\omega}
\newcommand{\bschi}{\boldsymbol\chi}
\newcommand{\bsSigma}{\boldsymbol\Sigma}
\newcommand{\bsA}{\boldsymbol A}
\newcommand{\bsN}{\boldsymbol N}

\numberwithin{equation}{section}
\newtheorem{theorem}{Theorem}[section]
\newtheorem{proposition}[theorem]{Proposition}
\newtheorem{corollary}[theorem]{Corollary}

\newtheorem{hypothesis}[theorem]{Hypothesis}
\theoremstyle{definition}
\newtheorem{definition}[theorem]{Definition}

\theoremstyle{remark}

\begin{document}
\maketitle

\begin{center}
  \emph{Ineluctable modality of the visible.}\\[0.5ex]
  --- James Joyce, \emph{Ulysses}, ``Proteus''
\end{center}

\begin{abstract}
We develop a stationary open-sector framework that organizes galactic
dynamics, cosmological matter balance, and infrared radiation as aspects of
one mesoscopic transport problem in an expanding universe.  Stationarity means
balance against Hubble dilution and therefore requires counted exchange
currents rather than an environmental trace alone.  At a boundary of fixed
physical radius, charge balance fixes the baryon return current.  A joint
energy ledger separates rest-mass recycling, horizon heat, and useful return
free energy, showing that entropy production alone does not determine the
available power.

Full counting statistics equips the boundary process with a
Berry--Sinitsyn--Nemenman connection.  Pulling its curvature back along three
spatial controls yields the free step-two tangent
$N_{3,2}=\R^3\oplus\Lambda^2\R^3$.  An autonomous finite-state horizon pump
realizes the clock and affinity, while an event-conditioned completely
positive channel selects the central character left undetermined by holonomy.
In a galactic disc this central response becomes an azimuthal connection.  We
derive its circular-orbit law, its inverse reconstruction from a rotation
curve, and a screened closure with a finite approximately flat branch.  The
baryonic Tully--Fisher normalization remains an explicit source-matching
hypothesis.

The same tangent has an exact Landau--longitudinal spectrum and associated sky
kernels.  At feature level these fit the measured TT locations, reproduce the
signs of the twelve measured TE extrema and the higher TE/EE interleaving, and
supply the first EE lobe without shifting TT.  Coherent log-frequency
translation preserves an exact Planck spectrum, whereas differential
translation is constrained by blackbody mixing.  The framework thus yields
exact structural relations and conditional galactic, thermodynamic, and
radiative closures, while leaving the common microscopic event law and full
light-cone covariance as the central open problem.
\end{abstract}

\tableofcontents

\section{Introduction}
\label{sec:introduction}

Flat galactic rotation curves and the baryonic Tully--Fisher relation point to
infrared gravitational structure beyond the directly inventoried baryons
\cite{McGaugh2000BTFR,LelliMcGaughSchombert2016BTFR}.  The standard account
uses nonbaryonic dark matter; Milgrom instead proposed that the low-acceleration
law itself changes and identified the associated acceleration scale
\cite{Milgrom1983MOND,Milgrom1983Galaxies}.  The construction below is
dynamically different from MOND: it retains the baryonic potential and adds an
open-system azimuthal connection whose source must be derived.  Its relation to
the Milgrom scaling is therefore an empirical and constitutive question, not a
change of notation.  The nearly perfect $2.7\,\mathrm K$ microwave blackbody
poses a complementary problem.  If the radiation is regarded as a maintained
late-time state rather than only as transported initial data, an exchange law
must preserve both the Planck spectrum and its anisotropy covariance.  This
paper asks whether those galactic and radiative questions can arise from one
stationary infrared transport framework while local general relativity and
quantum field theory remain the local dynamics.

The cosmological setting remains an expanding spacetime.  ``Stationary'' means
that selected observables of an observer-accessible open sector obey steady
balance laws on the timescales in question.  Physical number and energy
densities still carry the usual $3H$ and $4H$ dilution terms, so constant values
require compensating exchange.  This differs from the classical steady-state
cosmologies, which imposed a stronger global stationarity and continuous matter
creation \cite{BondiGold1948,Hoyle1948}.  Nothing below requires either a
past-infinite universe or a preferred finite initial surface; the results apply
whenever the stated stationary regime exists.

Operationally, divide the degrees of freedom into an algebra accessible to a
finite observer and an unresolved environment.  A causal horizon or a
black-hole exterior/interior split can motivate such a partition, but the
mathematics requires a declared subsystem and exchange channels, not a literal
membrane.  Nor does partial trace create a current.  The reduced generator must
contain a physical transition whose increments can be counted and whose charge
and energy debits appear in the complementary sector.

Counting here has its standard full-counting-statistics meaning.  One deforms
the reduced generator by a phase attached to each chosen transition and reads
off cumulants by differentiating at zero phase.  In an adiabatically driven
nonequilibrium system, a closed loop of controls can pump a quantity even when
the frozen current vanishes; the geometric part is governed by the
Berry--Sinitsyn--Nemenman (BSN) connection
\cite{SinitsynNemenman2007,Sinitsyn2009,RahavHorowitzJarzynski2008}.  The open
three-state ring in Section~\ref{subsec:open-ring-anchor} supplies a concrete
example:
one mixed weight--barrier cycle has nonzero first-cumulant curvature, positive
throughput, and a finite intensive response.

An externally prescribed cycle would leave the source of the pumping outside
the cosmological model.  Section~\ref{subsec:autonomous-horizon-pump} therefore
adjoins a dynamical horizon phase to the directional chart.  The horizon
affinity is then the only nonconservative cycle affinity: at zero affinity the
whole joint process is at detailed balance, while at nonzero affinity its
winding drives the same mixed weight--barrier loop.  In the slow,
fine-phase limit the ledger written per horizon winding converges to the BSN
holonomy.  In the physical proposal, horizon export is consequently not a
passive terminal sink or merely the denominator of a susceptibility; it is
the proposed source of the cycle orientation, cadence, and entropy production.
The finite generator proves consistency of that identification but does not
derive its rates from horizon quantum field theory.

Here and below, a \emph{tile insertion} is shorthand for a microscopic event
representing an increment of newly available physical volume; no literal
tessellation is assumed.  The term carries no independent energy debit unless
the event creates a separately identified gravitational excitation, as made
precise in Definition~\ref{def:extra-tile-energy}.

Thermodynamic affinity and available power must also be kept distinct.
Section~\ref{subsec:fixed-radius-recycling} derives $3Hn_B$ as the geometric
baryon flux through a constant-physical-radius horizon.  It therefore requires
one return per crossing when the complementary baryon charge is stationary,
rather than treating horizon crossing as a second sink.  The same section
identifies when tile insertion would double-count the ordinary dilution
balances.  Section~\ref{subsec:entropy-separated-return} then promotes baryon
number and energy to independent counting marks.  The energy-identifiability
proposition shows that $\mathcal A_HJ_H$ fixes entropy production but not pump
power.  Under an explicit charge-cyclic, entropy-open hypothesis, local
detailed balance instead calibrates the small retained heat and leaves the
remainder as return availability; a physical horizon model must still derive
the generalized energy marks and event-rate normalization.

The new step is to retain not merely a scalar total but the oriented second
level of the path.  We call this an \emph{infrared ledger}: it records the
antisymmetric area swept by ordered increments after their endpoint has been
forgotten.  Displacement together with accumulated area has the group law of a
step-two nilpotent space.  The term is therefore shorthand for a measurable
counting-statistics object, not an additional substance.  The BSN pullback
theorem below proves when this structure becomes spatial: the bracket of the
horizontal lifts is exactly the pulled-back first-cumulant curvature, giving
$N_{3,2}=\R^3\oplus\Lambda^2\R^3$ at full rank.

Writing a record is not yet dynamics.  Translation of the central coordinate
leaves every spectral function of its conjugate momentum invariant, so a pump
cannot select a physical central character by translation alone.  This
no-selection theorem is the central divide in the construction.  An
event-gated completely positive channel shows that the divide can be crossed:
Weyl kicks act on the conjugate variable and attenuation supplies contraction.
The result is an explicit stationary pseudovector response
$\bsomega$.  In a disc, its component normal to the plane enters a canonical
azimuthal connection.  That produces the exact rotation-curve relation and
inverse reconstruction developed in Section~\ref{sec:galactic-response}.  A
screened $1/r$ exterior response has a flat intermediate corridor; the mass
normalization needed for the baryonic Tully--Fisher law remains an explicit,
falsifiable matching hypothesis rather than a result of dimensional notation.

The same group determines a native harmonic analysis: its central blocks are
transverse Landau oscillators with a free longitudinal direction, and centered
scalar and spin-raised sky kernels follow exactly.  These results determine a
mode space, not its state.  For the CMB we therefore separate what can be tested
without a complete transfer theory.  The low band gives a compact calibration
of eight TT feature locations, while a separate one-pole carrier reproduces
the broad normalized height ladder only at toy-model accuracy.  Once the TT
phase is frozen, the stationary quadrature gives the observed TE sign sequence
and the higher EE interleaving without a polarization phase parameter.  A
minimal expansion-conditioned long-mode scattering term also supplies the
otherwise absent first EE lobe without moving TT.

The remaining polarization problem is dynamical rather than spectral.
Temperature and quadrupole marks of one return event define a positive mixed
noise matrix, and event-conditioned opacity is transverse to the leading
acoustic phase.  The simplest additive, single-age, and time-local
realizations nevertheless fail to reproduce the TE and EE amplitudes while
preserving the calibrated TT extrema.  We retain the analytic obstruction and
report the final quantitative comparison; the intermediate parameterizations
are not treated as physical models.  A joint TT--TE--EE prediction requires a
derived event chronology, photon-coupled opacity history, and unequal-time
matter--metric--photon projection.

Finally, cosmological propagation itself supplies strong consistency tests.  A
coherent translation in log frequency is an exact achromatic redshift and
waveform-dilation channel and maps the Planck family into itself.  Unresolved
differential translations instead mix blackbodies and are sharply bounded by
spectral-distortion data.  Liouville transport and reciprocity then recover the
usual conditional surface-brightness and distance-duality relations.  These
results do not replace metric redshift; they specify how any additional
infrared contribution must compose with it.

The paper keeps these applications in one chain.  Section~\ref{sec:balances}
states the stationary cosmological bookkeeping.  Sections~\ref{sec:bsn-geometry}
and \ref{sec:central-selection} construct and select the counted response.
Section~\ref{sec:galactic-response} derives its galactic consequences.
Sections~\ref{sec:n32-spectrum}--\ref{sec:cmb-checkpoint} give the exact
spectral theory and limited radiative application.  Section~\ref{sec:phase-soft}
sets propagation constraints.  The appendices record finite-generator and
numerical verifications.

\subsection{Logical status of the construction}
\label{subsec:logical-status}

The argument distinguishes exact structural statements from finite existence
models, constitutive hypotheses, and empirical calibrations.  The balance
identities, fixed-radius flux relation, BSN pullback theorem, central
no-selection theorem, galactic orbit relations, $N_{3,2}$ spectrum,
radiative-state underdetermination, and coherent-redshift identities are exact
within their displayed assumptions.  The finite horizon ring and the
completely positive selector likewise establish properties of specified open
systems; identifying their rates and marks with a semiclassical horizon is a
separate physical hypothesis.

The charge-cyclic return law, the screened galactic field equation, its source
matching, and the photon-coupled return process are constitutive.  In
particular, the horizon affinity does not fix the exported free-energy mark,
and the mode geometry does not fix a radiative covariance.  These omissions
are stated as explicit closure conditions rather than absorbed into fitted
normalizations.

The CMB calculations have a still narrower status.  TT feature locations are
calibrated; the TT height ladder and polarization comparisons are diagnostics
of low-dimensional carriers.  They are not likelihood fits or a replacement
for a continuous Boltzmann transfer calculation.  The numerical comparisons
are retained only where they distinguish a viable feature topology from a
structural incompatibility.

\section{Stationary cosmological balances}
\label{sec:balances}

The word \emph{stationary} carries conservation obligations.  Let $a(t)$ be
the scale factor of a homogeneous expanding chart and $H=\dot a/a$.  The
quantities below are physical densities in that chart.  A constant physical
density is not obtained by declaring its time derivative zero; the dilution
term must be balanced by a current.  This section states that bookkeeping
before introducing any geometric mechanism.

\subsection{The observer-accessible sector}
\label{subsec:accessible-sector}

Let $\mathcal A_{\rm acc}$ be an algebra of observables accessible to a finite
observer and $\mathcal A_{\rm env}$ the complementary degrees of freedom used
in a reduced description.  A causal exterior/interior split or a black-hole
exterior can furnish such an algebraic partition, but no particular horizon is
assumed by the theorems below.  ``Export'' means a transition from the declared
accessible sector to the declared environment; ``return'' means the reverse
transition in the reduced bookkeeping.  Both require a microscopic coupling.

The balance identities and the BSN pullback theorem are deliberately
horizon-independent.  The central cosmological hypothesis made later is
stronger: the export channel associated with a physical horizon supplies the
only nonequilibrium affinity that powers the ledger pump.  Keeping the generic
theorems separate from that identification makes the remaining burden clear:
one must derive the horizon transition rates and their matter coupling, not
infer a current from partial trace alone.

Three distinctions prevent the subsystem language from hiding a conservation
violation.
\begin{enumerate}[label=(\roman*)]
\item A partial trace changes the description, not the total charge.  It cannot
create an exterior gain term unless the reduced dilation contains a
charge-carrying exchange channel.
\item Entropy, energy, and baryon number are different currents.  A positive
entropy production rate does not determine the sign or magnitude of an energy
or charge transfer.
\item The orbital source radius $R_s$ and screening length $L_\omega$ introduced
in Section~\ref{sec:galactic-response} are response scales, not causal
horizons.  They carry no automatic entropy or charge flux.
\end{enumerate}

For late-time baryon number, neglecting any microscopic violation rate, write
the homogeneous exterior balance as
\begin{equation}
 \dot n_B+3Hn_B=-\Phi_B^{\rm extra}+\Gamma_B^{\rm ret},
 \label{eq:baryon-balance}
\end{equation}
where $\Phi_B^{\rm extra}\geq0$ is a declared local loss in addition to
ordinary Hubble transport and $\Gamma_B^{\rm ret}\geq0$ is the return rate per
physical volume.  The $3Hn_B$ term is itself the outward flux through a
constant-physical-radius control surface, as shown in
Section~\ref{subsec:fixed-radius-recycling}; that geometric flux must not also
be inserted into $\Phi_B^{\rm extra}$.  A stationary nonzero physical density
therefore requires
\begin{equation}
 \Gamma_B^{\rm ret}-\Phi_B^{\rm extra}=3Hn_B.
 \label{eq:baryon-stationary-balance}
\end{equation}
In a globally charge-conserving dilation, the complementary environment obeys
the opposite exchange balance.  Equation~\eqref{eq:baryon-stationary-balance}
is not baryon creation from nothing; it is a stringent demand on the unresolved
sector.

Energy and radiation have their own equations.  For a resolved component of
energy density $\rho$ and pressure $p$,
\begin{equation}
 \dot\rho+3H(\rho+p)=J_E^{\rm in}-J_E^{\rm out},
 \label{eq:energy-balance}
\end{equation}
while an approximately homogeneous photon bath satisfies
\begin{equation}
 \dot\rho_\gamma+4H\rho_\gamma
 =J_\gamma^{\rm in}-J_\gamma^{\rm out}.
 \label{eq:radiation-balance}
\end{equation}
Thus maintenance of a blackbody energy density requires net power
$4H\rho_\gamma$ in the physical-density convention.  The harder requirement is
spectral: the exchange must move the state approximately along the Planck
family, rather than inject arbitrary heat.  Section~\ref{subsec:planck-maintenance}
quantifies that statement.

An entropy balance may be written independently as
\begin{equation}
 \dot s+3Hs=J_S^{\rm in}-J_S^{\rm out}+\sigma_{\rm int},
 \qquad \sigma_{\rm int}\geq0,
 \label{eq:entropy-balance}
\end{equation}
but $J_S$ and $\sigma_{\rm int}$ do not replace the charge and energy entries in
Eqs.~\eqref{eq:baryon-balance}--\eqref{eq:radiation-balance}.  A proposed
microscopic event population must pass all of these balances with one set of
event rates and quantum numbers.

\subsection{Composition is a separate fixed-point condition}
\label{subsec:composition-balance}

Maintaining the total baryon density does not maintain its chemical
composition.  Let $\rho_a$ denote the diffuse baryonic mass density in species
$a$, let $\rho_d=\sum_a\rho_a$, and write
\begin{equation}
 \dot\rho_a+3H\rho_a=F_a,
 \qquad F_b:=\sum_aF_a,
 \qquad x_a:=\frac{\rho_a}{\rho_d}.
 \label{eq:species-balance}
\end{equation}
The expansion term cancels exactly from the fraction equation:
\begin{proposition}[Exact composition identity]
\label{prop:composition-identity}
Where $\rho_d>0$,
\begin{equation}
 \dot x_a=\frac{F_a-x_aF_b}{\rho_d}.
 \label{eq:composition-identity}
\end{equation}
Consequently a stationary composition $\bm x_*$ requires
\begin{equation}
 F_a(\bm\rho_*)=x_{a,*}F_b(\bm\rho_*)
 \quad\text{for every }a.
 \label{eq:composition-fixed-point}
\end{equation}
\end{proposition}

\begin{proof}
Differentiate $x_a=\rho_a/\rho_d$, use
$\dot\rho_d+3H\rho_d=F_b$, and cancel the two terms proportional to $H$.
\end{proof}

To expose the constraint, write the positive exchange current as
$J_B\bm y^{\rm ex}$ with $\sum_a y_a^{\rm ex}=1$, let $P_a$ be the net
nuclear-processing term with $\sum_aP_a=0$ at baryon-mass accuracy, and let
$L_a$ be a loss term.  Defining
\begin{equation}
 L_b=\sum_aL_a,
 \qquad \Delta L_a=L_a-x_aL_b,
 \label{eq:selective-loss}
\end{equation}
gives the exact exchange form
\begin{equation}
 \rho_d\dot x_a
 =J_B(y_a^{\rm ex}-x_a)+P_a-\Delta L_a.
 \label{eq:isotope-maintenance}
\end{equation}
Hence a stationary mixture demands
\begin{equation}
 y_{a,*}^{\rm ex}
 =x_{a,*}-\frac{P_{a,*}-\Delta L_{a,*}}{J_{B,*}}.
 \label{eq:required-return-composition}
\end{equation}
When exchange dominates processing and selective loss, the returned material
must already be close to the maintained composition.  Total baryon conservation
alone does not imply this isotope-resolved condition.

The helium component makes the scale of the restriction concrete.  For a
present-epoch normalization, take \cite{Planck2018Params}
\begin{equation}
 H_0=67.4\,\mathrm{km\,s^{-1}Mpc^{-1}},
 \qquad
 \rho_b\simeq6.2\times10^9\,M_\odot\,\mathrm{Mpc}^{-3}.
 \label{eq:helium-audit-inputs}
\end{equation}
Stationary physical density then requires at least
\begin{equation}
 J_B^{(0)}=3H_0\rho_b
 \simeq1.28\,M_\odot\,\mathrm{yr}^{-1}\mathrm{Mpc}^{-3}.
 \label{eq:baryon-replacement-audit}
\end{equation}
These standard-model-inferred values normalize the rate comparison; a completed
stationary model must infer its baryon density independently.
The present star-formation-rate density is about
$0.015\,M_\odot\,\mathrm{yr}^{-1}\mathrm{Mpc}^{-3}$
\cite{MadauDickinson2014}.  With a representative net helium yield
$p_4^{\rm new}\simeq0.022$, stellar processing supplies only
$P_4^\star\simeq3.3\times10^{-4}\,M_\odot\,\mathrm{yr}^{-1}\mathrm{Mpc}^{-3}$
\cite{WellerWeinbergJohnson2025}.  A hydrogen-only return current maintaining
$Y\simeq0.245$ would instead require
\begin{equation}
 P_4^\star\simeq YJ_B^{(0)}
 \simeq0.31\,M_\odot\,\mathrm{yr}^{-1}\mathrm{Mpc}^{-3},
 \label{eq:hydrogen-only-helium-audit}
\end{equation}
roughly three orders of magnitude too large.  The conclusion is narrow but
robust: any stationary baryon-return mechanism must be nearly
composition-preserving; ordinary stellar burning cannot rebuild the helium
inventory from a nearly pure-hydrogen return current.  This is a maintenance
constraint, not an alternative primordial-nucleosynthesis calculation.

\subsection{A conditional joint baryon--radiation energy ledger}
\label{subsec:joint-baryon-radiation-ledger}

The baryon and radiation balances can be compared without identifying their
microscopic channel.  At the present-epoch normalization used above, let
\begin{equation}
 n_b\simeq0.251\,\mathrm{m}^{-3},
 \qquad
 \rho_\gamma\simeq4.17\times10^{-14}\,\mathrm{J\,m}^{-3}.
 \label{eq:joint-ledger-inputs}
\end{equation}
In the minimal fixed-radius recycling balance, the horizon-crossing and return
currents are both $3Hn_b$, while maintenance of a constant blackbody energy
density requires power $4H\rho_\gamma$.  The radiative energy required per
horizon-crossing returned baryon is therefore
\begin{equation}
 \epsilon_{\gamma/B}
 :=\frac{4H\rho_\gamma}{3Hn_b}
 =\frac{4\rho_\gamma}{3n_b}
 \simeq1.38\,\mathrm{MeV}.
 \label{eq:radiative-energy-per-returned-baryon}
\end{equation}
The cancellation of $H$ makes this a stoichiometric comparison independent of
the expansion rate.

Consider, conditionally, a return event in which a helium mass fraction
$Y_4=0.245$ is assembled during the event rather than delivered as already
bound helium.  With the $^4\mathrm{He}$ nuclear binding energy
$B_4=28.296\,\mathrm{MeV}$ \cite{WangEtAl2021AME2020}, the binding-energy
ceiling per returned baryon is
\begin{equation}
 \epsilon_{4/B}^{\rm bind}
 =Y_4\frac{B_4}{4}
 \simeq1.73\,\mathrm{MeV}.
 \label{eq:helium-binding-ceiling-per-baryon}
\end{equation}
Thus the common ledger would require the fraction
\begin{equation}
 \eta_{\gamma}^{\rm ledger}
 :=\frac{\epsilon_{\gamma/B}}
 {\epsilon_{4/B}^{\rm bind}}
 \simeq0.80
 \label{eq:joint-ledger-efficiency}
\end{equation}
of this ceiling to appear in Planck-family radiation.  At
$T_0=2.72548\,\mathrm K$, the blackbody photon density is
\begin{equation}
 n_\gamma
 =\frac{2\zeta(3)}{\pi^2}
 \left(\frac{k_BT_0}{\hbar c_{\rm light}}\right)^3
 \simeq4.11\times10^8\,\mathrm{m}^{-3},
 \qquad
 \frac{n_\gamma}{n_b}\simeq1.64\times10^9.
 \label{eq:joint-ledger-photon-yield}
\end{equation}

Equations~\eqref{eq:radiative-energy-per-returned-baryon}--
\eqref{eq:joint-ledger-efficiency} show only that the proposed joint ledger is
not excluded by its gross energy budget.  The binding energy is an upper
bound: weak conversion, neutrino escape, kinetic losses, and incomplete
thermalization reduce the available electromagnetic energy, while helium
returned in an already bound state releases none of it.  The event must also
produce the isotope-resolved composition required by
Eq.~\eqref{eq:required-return-composition} and the enormous photon yield in
Eq.~\eqref{eq:joint-ledger-photon-yield} through number-changing reactions.
The nominal difference $1.73-1.38=0.35\,\mathrm{MeV}$ is therefore only an
ideal one-pass chemical margin.
Neither $Y_4$, the efficiency, nor a Planck-preserving reaction network is
derived here.  Moreover, the displayed binding energy is a one-pass conversion
scale, not automatically renewable fuel.  If the same isotope mixture leaves
and returns in a closed stationary cycle, dissociation or weak conversion on
the export leg pays back the energy released on reassembly.  A positive net
nuclear contribution requires the complementary sector to return the
reactants in a higher-free-energy state, and the cost of preparing that state
belongs to the horizon ledger below.  The calculation is therefore a
budget-compatibility test; prediction of the baryon abundance and identification
of an independent power source require the microscopic return law.

\subsection{Fixed-radius horizon recycling}
\label{subsec:fixed-radius-recycling}

The $3Hn_B$ term has a direct control-volume interpretation.  Work first in a
spatially flat homogeneous chart with areal radius $R=a(t)r$.  Let a spherical
observer-accessible domain have boundary $R=R_H(t)$ and physical volume
$V_H=4\pi R_H^3/3$.  The comoving baryon fluid crosses that boundary with
outward relative speed $HR_H-\dot R_H$.  Its geometric number flux is therefore
\begin{equation}
 \Phi_{B}^{H}
 =4\pi R_H^2n_B\bigl(HR_H-\dot R_H\bigr)
 =3V_Hn_B\left(H-\frac{\dot R_H}{R_H}\right).
 \label{eq:moving-horizon-baryon-flux}
\end{equation}
This is the integral form of the $3Hn_B$ dilution term, not an additional local
loss.  In particular, for the constant-physical-radius horizon used in the
stationary picture,
\begin{equation}
 \boxed{\frac{\Phi_B^H}{V_H}=3Hn_B.}
 \label{eq:fixed-horizon-baryon-flux}
\end{equation}

Let $\Gamma_B^{\rm ret}$ be the baryon return rate and
$\Phi_B^{\rm extra}$ any genuinely additional loss, both per physical volume.
Integrating Eq.~\eqref{eq:baryon-balance} over the fixed domain gives
\begin{equation}
 \frac{\dot N_B}{V_H}
 =\Gamma_B^{\rm ret}-\Phi_B^{\rm extra}
  -\frac{\Phi_B^H}{V_H}.
 \label{eq:fixed-volume-baryon-balance}
\end{equation}
Stationarity thus requires
\begin{equation}
 \boxed{\Gamma_B^{\rm ret}=3Hn_B+\Phi_B^{\rm extra}.}
 \label{eq:fixed-radius-return-rate}
\end{equation}
In the minimal model $\Phi_B^{\rm extra}=0$, so one baryon returned for every
baryon carried across the fixed-radius horizon maintains the exterior density:
\begin{equation}
 \boxed{\Gamma_B^{\rm ret}=\frac{\Phi_B^H}{V_H}=3Hn_B.}
 \label{eq:one-for-one-horizon-recycling}
\end{equation}
Counting the same horizon crossing once in $3Hn_B$ and again in
$\Phi_B^{\rm extra}$ would double-count the loss.

Conservation alone permits the complementary sector to accumulate the exported
baryons.  The one-for-one conclusion follows when its baryon charge is also
stationary.  With the signs above,
\begin{equation}
 \dot B_{\rm comp}
 =\Phi_B^H+V_H\Phi_B^{\rm extra}-V_H\Gamma_B^{\rm ret};
 \qquad
 \langle\dot B_{\rm comp}\rangle=0
 \Longrightarrow
 \Gamma_B^{\rm ret}=\frac{\Phi_B^H}{V_H}+\Phi_B^{\rm extra}.
 \label{eq:complementary-baryon-balance}
\end{equation}
Additional matched export--return cycles leave the mean density unchanged but
increase the gross event rate; they may therefore affect opacity or pump noise
without changing the minimal stoichiometric current.

The radiation comparison in
Eq.~\eqref{eq:radiative-energy-per-returned-baryon} is accordingly the energy
per minimal horizon-crossing return event,
\begin{equation}
 \epsilon_{\gamma/B}
 =\frac{4H\rho_\gamma}{\Gamma_B^{\rm ret}}
 =\frac{4\rho_\gamma}{3n_B}
 \qquad(\Phi_B^{\rm extra}=0).
 \label{eq:radiative-energy-per-horizon-baryon}
\end{equation}
Let $\overline m_B$ be the mean rest mass per baryon of the maintained bound
mixture.  At fixed composition,
\begin{equation}
 \rho_b=n_B\overline m_Bc_{\rm light}^2,
 \qquad
 \frac{P_{B,H}^{\rm rest}}{V_H}
 =3Hn_B\overline m_Bc_{\rm light}^2
 =3H\rho_b.
 \label{eq:horizon-rest-energy-ledger}
\end{equation}
Thus the outward and returned baryonic rest-energy streams cancel in a matched
cycle.  The dust term is the energy carried across the fixed boundary, not a
second rest-mass creation cost.

\begin{definition}[Extra tile energy]
\label{def:extra-tile-energy}
Let $P_{\rm tile}^{\rm extra}\geq0$ denote power stored or dissipated in a
physical gravitational excitation created by an insertion event, beyond the
matter and radiation terms already present in
Eqs.~\eqref{eq:energy-balance}--\eqref{eq:radiation-balance}.  Its debit per
minimal returned baryon is
\begin{equation}
 \epsilon_{\rm tile}^{\rm extra}
 :=\frac{P_{\rm tile}^{\rm extra}}{\Gamma_B^{\rm ret}}.
 \label{eq:extra-tile-energy}
\end{equation}
\end{definition}

If tile insertion is only the microscopic name for the ordinary expansion
whose dilution terms appear in those continuity equations, then by definition
\begin{equation}
 \boxed{\epsilon_{\rm tile}^{\rm extra}=0.}
 \label{eq:kinematic-tile-zero}
\end{equation}
The recycled baryonic rest-energy transfer accounts for the dust flux, while
Eq.~\eqref{eq:radiative-energy-per-horizon-baryon} accounts for the radiation
term.  Adding
a positive energy density times newly represented volume would count those
balances a second time.  If insertion instead creates a new propagating or
stored gravitational degree of freedom, then
$P_{\rm tile}^{\rm extra}$ must be derived from a canonical or quasi-local
boundary charge with stated boundary conditions.  The Einstein--Hilbert
action assigned to a region is not, by itself, such an energy.

\subsection{Entropy-separated horizon return}
\label{subsec:entropy-separated-return}

The fixed-radius identity determines the baryon current but not the quality of
the returned energy.  The intended thermodynamic closure is that conserved
charges nearly cycle while entropy need not.

\begin{hypothesis}[Charge-cyclic, entropy-open horizon return]
\label{hyp:charge-cyclic-entropy-open}
Over the stationary epoch, the complementary sector does not secularly store
baryon number and does not supply power by an undeclared secular depletion of
its mechanical horizon charges.  The common export--return process may,
however, leave positive entropy at the horizon together with the generalized
heat required by a specified horizon first law.  Returned baryons and useful
work are carried by the low-entropy part of the same energy current.
\end{hypothesis}

Normalize all entries below by the minimal return rate
$\Gamma_B^{\rm ret}=3Hn_B$.  Let $\epsilon_{\rm exp}$ be the generalized energy
carried outward per crossing baryon, $\epsilon_{B,\rm ret}$ the energy carried
back in the returned baryonic mixture, $\epsilon_H^{\rm draw}$ any declared
mechanical horizon-charge draw, $q_H$ the generalized heat retained at the
horizon, and $\epsilon_\nu,\epsilon_{\rm irr}\geq0$ neutrino and other
irreversible losses.  Conservation for one complete common event gives
\begin{equation}
 \boxed{
 \epsilon_{\rm exp}+\epsilon_H^{\rm draw}
 =\epsilon_{B,\rm ret}
  +\epsilon_{\gamma/B}
  +\epsilon_{\rm tile}^{\rm extra}
  +\epsilon_\nu+\epsilon_{\rm irr}+q_H.}
 \label{eq:complete-free-energy-ledger}
\end{equation}
This is a balance law, not a claim that an ordinary classical horizon supplies
the return channel.

Write
\begin{equation}
 \epsilon_{\rm exp}=\overline m_{B,\rm exp}c_{\rm light}^2+x_{\rm exp},
 \qquad
 \epsilon_{B,\rm ret}=\overline m_{B,\rm ret}c_{\rm light}^2+x_{\rm ret},
 \label{eq:rest-excess-energy-decomposition}
\end{equation}
where the mean masses include the nuclear mass defects of the specified
mixtures and $x$ contains kinetic, excitation, chemical, and other organized
energy not already included there.  If the isotope mixture is the same on
the two legs, the rest-mass terms cancel and Eq.~\eqref{eq:complete-free-energy-ledger}
becomes
\begin{equation}
 \boxed{
 x_{\rm exp}-x_{\rm ret}+\epsilon_H^{\rm draw}
 =\epsilon_{\gamma/B}+\epsilon_{\rm tile}^{\rm extra}
  +\epsilon_\nu+\epsilon_{\rm irr}+q_H.}
 \label{eq:matched-cycle-free-energy-ledger}
\end{equation}
If the compositions differ, their mass difference remains explicitly in
Eq.~\eqref{eq:complete-free-energy-ledger}.  In particular, the
$1.73\,\mathrm{MeV}$ helium ceiling is not an additional term in
Eq.~\eqref{eq:matched-cycle-free-energy-ledger} for a closed
composition-preserving cycle.  It is available on reassembly only if the
export or horizon step has already paid to prepare the higher-free-energy
reactants.

For a stationary black-hole horizon, the first law has the schematic form
\begin{equation}
 \delta E_H-\Omega_H\delta J_H^{\rm ang}-\Phi_H\delta Q_H
 =T_H\delta S_H
 \label{eq:black-hole-first-law}
\end{equation}
\cite{BardeenCarterHawking1973}, where $J_H^{\rm ang}$ is the horizon angular
momentum.  Once the mechanical charge increments and
orientation are separated, a quasistatic local-detailed-balance calibration
identifies the retained heat per event as
\begin{equation}
 \boxed{q_H=T_H\Delta s_H.}
 \label{eq:horizon-heat-calibration}
\end{equation}
For a cosmological horizon, the sign and thermodynamic energy depend on the
chosen static-patch ensemble and boundary normalization
\cite{GibbonsHawking1977,BanihashemiJacobsonSveskoVisser2023}.  Equation
\eqref{eq:horizon-heat-calibration} is therefore conditional on a declared
ensemble; the construction does not identify the black-hole and cosmological
horizon ledgers.

It is useful to isolate the energetic implication of entropy separation.
Define the return-availability fraction
\begin{equation}
 \eta_{\rm avail}
 :=1-
 \frac{T_H\Delta s_H+\epsilon_\nu+\epsilon_{\rm irr}}
 {\epsilon_{\rm exp}+\epsilon_H^{\rm draw}}.
 \label{eq:return-availability-fraction}
\end{equation}
When the ratio subtracted in Eq.~\eqref{eq:return-availability-fraction} is
small, nearly all generalized energy is available for low-entropy returned
matter, radiation maintenance, or declared tile work.  This is the precise
sense in which a cold entropy sink can support an almost entirely high-exergy
return.  The quantity is a cycle availability, not an environment-independent
thermodynamic potential; the chemical reference data are included in the
$x$-terms of Eq.~\eqref{eq:rest-excess-energy-decomposition}.
The Planck-family photon term is a declared thermal output, so it is one place
where part of that availability is deliberately degraded rather than returned
as organized work.

A large entropy gain alone does not prove that
$T_H\Delta s_H\ll\epsilon_{\rm exp}$.  Ordinary absorption can instead have
$T_H\Delta s_H$ comparable to the entire absorbed energy and return no work.
The desired process must divert organized energy into the return channel
before thermalization, leaving only the entropy-bearing heat at the horizon.
Hypothesis~\ref{hyp:charge-cyclic-entropy-open} asserts this kind of converter;
it is not a consequence of the horizon area law by itself.

The same separation can be made directly in full counting statistics.  Give
each microscopic transition $x\to y$ an oriented ledger mark $z_{xy}^A$, a
baryon-return mark $b_{xy}$, and an energy increment $q_{xy}$ into the
accessible sector.  Tilt its rate by
\begin{equation}
 k_{x\to y}^{\bschi,\chi_B,\chi_E}
 =k_{x\to y}
 \exp\!\left[
 \ii\chi_Az_{xy}^A+\ii\chi_Bb_{xy}+\ii\chi_Eq_{xy}
 \right].
 \label{eq:joint-ledger-energy-tilt}
\end{equation}
If $\lambda_0(\bschi,\chi_B,\chi_E)$ is the dominant eigenvalue, then
\begin{equation}
 J_B^{\rm ret}
 =\left.\partial_{(\ii\chi_B)}\lambda_0\right|_0,
 \qquad
 P_E
 =\left.\partial_{(\ii\chi_E)}\lambda_0\right|_0.
 \label{eq:baryon-energy-currents}
\end{equation}
In the minimal fixed-radius model, the same generator must produce
$J_B^{\rm ret}=\Gamma_B^{\rm ret}=3Hn_B$ and an energy current satisfying
Eq.~\eqref{eq:complete-free-energy-ledger}; neither follows from the ledger
curvature.

\begin{proposition}[Entropy affinity does not select energy current]
\label{prop:energy-current-underdetermination}
Fix an ergodic finite Markov generator and its stationary law $P^*$.  Suppose
at least one stationary edge current is nonzero.  Holding all transition rates,
ledger marks, and baryon marks fixed does not determine the mean energy current.
In particular, the horizon affinity $\mathcal A_H$, winding current $J_H$, and
entropy-production identity introduced by the autonomous phase ring in
Section~\ref{subsec:autonomous-horizon-pump} can remain unchanged while $P_E$
varies.
\end{proposition}

\begin{proof}
For an antisymmetric energy mark $q_{xy}=-q_{yx}$, differentiation of the
tilted eigenvalue gives
\begin{equation}
 P_E[q]
 =\frac12\sum_{x,y}
 \left(P_x^*k_{x\to y}-P_y^*k_{y\to x}\right)q_{xy}.
 \label{eq:stationary-energy-current}
\end{equation}
The stationary law, edge currents, cycle affinities, and entropy production
depend only on the untwisted rates.  Replacing a non-gradient energy mark by
$cq$ therefore rescales $P_E$ without changing any of them.  A gradient mark
$q_{xy}=u_y-u_x$ has zero stationary mean by current conservation, so only the
unfixed cycle component matters.  Thus the rate affinity alone cannot supply
an energy calibration.
\end{proof}

A physical calibration requires local detailed balance,
\begin{equation}
 \log\frac{k_{x\to y}}{k_{y\to x}}
 =\frac{\Delta s_{{\rm env},xy}}{k_B},
 \qquad
 \mathcal A_H
 =\sum_{\rm winding}\frac{\Delta s_{{\rm env},xy}}{k_B},
 \label{eq:horizon-local-detailed-balance}
\end{equation}
together with a decomposition of
$\Delta s_{{\rm env},xy}$ into heat, chemical, and mechanical horizon
increments \cite{Seifert2012StochasticThermodynamics}.  A single equilibrium
bath with no other free-energy bias gives zero cycle affinity.  Hence the
horizon export channel in Hypothesis~\ref{hyp:horizon-powered-ledger} must
represent the complete nonequilibrium boundary process---incoming
stress-energy, any mechanical charge draw, the return channel, and rejected
heat---rather than Hawking temperature alone.

\begin{corollary}[Conditional horizon-heat calibration]
\label{cor:horizon-heat-calibration}
Suppose the entropy production of the autonomous phase ring in
Section~\ref{subsec:autonomous-horizon-pump} is physically deposited at a
horizon of temperature
$T_H=\hbar H/(2\pi k_B)$, and suppose $J_H$ and
$\Gamma_B^{\rm ret}$ have been mapped to the same rate density.  Define the
number of horizon windings per returned baryon by
\begin{equation}
 \mathcal N_{H/B}:=\frac{J_H}{\Gamma_B^{\rm ret}}.
 \label{eq:horizon-windings-per-baryon}
\end{equation}
Then the retained horizon heat per returned baryon is
\begin{equation}
 \boxed{
 \epsilon_{H/B}^{\rm heat}
 =\frac{\hbar H}{2\pi}\,\mathcal A_H\mathcal N_{H/B}.}
 \label{eq:horizon-heat-per-returned-baryon}
\end{equation}
\end{corollary}

\begin{proof}
Local detailed balance gives entropy production rate
$k_B\mathcal A_HJ_H$.  Multiply by $T_H$, divide by
$\Gamma_B^{\rm ret}$, and substitute the stated horizon temperature.
\end{proof}

For the finite-ring value $\mathcal A_H=1.8$ in
Eq.~\eqref{eq:autonomous-horizon-reference}, the normalization
$\mathcal N_{H/B}=1$, and $H=H_0$ from
Eq.~\eqref{eq:helium-audit-inputs}, Eq.~\eqref{eq:horizon-heat-per-returned-baryon}
is only $4.12\times10^{-34}\,\mathrm{eV}$ per returned baryon.  The small
number illustrates the cold-sink mechanism but is not yet a prediction:
$\mathcal N_{H/B}$ is precisely the missing physical rate map between the
finite pump and a baryon-return event.  A macroscopic entropy mark or many
windings per baryon rescales the result.

The finite autonomous pump of
Section~\ref{subsec:autonomous-horizon-pump} supplies the event clock and
entropy affinity but deliberately leaves $b_{xy}$, $q_{xy}$,
$\mathcal N_{H/B}$, and Eq.~\eqref{eq:horizon-local-detailed-balance}
unspecified.  Without Hypothesis~\ref{hyp:charge-cyclic-entropy-open},
accessible-sector stationarity may coexist with secular baryon or mechanical
energy change in the complementary sector.  Under that hypothesis, conserved
charge and nearly all generalized energy return while horizon entropy and its
small calibrated heat debit may grow.

\section{Counted-current holonomy and the step-two tangent}
\label{sec:bsn-geometry}

\subsection{Tilted generators and the first-cumulant connection}

Let $\cU$ be a smooth manifold of slowly varying visible controls, with local
coordinates $U=(U^1,\ldots,U^m)$.  The unresolved sector is described, at
fixed $U$, by an ergodic Markov generator $L^0(U)$.  Let $N^\alpha_t$ count a
set of microscopic channel currents and choose a linear projection to a finite
dimensional ledger space $V_2$,
\begin{equation}
  \dd Z^A=M^A{}_{\alpha}\,\dd N^\alpha .
  \label{eq:ledger-projection}
\end{equation}
The projection $M$ is part of the coarse-graining data; counting statistics do
not identify a spatial bivector without such a map.

Introduce counting fields $\bschi=(\chi_A)$ for the ledger and a separate
field $\eta$ for a positive scalar export observable.  The tilted generator is
denoted $L^{\bschi,\eta}(U)$.  Near $(\bschi,\eta)=(0,0)$, assume its dominant
eigenvalue is simple and choose right and left eigenvectors
$r_0^{\bschi,\eta}$ and $\ell_0^{\bschi,\eta}$ with
\begin{equation}
 \langle\ell_0^{\bschi,\eta},r_0^{\bschi,\eta}\rangle=1.
\end{equation}
At $\eta=0$ the BSN connection and curvature are
\begin{align}
 A_i^{\bschi}(U)
 &=\left\langle \ell_0^{\bschi,0},\partial_i r_0^{\bschi,0}\right\rangle,
 \label{eq:bsn-connection}\\
 F_{ij}^{\bschi}
 &=\partial_iA_j^{\bschi}-\partial_jA_i^{\bschi}.
 \label{eq:bsn-curvature}
\end{align}
At zero ledger counting field the normalization of a probability vector makes
the ordinary curvature vanish.  The physical geometric first cumulant is the
derivative
\begin{equation}
 \cK^A_{ij}(U)
 :=\left.
 \frac{\partial F_{ij}^{\bschi}(U)}{\partial(\ii\chi_A)}
 \right|_{\bschi=0}.
 \label{eq:first-cumulant-curvature}
\end{equation}
For an adiabatic control loop $\Gamma=\partial\Sigma$,
\begin{equation}
 \E\Delta Z^A_{\mathrm{geom}}
 =\int_\Sigma \cK^A.
 \label{eq:geometric-ledger}
\end{equation}
This is the counted-current analogue of a Berry curvature.  It is independent
of the parametrization of the slow loop and odd under reversal of its
orientation.

Differentiate the connection itself at zero counting field:
\begin{equation}
 \mathscr A_i^A(U)
 :=\left.
 \frac{\partial A_i^{\bschi}(U)}{\partial(\ii\chi_A)}
 \right|_{\bschi=0}.
 \label{eq:first-cumulant-connection}
\end{equation}
Then
\begin{equation}
 \partial_i\mathscr A_j^A-\partial_j\mathscr A_i^A=\cK^A_{ij}.
 \label{eq:first-cumulant-curvature-identity}
\end{equation}
A change of counting-field gauge changes $\mathscr A^A$ by an exact one-form
and leaves $\cK^A$ unchanged.

\subsection{A concrete open ring}
\label{subsec:open-ring-anchor}

Before pulling the construction into space, it is useful to see every object
in one finite generator.  Consider three states $0,1,2$ on the oriented ring
$0\to1\to2\to0$.  For an oriented edge $i\to j$, take
\begin{align}
 k_{i\to j}&=c\exp\!\left(E_i-W_{ij}+\frac{a_{ij}}2\right),
 \label{eq:ring-forward}\\
 k_{j\to i}&=c\exp\!\left(E_j-W_{ij}-\frac{a_{ij}}2\right).
 \label{eq:ring-backward}
\end{align}
The node variables $E_i$ and symmetric barriers $W_{ij}$ are controls; the
antisymmetric $a_{ij}$ maintain the nonequilibrium affinity
\begin{equation}
 \mathcal A=a_{01}+a_{12}+a_{20}.
 \label{eq:ring-affinity}
\end{equation}
Set $c=1$ and choose
\begin{align}
 (E_0,E_1,E_2)&=(0.2,-0.1,0),\\
 (W_{01},W_{12},W_{20})&=(0.4,0.2,0.3),\\
 (a_{01},a_{12},a_{20})&=(0.7,0.6,0.5).
 \label{eq:ring-base-point}
\end{align}
The six directed rates are then
\begin{equation}
 (k_{01},k_{10};k_{12},k_{21};k_{20},k_{02})
 \simeq(1.16183,0.427415;1,0.606531;0.951229,0.704688).
 \label{eq:ring-six-rates}
\end{equation}
Solving $Lp^*=0$ with $\mathbf1^{\mathsf T}p^*=1$ gives
\begin{equation}
 p^*\simeq(0.268521,0.371515,0.359964),
 \label{eq:ring-stationary}
\end{equation}
and the common oriented edge current is $J\simeq0.153185$.  Since
$\mathcal A=1.8$, the scalar stationary export is
\begin{equation}
 \Phi_{\rm exp}=J\mathcal A\simeq0.275734>0.
 \label{eq:ring-export}
\end{equation}

Place the ledger field on the $0\leftrightarrow1$ current and vary the mixed
control pair $(E_0,W_{01})$.  The calculation can be done without choosing a
smooth eigenvector gauge.  If $L^\#$ is the group inverse of $L$ and
\begin{equation}
 j^{\mathsf T}
 :=\mathbf1^{\mathsf T}
 \left.\partial_{(\ii\chi)}L^\chi\right|_{\chi=0},
 \qquad
 \mathscr A_i^{\rm ex}=-j^{\mathsf T}L^\#\partial_i p^*,
 \label{eq:ring-group-inverse-connection}
\end{equation}
then the curl of $\mathscr A^{\rm ex}$ agrees with the spectral
first-cumulant curvature, with the sign fixed by the chosen current
orientation.  Direct differentiation gives
\begin{equation}
 \cK_{E_0W_{01}}\simeq-0.0302433,
 \qquad
 \gamma_{\rm rel}\simeq2.42585,
 \qquad
 \frac{\gamma_{\rm rel}\cK_{E_0W_{01}}}{\Phi_{\rm exp}}
 \simeq-0.266074.
 \label{eq:ring-xi}
\end{equation}

This one ring already displays the three quantities used below: an oriented
geometric ledger, positive scalar throughput, and a relaxation clock.  It also
shows why none can be substituted for another.  The curvature is a two-form on
the control manifold, whereas the export and gap are scalars.  What is still
missing is a map from the control-space two-form to resolved spatial transport;
that is the purpose of the pullback.

\subsection{An autonomous horizon-export pump}
\label{subsec:autonomous-horizon-pump}

The preceding BSN loop is an open-system response calculation with externally
prescribed controls.  By itself it does not say what makes those controls wind.
For the cosmological application we impose the following stronger physical
identification.

\begin{hypothesis}[Horizon-powered ledger]
\label{hyp:horizon-powered-ledger}
The observer-relative horizon export channel is the sole nonconservative
affinity that powers the infrared-ledger cycle.  Its internal export phase
supplies the orientation and cadence of a common event, while the event
modulates direction-resolved transition weights and barriers.  The
complementary sector carries the associated energy, charge, and entropy
debits; a passive partial trace is not a pump.  Hypothesis
\ref{hyp:charge-cyclic-entropy-open} separately requires the conserved charge
and usable-energy parts to re-enter the accessible sector while the horizon
retains entropy-bearing heat.
\end{hypothesis}

Here is a finite-state witness for this hypothesis.  Adjoin to the chart state
$i\in\{0,1,2\}$ a horizon phase $a\in\mathbb Z_N$, with
\begin{equation}
 \phi_a=\frac{2\pi a}{N}.
 \label{eq:horizon-phase-grid}
\end{equation}
At fixed $a$, take the chart energies and symmetric barriers
\begin{align}
 (E_0^{(a)},E_1^{(a)},E_2^{(a)})
 &=\bigl(0.2+r_E\cos\phi_a,-0.1,0\bigr),
 \label{eq:autonomous-chart-energies}\\
 (W_{01}^{(a)},W_{12}^{(a)},W_{20}^{(a)})
 &=\bigl(0.4+r_W\sin\phi_a,0.2,0.3\bigr).
 \label{eq:autonomous-chart-barriers}
\end{align}
For every unoriented chart edge $\{i,j\}$, define
\begin{equation}
 k_{i\to j}^{(a)}=\exp\!\bigl(E_i^{(a)}-W_{ij}^{(a)}\bigr).
 \label{eq:autonomous-chart-rates}
\end{equation}
The conditional chart is therefore detailed balanced at every frozen phase,
with
\begin{equation}
 \pi_i^{(a)}=\frac{\exp(-E_i^{(a)})}{Z_a},
 \qquad
 \pi_i^{(a)}k_{i\to j}^{(a)}
 =\pi_j^{(a)}k_{j\to i}^{(a)}
 =\frac{\exp(-W_{ij}^{(a)})}{Z_a}.
 \label{eq:frozen-chart-detailed-balance}
\end{equation}
Thus the directional current vanishes at every frozen value of the horizon
phase.  The sine--cosine offset nevertheless encloses a mixed energy--barrier
area.

The horizon edge $(a,i)\leftrightarrow(a+1,i)$ has one symmetric barrier
\begin{equation}
 B_{a,i}=B_H+E_i^{(a)}
 \label{eq:horizon-edge-barrier}
\end{equation}
and forward and reverse rates
\begin{align}
 h_{a,i}^{+}
 &=\kappa\exp\!\left(
 E_i^{(a)}-B_{a,i}+\frac{\mathcal A_H}{2N}\right),
 \label{eq:horizon-forward-rate}\\
 h_{a+1,i}^{-}
 &=\kappa\exp\!\left(
 E_i^{(a+1)}-B_{a,i}-\frac{\mathcal A_H}{2N}\right).
 \label{eq:horizon-reverse-rate}
\end{align}
The forward rate is independent of the chart state.  More importantly,
\begin{equation}
 \log\frac{h_{a,i}^{+}}{h_{a+1,i}^{-}}
 =E_i^{(a)}-E_i^{(a+1)}+\frac{\mathcal A_H}{N}.
 \label{eq:horizon-local-affinity}
\end{equation}
The energy terms telescope around the phase ring, so one positive winding has
affinity $\mathcal A_H$.

Let $P_{a,i}$ be the unique stationary law of the joint $3N$-state process.
The net horizon winding current through phase cut $a$ and the aggregate chart
current are
\begin{align}
 J_H^{(a)}
 &:=\sum_i\left(
 P_{a,i}h_{a,i}^{+}-P_{a+1,i}h_{a+1,i}^{-}\right)=J_H,
 \label{eq:horizon-winding-current}\\
 J_{ij}
 &:=\sum_a\left(
 P_{a,i}k_{i\to j}^{(a)}-P_{a,j}k_{j\to i}^{(a)}\right).
 \label{eq:aggregate-chart-current}
\end{align}
Stationarity makes $J_H^{(a)}$ independent of the cut and gives
$J_{01}=J_{12}=J_{20}=:J_Z$ for the oriented chart ring.

\begin{proposition}[Autonomous horizon-powered BSN limit]
\label{prop:autonomous-horizon-pump}
For the joint generator in
Eqs.~\eqref{eq:autonomous-chart-rates}--\eqref{eq:horizon-reverse-rate}:
\begin{enumerate}[label=(\roman*)]
\item if $\mathcal A_H=0$, the full process satisfies detailed balance with
\begin{equation}
 P_{a,i}^{(0)}=\frac{\exp(-E_i^{(a)})}
 {\sum_{b,j}\exp(-E_j^{(b)})};
 \label{eq:joint-horizon-equilibrium}
\end{equation}
\item for $\mathcal A_H\neq0$, the horizon winding is the only
non-gradient thermodynamic cycle and the stationary entropy production is
\begin{equation}
 \boxed{\dot S_{\rm prod}=\mathcal A_HJ_H>0};
 \label{eq:horizon-entropy-production}
\end{equation}
\item suppose the frozen chart has a uniform relaxation gap
$\gamma_{\rm rel}>0$.  For fixed nonzero $\mathcal A_H$, first take the
phase motion slow relative to this gap and then refine the phase ring.  The
ledger written per net horizon winding obeys
\begin{equation}
 \lim_{N\to\infty}\lim_{\kappa/\gamma_{\rm rel}\to0}
 \frac{J_Z}{J_H}
 =\oint_\Gamma\mathscr A^{01}
 =\int_\Sigma\cK^{01},
 \qquad \Gamma=\partial\Sigma.
 \label{eq:autonomous-to-bsn-limit}
\end{equation}
\end{enumerate}
Hence horizon export can power a frozen-current-free geometric ledger without
an external protocol.
\end{proposition}

\begin{proof}
Equation~\eqref{eq:frozen-chart-detailed-balance} proves detailed balance on
every chart edge.  At $\mathcal A_H=0$, the unnormalized fluxes on a horizon
edge are also equal:
\begin{equation}
 \e^{-E_i^{(a)}}h_{a,i}^{+}
 =\e^{-E_i^{(a+1)}}h_{a+1,i}^{-}
 =\kappa\e^{-B_H-E_i^{(a)}}.
\end{equation}
This proves (i).  For nonzero affinity, summing
Eq.~\eqref{eq:horizon-local-affinity} around a winding cancels the energy
differences and leaves $\mathcal A_H$.  All chart affinities and all
zero-winding phase loops are gradients.  In the stationary entropy-production
sum, the probability-potential and energy-gradient contributions cancel by
current conservation; only the net winding remains, giving
Eq.~\eqref{eq:horizon-entropy-production}.  Irreducibility and
$\mathcal A_H\neq0$ exclude detailed balance, so the inequality is strict.

Finally, in the slow-phase limit the chart relaxes to $\pi^{(a)}$ between
successive phase changes.  Its first lag is the group-inverse response to
$\pi^{(a+1)}-\pi^{(a)}$, and the counted chart current is the corresponding
discrete first-cumulant connection.  Forward and reverse traversals contribute
with opposite orientation.  Dividing by the net winding current therefore
gives the discrete holonomy per cycle.  As $N\to\infty$, its Riemann sum
converges to $\oint_\Gamma\mathscr A^{01}$; Stokes' theorem and
Eq.~\eqref{eq:first-cumulant-curvature-identity} give the final equality.
\end{proof}

The numerical witness uses
\begin{equation}
 N=3,\quad \kappa=0.02,\quad B_H=1.2,\quad
 r_E=0.45,\quad r_W=0.30,\quad |\mathcal A_H|=1.8.
 \label{eq:autonomous-horizon-reference}
\end{equation}
Every frozen chart current is below $5.6\times10^{-17}$.  The joint stationary
calculation gives:
\begin{table}[H]
\centering
\small
\begin{tabular}{rrrrrr}
\toprule
$\mathcal A_H$ & $r_E$ & $r_W$ & $J_H$ & $J_Z$ & $\dot S_{\rm prod}$\\
\midrule
$ 1.8$ & $0.45$ & $0.30$ & $ 1.21706\times10^{-3}$ &
$ 7.46219\times10^{-6}$ & $2.19070\times10^{-3}$\\
$-1.8$ & $0.45$ & $0.30$ & $-1.21227\times10^{-3}$ &
$-7.47488\times10^{-6}$ & $2.18208\times10^{-3}$\\
$ 0$   & $0.45$ & $0.30$ & $<4\times10^{-17}$ &
$<2\times10^{-16}$ & $1.84\times10^{-30}$\\
$ 1.8$ & $0$    & $0.30$ & $ 1.22293\times10^{-3}$ &
$<9\times10^{-17}$ & $2.20127\times10^{-3}$\\
$ 1.8$ & $0.45$ & $0$    & $ 1.21706\times10^{-3}$ &
$<1\times10^{-16}$ & $2.19072\times10^{-3}$\\
\bottomrule
\end{tabular}
\caption{Autonomous horizon-pump checks.  Reversing the horizon affinity
reverses both currents while entropy production remains positive.  With
either half of the mixed modulation removed, horizon export continues but the
ledger pump vanishes.}
\label{tab:autonomous-horizon-pump}
\end{table}
For the first two rows, $J_Z/J_H=0.006131$ and $0.006166$, respectively, and
Eq.~\eqref{eq:horizon-entropy-production} holds to better than
$3\times10^{-18}$ in absolute units.

As a separate convergence test, set $\kappa=10^{-4}$.  Direct integration of
the continuous BSN connection gives
\begin{equation}
 Q_{\rm BSN}=\oint_\Gamma\mathscr A^{01}=0.0152345324.
 \label{eq:continuous-bsn-value}
\end{equation}
The autonomous ratios are
\begin{center}
\begin{tabular}{rr}
\toprule
$N$ & $J_Z/J_H$\\
\midrule
$3$  & $0.00620058$\\
$6$  & $0.01257781$\\
$12$ & $0.01453393$\\
$24$ & $0.01505362$\\
$48$ & $0.01518726$\\
\bottomrule
\end{tabular}
\end{center}
The $N=48$ value is within $0.31\%$ of
Eq.~\eqref{eq:continuous-bsn-value}.  The stationary laws, edge currents,
entropy production, no-pumping controls, continuous BSN integral, and
refinement table follow directly from
Eqs.~\eqref{eq:autonomous-chart-rates}--\eqref{eq:horizon-reverse-rate}; no
ancillary source file is required.

One chart writes one oriented ledger component.  Three rotated copies with
marks $e_2\wedge e_3$, $e_3\wedge e_1$, and $e_1\wedge e_2$ give a full-rank
three-dimensional witness whenever the common scalar holonomy is nonzero.
An isotropic ensemble of such local frames has zero mean signed direction but
can retain nonzero invariant ledger magnitude.  This establishes the
autonomous mechanism in kind; the spatial control map is still the additional
input treated next.

The proposition establishes the autonomous source of the cycle orientation
and entropy production.  A quantitative horizon model must derive the
dimensionless affinity $\mathcal A_H$ and rate $\kappa$ from a specified
semiclassical exterior algebra, and
their physical energy current must satisfy the baryon and radiation balances
of Section~\ref{sec:balances}.  What is established is narrower and central:
once horizon export supplies the nonconservative affinity, it can generate the
ledger holonomy itself rather than merely accompany or absorb an independently
driven pump.  Proposition~\ref{prop:energy-current-underdetermination} shows
that this entropy affinity does not yet select the energy delivered per
winding.  Under Hypothesis~\ref{hyp:charge-cyclic-entropy-open}, local detailed
balance converts the entropy current into the retained-heat term in
Corollary~\ref{cor:horizon-heat-calibration}; the baryon and energy marks and
the normalization $\mathcal N_{H/B}$ remain the physical inputs identified in
Section~\ref{subsec:entropy-separated-return}.

\subsection{Spatial pullback}

Let $\cX$ be the resolved spatial transport manifold and let
\begin{equation}
 U:\cX\longrightarrow\cU
 \label{eq:control-map}
\end{equation}
assign slow controls to a spatial point.  Pull back the first-cumulant
connection and curvature:
\begin{align}
 a^A&:=U^*\mathscr A^A,
 &a_\mu^A(x)&=\mathscr A_i^A(U(x))\,\partial_\mu U^i(x),
 \label{eq:pulled-connection}\\
 \Omega^A&:=\dd a^A=U^*\cK^A,
 &\Omega_{\mu\nu}^A(x)&=\cK_{ij}^A(U(x))
 \partial_\mu U^i\partial_\nu U^j.
 \label{eq:pulled-curvature}
\end{align}
For a physical loop $\gamma=\partial S$ in $\cX$,
\begin{equation}
 \E\Delta Z^A_{\mathrm{geom}}=\oint_\gamma a^A=\int_S\Omega^A.
 \label{eq:spatial-loop-holonomy}
\end{equation}
On the augmented bundle $\cX\times V_2$, with fiber coordinates $z^A$, define
the horizontal lifts
\begin{equation}
 \widetilde X_\mu
 =\partial_{x^\mu}+a_\mu^A(x)\partial_{z^A}.
 \label{eq:horizontal-lifts}
\end{equation}

\begin{theorem}[BSN pullback and nilpotent tangent]
\label{thm:bsn-pullback}
Assume that the dominant eigenspace of $L^{\bschi,0}(U)$ is simple and smooth
near $\bschi=0$, that $U$ is $C^2$, and that $V_2$ is finite dimensional.
Then:
\begin{enumerate}[label=(\roman*)]
\item the lifted fields satisfy the exact bracket identity
\begin{equation}
 [\widetilde X_\mu,\widetilde X_\nu]
 =\Omega_{\mu\nu}^A\partial_{z^A};
 \label{eq:lifted-bracket}
\end{equation}
\item at $x\in\cX$, the step-two nilpotent approximation has graded Lie
algebra
\begin{equation}
 \mathfrak n_x=T_x\cX\oplus\im\Omega_x,
 \qquad [v,w]_x=\Omega_x(v,w),
 \label{eq:nilpotent-algebra}
\end{equation}
with $\im\Omega_x$ central;
\item every retained central direction is generated at $x$ if and only if
\begin{equation}
 \Omega_x:\Lambda^2T_x\cX\longrightarrow V_2
 \quad\text{is surjective};
 \label{eq:surjective-curvature}
\end{equation}
\item if $\dim\cX=3$, $V_2=\Lambda^2T_x\cX$, and $\Omega_x$ is an
isomorphism, then $\mathfrak n_x$ is isomorphic to the free step-two algebra
\begin{equation}
 \mathfrak n_{3,2}=T_x\cX\oplus\Lambda^2T_x\cX,
 \qquad[v,w]=v\wedge w.
 \label{eq:free-step-two}
\end{equation}
If the identification and $\Omega_x$ are $SO(3)$-equivariant, then
$\Omega_x=c_x\operatorname{id}_{\Lambda^2T_x\cX}$ for a nonzero scalar
$c_x$.
\end{enumerate}
\end{theorem}

\begin{proof}
The coefficients $a_\mu^A$ depend only on base variables and the fiber
derivatives commute.  Therefore
\[
 [\widetilde X_\mu,\widetilde X_\nu]
 =(\partial_\mu a_\nu^A-\partial_\nu a_\mu^A)\partial_{z^A}
 =\Omega_{\mu\nu}^A\partial_{z^A}.
\]
Choose a fiber gauge with $a(x)=0$ and privileged coordinates at $x$.  Under
the parabolic dilation assigning weight one to base coordinates and weight two
to fiber coordinates, the leading bracket is the constant bilinear map
$\Omega_x$.  The fiber directions commute with every leading field, which
gives \eqref{eq:nilpotent-algebra}.  First commutators span exactly
$\im\Omega_x$, proving the rank statement.  In the three-dimensional
full-rank case, use $\Omega_x$ to identify the center with
$\Lambda^2T_x\cX$; the bracket becomes the universal wedge bracket.  Finally,
$\Lambda^2\R^3\simeq\R^3$ is an irreducible real $SO(3)$ representation, so an
equivariant endomorphism is scalar.
\end{proof}

\begin{corollary}[Hypoelliptic augmented diffusion]
\label{cor:hormander}
Suppose that the principal second-order part of the augmented slow generator is
\begin{equation}
 \mathscr L_2=D^{\mu\nu}(x)\widetilde X_\mu\widetilde X_\nu,
 \label{eq:augmented-generator}
\end{equation}
with $D^{\mu\nu}$ positive definite on $T\cX$.  If $\Omega_x$ is surjective,
the lifted fields satisfy H\"ormander's condition at step two.  In the
three-dimensional full-rank case, the nilpotent tangent is the sub-Laplacian on
$N_{3,2}$ up to the horizontal metric and central normalization.
\end{corollary}

This is the step-two instance of the nilpotent approximation underlying
hypoelliptic regularity \cite{Hormander1967,RothschildStein1976}.

Theorem~\ref{thm:bsn-pullback} separates three statements often conflated in
effective models.  Nonzero BSN curvature on $\cU$ does not imply a spatial
bracket unless its pullback is nonzero.  A nonzero spatial bracket need not have
full central rank.  Finally, a bracket tensor does not by itself specify the
second-order diffusion symbol.  In particular,
\begin{equation}
 \cK=0\ \Longrightarrow\ \Omega=0\ \Longrightarrow\
 [\widetilde X_\mu,\widetilde X_\nu]=0.
 \label{eq:no-pump-no-bracket}
\end{equation}
Even if $\cK\neq0$, the pullback can vanish when the image of $U$ has rank less
than two or is tangent to a null two-plane of the curvature.

\subsection{No-pumping restrictions}

In thermally activated stochastic pumps satisfying detailed balance, varying
only well depths at fixed barriers produces no directed adiabatic current
\cite{RahavHorowitzJarzynski2008,ChernyakSinitsyn2008}.  Pure barrier
modulation is likewise degenerate in the minimal two-link example.  A nonzero
ledger requires a mixed cycle.  If $E_a(U)$ are channel weights and
$W_{ab}(U)$ are barriers or conductances, a typical component has the form
\begin{equation}
 \cK^A_{ij}\sim C^{ab}
 \left(\partial_iE_a\,\partial_jW_{ab}
      -\partial_jE_a\,\partial_iW_{ab}\right).
 \label{eq:mixed-curvature}
\end{equation}
Proportional $E$ and $W$ gradients enclose no mixed area.  Independent
gradients can produce a nonzero geometric current.  The ring in
Section~\ref{subsec:open-ring-anchor} is an open example;
Appendix~\ref{app:bsn-model} gives a closed no-pumping control.
The autonomous witness in
Section~\ref{subsec:autonomous-horizon-pump} makes the same restriction
especially transparent: setting either $r_E=0$ or $r_W=0$ leaves
$J_H>0$ but gives $J_Z=0$ to numerical precision.

The physical application therefore requires a generator outside the scalar
no-pumping class.  This condition is sharper than positive entropy export:
one may have a positive scalar throughput and zero oriented ledger, or a closed
geometric pump with no positive stationary export denominator.

\subsection{An intensive susceptibility}

Let $\Phi_{\mathrm{exp}}$ be the separately counted scalar export rate,
\begin{equation}
 \Phi_{\mathrm{exp}}(U)
 :=\left.
 \frac{\partial\lambda_0^{0,\eta}(U)}{\partial(\ii\eta)}
 \right|_{\eta=0}>0,
 \label{eq:export-rate}
\end{equation}
and let $\gamma_{\mathrm{rel}}$ be the smallest nonzero decay rate of the
untwisted generator.  The export accumulated during one correlation time is
\begin{equation}
 s_{\mathrm{corr}}=\frac{\Phi_{\mathrm{exp}}}{\gamma_{\mathrm{rel}}}.
 \label{eq:export-per-correlation}
\end{equation}
After choosing dimensionless controls and ledger units, define the normalized
BSN susceptibility
\begin{equation}
 \Xi^A_{ij}
 :=\frac{\cK^A_{ij}}{s_{\mathrm{corr}}}
 =\frac{\gamma_{\mathrm{rel}}\cK^A_{ij}}
 {\Phi_{\mathrm{exp}}}.
 \label{eq:normalized-xi}
\end{equation}
Equivalently, a control metric $G$ and ledger metric $h$ define the invariant
norm
\begin{equation}
 \|\Xi\|_{G,h}^2
 =\frac12h_{AB}G^{ik}G^{jl}\Xi^A_{ij}\Xi^B_{kl}.
 \label{eq:xi-norm}
\end{equation}
The metrics are constitutive normalization data and are fixed before comparing
different systems.

\begin{proposition}[Intensivity]
\label{prop:intensivity}
Replace the unresolved sector by $N$ independent identical copies and sum the
ledger and scalar export over copies.  If the single-copy generator is
ergodic, then
\begin{equation}
 \cK_N=N\cK,\qquad
 \Phi_{\mathrm{exp},N}=N\Phi_{\mathrm{exp}},\qquad
 \gamma_{\mathrm{rel},N}=\gamma_{\mathrm{rel}},\qquad
 \Xi_N=\Xi.
 \label{eq:replication-law}
\end{equation}
Under a uniform rate rescaling $L^{\bschi,\eta}\mapsto
cL^{\bschi,\eta}$ with $c>0$,
\begin{equation}
 \cK\mapsto\cK,\qquad
 \Phi_{\mathrm{exp}}\mapsto c\Phi_{\mathrm{exp}},\qquad
 \gamma_{\mathrm{rel}}\mapsto c\gamma_{\mathrm{rel}},\qquad
 \Xi\mapsto\Xi.
 \label{eq:rate-rescaling-law}
\end{equation}
Thus $\Xi$ is independent of unresolved-sector size and of a common microscopic
clock rescaling.
\end{proposition}

\begin{proof}
For independent copies, the tilted generator is a Kronecker sum.  Its dominant
right and left eigenvectors are tensor products, so the BSN connection is a sum
of the single-copy connections.  Its curvature and first counting-field
derivative scale by $N$.  Scalar currents add, while the smallest nonzero decay
rate of the Kronecker sum remains the single-copy gap.  This gives
\eqref{eq:replication-law}.

Uniform multiplication of the generator changes no instantaneous eigenvector,
and therefore changes neither the BSN connection nor $\cK$.  Every eigenvalue
and every current per unit time is multiplied by $c$.  The factors cancel in
\eqref{eq:normalized-xi}.
\end{proof}

The proposition does not determine a central response.  It only identifies a
dimensionless intensive object on the positive-throughput branch.  The map from
$\Xi$ to a conjugate central momentum is the subject of the next section.

\section{From a pumped ledger to a central response}
\label{sec:central-selection}

Section~\ref{sec:bsn-geometry} established that a counted current can write an
intensive, bracketed ledger.  It did not establish that the ledger acts as a
force or selects its conjugate central momentum.  This section is the second
half of that story: first isolate the most general isotropic response, then
identify the exact obstruction, and finally exhibit the smallest additional
open-system channel that overcomes it.

\subsection{Protocol contraction and rotational covariance}

In the full-rank spatial case the ledger space is
$V_2=\Lambda^2V$, where $V$ is the oriented three-dimensional spatial tangent.
The susceptibility $\Xi$ still carries two control indices.  A local physical
protocol must supply an oriented control bivector
\begin{equation}
 \Pi_U\in\Lambda^2T_U\cU .
 \label{eq:protocol-bivector}
\end{equation}
The contracted ledger response is
\begin{equation}
 \bsxi_U:=\Xi(U)\mathbin{\lrcorner}\Pi_U\in V_2,
 \qquad
 \xi_U^A=\frac12\Xi^A_{ij}\Pi_U^{ij}.
 \label{eq:contracted-xi}
\end{equation}
It is invariant under a reparametrization of control space and changes sign
when the protocol orientation is reversed.

Once a Euclidean metric $h$ on $V_2$ is fixed, rotational covariance leaves
only one scalar constitutive freedom.

\begin{proposition}[Isotropic local response]
\label{prop:isotropic-response}
Let $\mathfrak C_U:V_2\to V_2^*$ be smooth, odd, and $SO(3)$-equivariant.
Then there is a smooth scalar function $c_U:[0,\infty)\to\R$ such that
\begin{equation}
 \mathfrak C_U(\bsxi)
 =c_U(\|\bsxi\|_h^2)\,\bsxi^{\flat_h}.
 \label{eq:isotropic-normal-form}
\end{equation}
If $\mathfrak C_U$ is linear, $c_U$ is constant.
\end{proposition}

\begin{proof}
For $\bsxi\neq0$, its stabilizer is an $SO(2)$ subgroup.  The only covectors
fixed by this stabilizer lie on the line spanned by $\bsxi^{\flat_h}$.
Equivariance therefore gives a scalar multiple of that covector.  Rotations act
transitively on spheres, so the scalar depends only on $\|\bsxi\|_h^2$.
Oddness makes the coefficient even in $\bsxi$, and smoothness extends the
formula through the origin.
\end{proof}

If completed events occur at rate $\lambda_{\rm ev}$, the corresponding
frequency-valued central response has the form
\begin{equation}
 \bsomega_{\rm eq}
 =\lambda_{\rm ev}\,c_U(\|\bsxi_U\|_h^2)
 \bsxi_U^{\flat_h},
 \qquad \sigma:=\|\bsomega_{\rm eq}\|_h.
 \label{eq:constitutive-central-response}
\end{equation}
Equation~\eqref{eq:constitutive-central-response} is the most general local,
smooth, odd, isotropic closure once $\Pi_U$ is supplied.  It is not yet an
existence theorem: the counted ledger and its conjugate character are
different dynamical objects.

\subsection{Translation alone cannot select a central character}

Let $z\in V_2$ be the central coordinate and
\begin{equation}
 \widehat Q_A=-\ii\frac{\partial}{\partial z^A}
 \label{eq:central-momenta}
\end{equation}
its commuting conjugate momenta.  Suppose a counted jump $a\to b$, carrying
increment $d_{ba}\in V_2$, acts only by translation of the coordinate,
\begin{equation}
 T_{d_{ba}}=\exp(-\ii d_{ba}^A\widehat Q_A),
 \qquad
 J_{ba}=\sqrt{k_{a\to b}}\,|b\rangle\langle a|\otimes T_{d_{ba}}.
 \label{eq:translation-jumps}
\end{equation}
We use the standard completely positive semigroup convention
\cite{GoriniKossakowskiSudarshan1976,Lindblad1976}.

\begin{theorem}[No central selection from translation alone]
\label{thm:no-central-selection}
For the GKSL generator constructed from \eqref{eq:translation-jumps}, with an
arbitrary time-dependent protocol $U(t)$,
\begin{equation}
 \mathscr G_{U(t)}^*\bigl(I\otimes F(\bm Q)\bigr)=0
 \label{eq:q-conservation}
\end{equation}
for every bounded Borel function $F$ of the commuting central momenta.  The
complete central-momentum spectral measure is conserved.  In particular,
nonzero BSN curvature and a nonzero pumped ledger do not select a nonzero
central character from a zero-mean initial sector.
\end{theorem}

\begin{proof}
Each $T_{d_{ba}}$ is a function of the $\widehat Q_A$ and commutes with
$F(\bm Q)$.  In the adjoint GKSL generator the gain term
$J_{ba}^*(I\otimes F)J_{ba}$ is therefore cancelled exactly by the
anticommutator term.  The cancellation holds jump by jump and at every value
of the controls.
\end{proof}

Thus
\begin{equation}
 \Xi\neq0\quad\not\!\Longrightarrow\quad\bsomega_{\rm eq}\neq0.
 \label{eq:xi-not-omega}
\end{equation}
This is the precise point at which the geometric construction stops.  Moving a
central coordinate and selecting a Fourier character of that coordinate are
different operations.  A selector must act on central momentum itself and
must contain a contraction or relaxation channel.

\subsection{An event-gated completely positive selector}
\label{subsec:event-selector}

We now give a finite existence model.  Let $C=\{0,1,2\}$ label three reversible
processing charts and let $\partial$ be a completed-event state.  On $C$ set
\begin{equation}
 k_{i\to j}=c\exp(E_i-W_{ij}),
 \qquad W_{ij}=W_{ji},
 \qquad
 p_i^*=\frac{\e^{-E_i}}{\sum_{a\in C}\e^{-E_a}}.
 \label{eq:chart-rates}
\end{equation}
Fresh records enter in the equilibrium chart mixture and completed records
leave at a chart-independent rate,
\begin{equation}
 k_{\partial\to i}=\lambda p_i^*,
 \qquad k_{i\to\partial}=\mu.
 \label{eq:renewal-rates}
\end{equation}
The state $\partial$ is a renewal reservoir, not the return of the same record.

\begin{proposition}[Throughput without frozen housekeeping current]
\label{prop:event-throughput}
The four-state generator \eqref{eq:chart-rates}--\eqref{eq:renewal-rates} has
stationary law
\begin{equation}
 \pi_\partial^*=\frac{\mu}{\lambda+\mu},
 \qquad
 \pi_i^*=\frac{\lambda}{\lambda+\mu}p_i^*.
 \label{eq:event-stationary-law}
\end{equation}
Every frozen oriented chart current vanishes, while the gross completed-event
activity is positive:
\begin{equation}
 J_{i\to j}^{\rm hk}=0,
 \qquad
 \Phi_{\rm ev}=\sum_i\mu\pi_i^*
 =\frac{\lambda\mu}{\lambda+\mu}>0.
 \label{eq:event-throughput}
\end{equation}
If a completed record exports an amount $S_{\rm reg}$, then
$\Phi_{\rm exp}=S_{\rm reg}\Phi_{\rm ev}$.
\end{proposition}

\begin{proof}
The chart rates satisfy detailed balance with $p^*$, and the renewal edges obey
$\pi_\partial^*\lambda p_i^*=\mu\pi_i^*$.  These identities give the
stationary law and zero frozen chart currents.  Summing the terminal rates
gives \eqref{eq:event-throughput}.
\end{proof}

A mixed loop in an $E$ and a $W$ control has nonzero BSN curvature despite the
vanishing frozen current.  The orientation-odd integrated current converges to
the curvature flux in the adiabatic limit; the orientation-even terminal
transient vanishes as the inverse cycle time.  A numerical instance is given
in Appendix~\ref{app:event-model}.

Introduce a canonical pair
$[\widehat Z^A,\widehat Q_B]=\ii\delta^A{}_B$.  A chart transition carries a
logical-blind Weyl kick
\begin{equation}
 B_d=\exp(\ii g d_A\widehat Z^A),
 \qquad
 B_d^*\widehat Q_A B_d=\widehat Q_A+g d_A.
 \label{eq:weyl-kick}
\end{equation}
Unlike a translation in $Z$, this acts on $Q$ and evades
Theorem~\ref{thm:no-central-selection}.  Let $\tau_0$ be a fixed central state
with vanishing first moment and finite covariance $C_0$.  For
$0\leq\eta_{\rm sel}<1$
define the reset-or-retain channel
\begin{equation}
 \mathcal D_{\eta_{\rm sel}}(\rho)
 =\eta_{\rm sel}\rho+(1-\eta_{\rm sel})\tau_0\Tr\rho.
 \label{eq:reset-channel}
\end{equation}
It is CPTP and
$\mathcal D_{\eta_{\rm sel}}^*(\widehat Q_A)=\eta_{\rm sel}\widehat Q_A$.

Let $r$ be the record accumulated between renewal entry and terminal export,
with probability $p_\Gamma(r)$ under protocol $\Gamma$ and total normalized
increment $d(r)$.  Define
\begin{align}
 \mathcal U_\Gamma(\rho)
 &=\sum_r p_\Gamma(r)B_{d(r)}\rho B_{d(r)}^*,
 \label{eq:random-kick-channel}\\
 \mathcal M_\Gamma
 &=\mathcal U_\Gamma\circ\mathcal D_{\eta_{\rm sel}},
 \qquad
 \overline d_A(\Gamma)=\sum_rp_\Gamma(r)d_A(r).
 \label{eq:completed-event-map}
\end{align}
The adjoint map has the exact affine action
\begin{equation}
 \mathcal M_\Gamma^*(\widehat Q_A)
 =\eta_{\rm sel}\widehat Q_A+g\overline d_A(\Gamma)I.
 \label{eq:affine-selector}
\end{equation}

Suppose independent completed records arrive as a Poisson process of rate
$\lambda_{\rm ev}$.  The central state obeys
\begin{equation}
 \dot\rho=\lambda_{\rm ev}(\mathcal M_\Gamma-I)\rho.
 \label{eq:poisson-selector}
\end{equation}
For the mean $m_A=\Tr(\rho\widehat Q_A)$,
\begin{equation}
 \dot m_A=-\gamma_{\rm sel}m_A
 +\lambda_{\rm ev}g\overline d_A,
 \qquad
 \gamma_{\rm sel}=\lambda_{\rm ev}(1-\eta_{\rm sel}).
 \label{eq:selector-mean}
\end{equation}
Thus
\begin{equation}
 m_A^*=\frac{g}{1-\eta_{\rm sel}}\overline d_A,
 \qquad
 \omega_A^*:=\gamma_{\rm sel}m_A^*
 =\lambda_{\rm ev}g\overline d_A.
 \label{eq:selector-response}
\end{equation}
The retention parameter controls memory and the stored mean but cancels from
the stationary frequency-valued response.

\begin{theorem}[Event-gated selector and BSN limit]
\label{thm:event-selector}
The repeated-interaction process
\eqref{eq:chart-rates}--\eqref{eq:poisson-selector} is completely positive,
has positive completed-record throughput and zero frozen oriented chart
current, and admits a unique stationary central law.  Differences of equal
trace contract at least as
$\exp[-\lambda_{\rm ev}(1-\eta_{\rm sel})t]$, and the stationary mean obeys
\eqref{eq:selector-response}.

For a slow periodic chart protocol $\Gamma_T$, let
\begin{equation}
 \overline{\bm d}_{\rm odd}(\Gamma_T)
 :=\frac12\left[
 \overline{\bm d}(\Gamma_T)-
 \overline{\bm d}(\Gamma_T^{-1})\right].
 \label{eq:odd-record}
\end{equation}
Let $\Sigma_T$ be a spanning surface in control space and use the reference
chart gap and export rate to define
\begin{equation}
 \bsxi_{\rm corr}(\Gamma_T)
 :=\frac1{s_{\rm corr}}\int_{\Sigma_T}\cK,
 \qquad
 s_{\rm corr}:=\frac{\Phi_{\rm exp}}{\gamma_4}.
 \label{eq:loop-normalized-response}
\end{equation}
If the chart increment uses the same ledger and scalar-export normalization,
the adiabatic expansion gives
\begin{equation}
 \overline{\bm d}_{\rm odd}(\Gamma_T)
 =\bsxi_{\rm corr}^{\flat_h}(\Gamma_T)+\bm\varepsilon_T,
 \qquad \bm\varepsilon_T=O((\gamma_4T)^{-1}),
 \label{eq:adiabatic-bsn-identification}
\end{equation}
where $\gamma_4$ is the internal chart gap.  Consequently
\begin{equation}
 \bsomega_{\rm odd}^*
 =\lambda_{\rm ev}g
 \left(\bsxi_{\rm corr}^{\flat_h}+\bm\varepsilon_T\right).
 \label{eq:linear-bsn-bridge}
\end{equation}
For a sufficiently small local cycle with oriented area bivector $\Pi_U$,
$\bsxi_{\rm corr}=\Xi\mathbin{\lrcorner}\Pi_U$ to leading order in the loop
area.
\end{theorem}

\begin{proof}
Proposition~\ref{prop:event-throughput} supplies the stationary chart law and
the current statements.  The reset map is a convex combination of the identity
and a replacement channel, while $\mathcal U_\Gamma$ is random unitary; their
composition is CPTP.  For equal-trace states,
\[
 \|\mathcal M_\Gamma(\rho)-\mathcal M_\Gamma(\rho')\|_1
 \leq\eta_{\rm sel}\|\rho-\rho'\|_1.
\]
Poisson averaging gives the stated contraction and uniqueness.  Conjugation by
$B_d$ shifts $Q$ by $gd$, so \eqref{eq:affine-selector} and
\eqref{eq:selector-response} follow directly.

At frozen controls the chart current vanishes.  The adiabatic current expansion
therefore begins with the BSN first-cumulant curvature.  Loop reversal removes
the orientation-even finite-rate term.  Dividing by scalar export per
correlation time gives \eqref{eq:adiabatic-bsn-identification}, with the stated
standard first-order adiabatic remainder; substitution in the exact affine law
gives \eqref{eq:linear-bsn-bridge}.
\end{proof}

The selector has three conceptually distinct rates.  The internal gap
$\gamma_4$ controls the adiabatic error, $\lambda_{\rm ev}$ fixes the response
amplitude, and $\gamma_{\rm sel}$ fixes central-memory loss.  A spectral gap in
the internal generator cannot be identified with the completed-event cadence
without additional microscopic information.

\subsection{Stationary fluctuations and the selector clock}

Let
\begin{equation}
 D_d=\sum_rp_\Gamma(r)\bm d(r)\bm d(r)^{\mathsf T},
 \qquad
 \Sigma_d=D_d-\overline{\bm d}\,\overline{\bm d}^{\mathsf T}.
 \label{eq:record-moments}
\end{equation}
Direct iteration of the reset--kick channel gives the stationary symmetric
covariance
\begin{equation}
 V_*=C_0+
 \frac{g^2}{1-\eta_{\rm sel}}\Sigma_d+
 \frac{g^2\eta_{\rm sel}}{(1-\eta_{\rm sel})^2}
 \overline{\bm d}\,\overline{\bm d}^{\mathsf T}.
 \label{eq:selector-covariance}
\end{equation}
The centered two-time covariance and spectrum of the continuous-time Poisson
selector are
\begin{align}
 C_Q(t)&=\e^{-\gamma_{\rm sel}|t|}V_*,
 \label{eq:selector-time-covariance}\\
 S_Q(\omega)&=
 \frac{2\gamma_{\rm sel}}
 {\gamma_{\rm sel}^2+\omega^2}V_*.
 \label{eq:selector-lorentzian}
\end{align}
The model therefore derives a matrix-valued central covariance but only a
one-scale Lorentzian temporal factor.

The chart process does not secretly contain a long serial clock.  Since every
chart exports at the same rate $\mu$, its pre-export transient generator is
$Q_C-\mu I$, where $Q_C\bm1=0$.  From any initial chart distribution $\alpha$,
\begin{equation}
 \Pr(T>t)=\alpha\e^{(Q_C-\mu I)t}\bm1=\e^{-\mu t},
 \qquad
 \E T=\mu^{-1},\quad \Var T=\mu^{-2}.
 \label{eq:selector-first-passage}
\end{equation}
Internal chart motion changes the geometric mark written before export, but it
does not narrow the residence-time distribution.  This fact will matter for
the stationary-radiation application.

\subsection{Physical interpretation}

The two finite constructions now separate power from readout.  Proposition
\ref{prop:autonomous-horizon-pump} shows that a horizon-export affinity can
supply the pump clock and write counted-current holonomy without a frozen
chart current or an externally prescribed protocol.  The event-gated selector
then shows that such holonomy can be converted into a stationary central
response without violating complete positivity.  A specified exterior algebra
and its microscopic interaction must supply the horizon event population,
$\mathcal A_H$, the Weyl coupling $g$, the retention factor $\eta_{\rm sel}$,
and the physical direction-resolved rate map.  In particular, the autonomous
finite-state witness establishes a possible thermodynamic architecture, not
the available power of a cosmological horizon.

Statistical isotropy is compatible with a nonzero local scale.  An isotropic
ensemble can satisfy
\begin{equation}
 \E\bsomega=0,
 \qquad
 \E\|\bsomega\|^2>0.
 \label{eq:isotropic-central-ensemble}
\end{equation}
The invariant scalar $\sigma=\|\bsomega\|$ can therefore characterize local
domains without selecting a cosmological direction.  Any observable linear in
a signed component of $\bsomega$, however, requires a physical local
orientation and cannot be recovered by averaging that component to zero.

\section{Galactic trajectories and a screened central response}
\label{sec:galactic-response}

The selected central character has a direct one-body consequence.  A thin
galactic disc supplies a local orientation, its unit normal $\bm n$, and hence
selects the scalar component
\begin{equation}
 \varpi=\langle\bsomega,\bm n\rangle.
 \label{eq:disc-central-component}
\end{equation}
With the group-law convention in Eq.~\eqref{eq:n32-group-law}, the
group-Fourier covector is normalized as $B=2\bsomega$, so equivalently
$\varpi=\tfrac12\langle B,\bm n\rangle$.  The factor is conventional and can be
absorbed into the scalar response coefficient in
Eq.~\eqref{eq:isotropic-normal-form}; stating it fixes the normalization of the
galactic Hamiltonian relative to the later Landau block.
This section first derives the exact orbit kinematics associated with a radial
profile $\varpi(r)$.  It then asks what minimal constitutive equation could
produce the profile required by a flat rotation curve.  The order matters:
the orbit relations are exact within the displayed Hamiltonian, whereas the
bulk response equation and its source normalization are additional physics.

\subsection{Planar reduction and the azimuthal connection}
\label{subsec:planar-reduction}

Under the Hodge identification
$\Lambda^2\R^3\simeq\R^3$, the disc-normal central block of $N_{3,2}$ reduces
to the planar Heisenberg algebra generated by two horizontal translations and
their oriented-area commutator.  A fixed central character therefore acts on
planar motion like an azimuthal connection.  For a radially varying environment
we do not let a conserved coadjoint momentum vary along one orbit.  Instead we
introduce the canonical effective Hamiltonian per unit test mass
\begin{equation}
 H_{\rm eff}(r,\theta;p_r,L)
 =\frac12\left[
 p_r^2+\frac{\bigl(L+\mathcal A_\theta(r)\bigr)^2}{r^2}
 \right]+V(r),
 \qquad
 \mathcal A_\theta(r):=r^2\varpi(r).
 \label{eq:galactic-effective-hamiltonian}
\end{equation}
Here $(r,p_r)$ and $(\theta,L)$ are canonical coordinate--momentum pairs,
$V(r)$ is the ordinary
baryonic gravitational potential, and $\mathcal A_\theta$ is a constitutive
background connection.  For constant $\varpi$, Eq.~\eqref{eq:galactic-effective-hamiltonian}
is the usual fixed-central-character planar Hamiltonian.  Allowing
$\mathcal A_\theta(r)$ to vary is a genuine field profile, not a relabelling of
that constant representation.

For a circular orbit, $p_r=0$ and $\dot p_r=0$.  Define the signed tangential
velocity, baryonic circular speed, and connection response by
\begin{align}
 v(r)&:=r\dot\theta
 =\frac{L+\mathcal A_\theta(r)}{r},
 \label{eq:galactic-tangential-velocity}\\
 v_{\rm N}^2(r)&:=rV'(r),
 \label{eq:baryonic-circular-speed}\\
 q_\varpi(r)&:=\mathcal A_\theta'(r)
 =\bigl(r^2\varpi(r)\bigr)'.
 \label{eq:response-velocity}
\end{align}

\begin{proposition}[Circular branches]
\label{prop:galactic-circular-branches}
The circular orbits of Eq.~\eqref{eq:galactic-effective-hamiltonian} obey
\begin{equation}
 \boxed{
 v^2-q_\varpi v-v_{\rm N}^2=0,
 \qquad
 v_\pm(r)=\frac{q_\varpi(r)\pm
 \sqrt{q_\varpi^2(r)+4v_{\rm N}^2(r)}}{2}.}
 \label{eq:galactic-circular-branches}
\end{equation}
Under orientation reversal $q_\varpi\mapsto-q_\varpi$, the branches satisfy
$v_\pm(-q_\varpi)=-v_\mp(q_\varpi)$.
\end{proposition}

\begin{proof}
Differentiate $H_{\rm eff}$ with respect to $r$ at fixed canonical $L$ and use
$L+\mathcal A_\theta=rv$.  The condition $\dot p_r=-\partial_rH_{\rm eff}=0$
becomes $v^2-q_\varpi v-rV'=0$.  Solving the quadratic gives the two branches;
the reversal identity follows by substitution.
\end{proof}

The connection enters through its radial derivative, not through $\varpi$
alone.  In particular, a spatially constant $\varpi$ gives
$q_\varpi=2r\varpi$ and therefore a rising response-dominated rotation curve.
It cannot explain an extended flat branch.

\begin{proposition}[Inverse rotation--response relation]
\label{prop:inverse-rotation-response}
Let $v(r)$ be a nonzero signed circular-speed branch in a specified baryonic
potential.  The azimuthal connections producing that branch are precisely
\begin{align}
 \mathcal A_\theta(r)
 &=C+\int_{r_0}^{r}
 \left[v(s)-\frac{v_{\rm N}^2(s)}{v(s)}\right]\dd s,
 \label{eq:inverse-azimuthal-connection}\\
 \varpi(r)
 &=\frac1{r^2}\left\{C+\int_{r_0}^{r}
 \left[v(s)-\frac{v_{\rm N}^2(s)}{v(s)}\right]\dd s\right\}.
 \label{eq:inverse-rotation-response}
\end{align}
The constant $C$ is kinematically degenerate with a constant shift of $L$ and
does not affect a circular speed.
\end{proposition}

\begin{proof}
Equation~\eqref{eq:galactic-circular-branches} gives
$q_\varpi=v-v_{\rm N}^2/v$.  Integrate
$q_\varpi=\mathcal A_\theta'$ and use
$\mathcal A_\theta=r^2\varpi$.  The converse follows by differentiation and
substitution.
\end{proof}

This inverse formula turns a rotation curve and a baryonic mass model into a
direct reconstruction test.  In a response-dominated interval with
$v_{\rm N}^2\ll v_f^2$, a flat branch $v\simeq v_f$ requires
\begin{equation}
 \varpi(r)
 =\frac{v_f}{r}+\frac{C}{r^2}
 +O\!\left(\frac1{r^2}\int^r
 \frac{v_{\rm N}^2(s)}{v_f}\,\dd s\right).
 \label{eq:flat-branch-required-response}
\end{equation}
Thus the needed leading profile is $1/r$, not a constant central frequency.

\subsection{A minimal screened bulk closure}
\label{subsec:screened-bulk-response}

The inverse relation says what the orbit needs; it does not explain why the
environment supplies it.  The lowest-order isotropic dissipative closure for
the central pseudovector is
\begin{equation}
 \partial_t\bsomega
 =D_\omega\Delta_3\bsomega
 -\tau_\omega^{-1}\bsomega
 +\bm S_{\rm pump}
 +\bm{\mathcal N}(\bsomega,\nabla\bsomega;U),
 \label{eq:screened-central-dynamics}
\end{equation}
with $D_\omega>0$, $\tau_\omega>0$, an oriented pump source, and nonlinear
saturation or material couplings collected in $\bm{\mathcal N}$.  The source
must reverse with the control-loop orientation and vanish in a no-pumping
class.  Equation~\eqref{eq:screened-central-dynamics} is a constitutive
hypothesis; BSN curvature alone does not imply diffusion, relaxation, or this
source.

Outside a compact active region and at small amplitude, stationarity gives
\begin{equation}
 (\Delta_3-L_\omega^{-2})\bsomega=0,
 \qquad
 L_\omega:=\sqrt{D_\omega\tau_\omega}.
 \label{eq:screened-helmholtz}
\end{equation}

\begin{proposition}[Screened exterior mode]
\label{prop:screened-exterior-mode}
Suppose the source lies in $r\leq R_s$ and the leading exterior mode has fixed
orientation $\bm n$, so that $\bsomega(r)=\varpi(r)\bm n$.  The solution of
Eq.~\eqref{eq:screened-helmholtz} with spherically symmetric coefficient that
decays at infinity is
\begin{equation}
 \varpi(r)=\frac{A_\omega}{r}\e^{-r/L_\omega}.
 \label{eq:screened-response-profile}
\end{equation}
For $R_s\ll r\ll L_\omega$ it has the intermediate asymptotic
$\varpi=A_\omega/r+O(A_\omega/L_\omega)$.
\end{proposition}

\begin{proof}
For a radial coefficient,
$(r^2\varpi')'-r^2L_\omega^{-2}\varpi=0$.  Hence
$r\varpi=C_-\e^{-r/L_\omega}+C_+\e^{r/L_\omega}$; decay sets $C_+=0$.
\end{proof}

For a projected stationary source
$S_n=\langle\bm S_{\rm pump},\bm n\rangle$, the linear Green representation is
\begin{equation}
 \varpi(\bm x)=\frac1{4\pi D_\omega}
 \int_{\R^3}\frac{\e^{-|\bm x-\bm y|/L_\omega}}
 {|\bm x-\bm y|}S_n(\bm y)\,\dd^3y.
 \label{eq:screened-response-green}
\end{equation}
The exterior equation fixes the radial form, while $A_\omega$ is a source
moment determined by the pump and boundary matching.  If the oriented leading
moment vanishes, the monopole is absent and a faster-decaying multipole takes
its place.

For Eq.~\eqref{eq:screened-response-profile}, the observable response velocity
is
\begin{equation}
 q_\varpi(r)=A_\omega\e^{-r/L_\omega}
 \left(1-\frac r{L_\omega}\right).
 \label{eq:screened-response-velocity}
\end{equation}
Thus, on the co-oriented branch in the controlled corridor
$R_s\ll r\ll L_\omega$,
\begin{equation}
 v_+(r)=\frac{A_\omega+
 \sqrt{A_\omega^2+4v_{\rm N}^2(r)}}2
 +O\!\left(|A_\omega|\frac r{L_\omega}\right).
 \label{eq:screened-flat-branch}
\end{equation}
When the baryonic term is subdominant, $v_+\simeq A_\omega$.  At
$r=O(L_\omega)$ the response turns over and is screened.  A galaxy can display
an extended flat interval in this closure only if its active source is compact
relative to $L_\omega$ and has a nonzero oriented leading moment.

\subsection{Conditional baryonic Tully--Fisher matching}
\label{subsec:conditional-btfr}

The empirical baryonic Tully--Fisher relation has approximately fourth-power
slope and small intrinsic scatter
\cite{McGaugh2000BTFR,LelliMcGaughSchombert2016BTFR}.  The screened field
equation does not derive that scaling because it leaves $A_\omega$ free.  The
missing statement can be isolated cleanly.

Milgrom's deep-acceleration law gives the asymptotic relation
$v_f^4=GM_ba_0$ and treats $a_0$ as a new acceleration scale
\cite{Milgrom1983MOND,Milgrom1983Galaxies}.  The present orbit equation is not
that modified force law: its additional term is linear in the signed velocity
and arises from an azimuthal connection.  Agreement with the same fourth-power
scaling therefore requires the following source law.

\begin{hypothesis}[Scale-free source matching]
\label{hyp:btfr-source-matching}
For coherent discs in the intermediate corridor, suppose the magnitude of the
oriented source moment is
\begin{equation}
 |A_\omega|=\eta_{\rm BTF}\bigl(GM_ba_\sigma\bigr)^{1/4},
 \qquad a_\sigma:=c_{\rm light}\sigma_*,
 \label{eq:btfr-source-matching}
\end{equation}
where $M_b$ is total baryonic mass, $c_{\rm light}$ is the speed of light,
$\sigma_*$ is an independently selected stationary background frequency, and
the dimensionless matching coefficient $\eta_{\rm BTF}$ is mass-independent up to
measurable structural and environmental scatter.
\end{hypothesis}

In a response-dominated flat corridor this hypothesis gives
\begin{equation}
 v_f^4=\eta_{\rm BTF}^4GM_bc_{\rm light}\sigma_*=GM_ba_0,
 \qquad a_0:=\eta_{\rm BTF}^4c_{\rm light}\sigma_*.
 \label{eq:conditional-btfr}
\end{equation}
The fourth root is the unique velocity monomial formed from $GM_b$ and an
acceleration, but dimensional uniqueness is not a microscopic derivation.
The model must calculate $A_\omega$ from
Eq.~\eqref{eq:screened-response-green}, explain the small scatter in
$\eta_{\rm BTF}$,
and determine $\sigma_*$ independently of the rotation curves.  In particular,
$\sigma_*/H=O(1)$ is not assumed here; it remains a constitutive ratio until the
event cadence and proper-time calibration are derived.

The galactic closure is therefore falsifiable in three logically independent
ways.  First, Eq.~\eqref{eq:inverse-rotation-response} tests whether an observed
curve and baryonic potential admit a smooth response.  Second, the reconstructed
profile can be tested for a common $1/r$ corridor and screened turnover.  Third,
$A_\omega^4/(GM_b)$ can be tested against an independently measured background
scale and against morphology and environment.  Passing the kinematic inverse
test does not establish the field equation, and passing the screened-profile
test does not establish the mass matching.

Finally, Eq.~\eqref{eq:galactic-effective-hamiltonian} is a massive-particle
effective Hamiltonian.  Rotation curves alone do not determine gravitational
lensing.  A viable cosmology must embed the response in a covariant spacetime
law, derive the null Hamiltonian and Jacobi map, and recover the observed
relation between dynamical and lensing effects.  That covariant completion is
one of the principal open problems of the construction.

\section{The spatial group \texorpdfstring{$N_{3,2}$}{N3,2}}
\label{sec:n32-spectrum}

Sections~\ref{sec:bsn-geometry} and \ref{sec:central-selection} supplied a
conditional tangent geometry and an explicit selector for its central
character.  Once that central block is given, the spatial spectral problem
becomes exact.  The remaining distinction is crucial: a mode basis and its
projection kernels determine what can propagate, not how those modes are
occupied.

\subsection{Covariant parent and spatial reduction}

The spatial algebra has a simple Lorentz-covariant parent.  Let $M$ be
Minkowski space and define the free step-two algebra
\begin{equation}
 \mathfrak n_{4,2}=M\oplus\Lambda^2M,
 \qquad [u,v]=u\wedge v,
 \qquad [\Lambda^2M,\mathfrak n_{4,2}]=0.
 \label{eq:n42-parent}
\end{equation}
The Lorentz group acts by automorphisms on both layers.  A choice of observer
splits
\begin{equation}
 M=\R u\oplus u^\perp
 \label{eq:observer-splitting}
\end{equation}
and the purely spatial subalgebra generated by $u^\perp$ is
\begin{equation}
 u^\perp\oplus\Lambda^2u^\perp
 \simeq\R^3\oplus\Lambda^2\R^3
 =\mathfrak n_{3,2}.
 \label{eq:n42-spatial-restriction}
\end{equation}
Thus $N_{3,2}$ is not introduced as a preferred global foliation.  It is the
spatial reduction seen by a timelike observer of the covariant free step-two
tangent.

The same algebra arises as a weighted tangent of a Lorentzian symmetric pair.
Let
\begin{equation}
 \mathfrak g=\mathfrak h\oplus\mathfrak p,
 \qquad
 [\mathfrak h,\mathfrak h]\subset\mathfrak h,
 \quad[\mathfrak h,\mathfrak p]\subset\mathfrak p,
 \quad[\mathfrak p,\mathfrak p]\subset\mathfrak h,
 \label{eq:symmetric-pair}
\end{equation}
where $\mathfrak p\simeq M$ and
$\mathfrak h\simeq\mathfrak{so}(1,3)\simeq\Lambda^2M$.  Assign weight one to
$\mathfrak p$ and weight two to $\mathfrak h$ by
\begin{equation}
 \delta_\varepsilon(X+Z)=\varepsilon X+\varepsilon^2Z,
 \qquad
 [A,B]_\varepsilon
 :=\delta_\varepsilon^{-1}
 [\delta_\varepsilon A,\delta_\varepsilon B].
 \label{eq:weighted-contraction}
\end{equation}
Then
\begin{align}
 [X,Y]_\varepsilon&=[X,Y]_{\mathfrak h},
 &X,Y&\in\mathfrak p,\\
 [Z,X]_\varepsilon&=\varepsilon^2[Z,X],
 &Z&\in\mathfrak h,\ X\in\mathfrak p,\\
 [Z,W]_\varepsilon&=\varepsilon^2[Z,W],
 &Z,W&\in\mathfrak h.
 \label{eq:contraction-brackets}
\end{align}
The limit $\varepsilon\to0$ is step two, with $\mathfrak h$ central.  For the
de Sitter and anti-de Sitter symmetric pairs this produces the same vector
space $M\oplus\Lambda^2M$, up to the normalization and real-form convention
of the bracket.

This contraction is a tangent statement.  It does not make the central layer
a propagating Maxwell field, select a central character, or determine a
stationary covariance.  Those are representation and state questions treated
separately below.

\subsection{Group law and horizontal sub-Laplacian}

The free step-two group in three horizontal dimensions is
\begin{equation}
 N_{3,2}=\R_x^3\oplus\R_z^3,
 \qquad
 (x,z)(x',z')=
 \left(x+x',z+z'+\frac12x\times x'\right).
 \label{eq:n32-group-law}
\end{equation}
Identifying $\R_z^3\simeq\Lambda^2\R_x^3$ by the chosen orientation, the
left-invariant fields are
\begin{equation}
 X_i=\partial_{x_i}-\frac12\epsilon_{ijk}x_j\partial_{z_k},
 \qquad Z_i=\partial_{z_i},
 \label{eq:n32-fields}
\end{equation}
and satisfy
\begin{equation}
 [X_i,X_j]=\epsilon_{ijk}Z_k,
 \qquad[Z_i,\cdot]=0.
 \label{eq:n32-brackets}
\end{equation}
The positive horizontal sub-Laplacian is
\begin{equation}
 \cL_{3,2}=-\sum_{i=1}^3X_i^2.
 \label{eq:n32-sublaplacian}
\end{equation}
Theorem~\ref{thm:bsn-pullback} identifies this operator as the nilpotent
principal model of a full-rank counted-current diffusion, after fixing the
horizontal metric and central normalization.

The central dual variable has two related interpretations.  In group Fourier
analysis it labels an irreducible central-character sector.  In the classical
coadjoint description it is the conserved vertical momentum.  The selector of
Section~\ref{sec:central-selection} acts on this variable; it does not create a
new position-space matter density.

\subsection{Landau--longitudinal spectrum}

Fourier transformation in $z$ at central covector $B\in(\R_z^3)^*$ gives
\begin{equation}
 \cL_B=-\sum_{i=1}^3
 \left(\partial_{x_i}+\frac{\ii}{2}(B\times x)_i\right)^2.
 \label{eq:magnetic-block}
\end{equation}
Rotate to $B=be_3$, where $b=|B|$, and write $x=(u,v,s)$.  Then
\begin{equation}
 \cL_B=
 -\left(\partial_u-\frac{\ii b}{2}v\right)^2
 -\left(\partial_v+\frac{\ii b}{2}u\right)^2
 -\partial_s^2.
 \label{eq:landau-block}
\end{equation}
Thus the transverse operator is a Landau Hamiltonian and the longitudinal
direction is free.  In the centered axisymmetric sector its spectrum is
\begin{equation}
 \boxed{\lambda_{b,n,\xi}=\xi^2+b(2n+1)},
 \qquad n=0,1,2,\ldots,
 \label{eq:n32-spectrum}
\end{equation}
where $\xi$ is longitudinal momentum.  This identity is exact for
$\cL_{3,2}$; after fixing a nonzero central character, it is the standard
Heisenberg-block harmonic analysis \cite{Folland1989}.

A normalized centered transverse mode is
\begin{equation}
 \psi_{n,b}(r)=
 \left(\frac{b}{2\pi}\right)^{1/2}
 \e^{-x/2}L_n(x),
 \qquad x=\frac{br^2}{2}.
 \label{eq:centered-landau-mode}
\end{equation}
Each Landau level also carries guiding-centre multiplicity.  The centered mode
is a convenient representative, not the complete homogeneous sector.  A
translation-invariant state is a positive operator on this multiplicity space;
its trace cannot generally be replaced by a scalar degeneracy factor.

\subsection{Centered sky kernel}

Restrict the mode $\e^{\ii\xi s}\psi_{n,b}(r)$ to a Euclidean shell
$|x|=\chi$.  With $\mu=s/\chi$ define
\begin{equation}
 q=\xi\chi,
 \qquad
 \alpha=\frac{b\chi^2}{4},
 \qquad
 r^2=\chi^2(1-\mu^2).
 \label{eq:kernel-variables}
\end{equation}
The centered scalar angular kernel is
\begin{equation}
 \boxed{
 \cK_\ell^{(n)}(\alpha,q)
 =\frac12\int_{-1}^{1}\dd\mu\,
 P_\ell(\mu)\e^{\ii q\mu}
 \e^{-\alpha(1-\mu^2)}
 L_n\!\left(2\alpha(1-\mu^2)\right).}
 \label{eq:centered-sky-kernel}
\end{equation}
Its Abelian limit is
\begin{equation}
 \cK_\ell^{(0)}(0,q)=\ii^\ell j_\ell(q).
 \label{eq:bessel-limit}
\end{equation}
Equation~\eqref{eq:centered-sky-kernel} contains genuine non-Abelian angular
structure.  For $b\neq0$ it generally cannot be factorized into a spherical
Bessel function and a scalar envelope.

For visibility $g_{\rm vis}(\chi)$, source
$S_n(b,\xi,\mathfrak m;\chi)$, and guiding-centre label $\mathfrak m$, the
formal full transfer target is
\begin{align}
 \Theta_\ell(b,n,\xi,\mathfrak m)
 &=\int\dd\chi\;g_{\rm vis}(\chi)S_n(b,\xi,\mathfrak m;\chi)
 \cK_{\ell,\mathfrak m}^{(n)}
 \left(\frac{b\chi^2}{4},\xi\chi\right),
 \label{eq:exact-transfer-target}\\
 C_\ell
 &=\sum_{n\geq0}\int\dd b\,\dd\xi\,\dd\mathfrak m\;
 \rho(b,n,\xi,\mathfrak m)\mathcal P(b,n,\xi,\mathfrak m)
 |\Theta_\ell(b,n,\xi,\mathfrak m)|^2.
 \label{eq:exact-spectrum-target}
\end{align}
The centered reduction of the kernel is
Eq.~\eqref{eq:centered-sky-kernel}; the general multiplicity-resolved kernel is
the corresponding matrix element of the spherical restriction map.  Geometry
fixes those maps; the source, visibility, and positive occupation operator
$\mathcal P$ are independent state data.

There is a useful endpoint form.  For $b>0$ define the invariant ratio
\begin{equation}
 \varrho:=\frac{\xi^2}{b},
 \qquad
 \alpha=\frac{q^2}{4\varrho}.
 \label{eq:endpoint-ratio}
\end{equation}
Then
\begin{equation}
 s=\frac{\varrho(\ell+1/2)^2}{q^2},
 \label{eq:endpoint-variable}
\end{equation}
the large-$\ell$ saddle gives, up to a common normalization and phase,
\begin{equation}
 \cK_\ell^{(n)}\sim
 s\e^{-s}(-1)^nL_n(2s).
 \label{eq:kernel-endpoint-form}
\end{equation}
In particular, at the matched value $\varrho=1$ the ground-state ridge has
$q/(\ell+1/2)\to1$.  Higher event weights produce only constant ridge
rescalings.

\subsection{Centered spin-raised partner}
\label{subsec:spin-raised-kernel}

The centered scalar kernel has an exact spin-two partner for the particular
angular screen-Hessian analyzer.  Write
\begin{equation}
 F_n(\mu;\alpha,q)
 :=\e^{\ii q\mu}\e^{-\alpha(1-\mu^2)}
 L_n\!\left(2\alpha(1-\mu^2)\right),
 \qquad
 \Lambda_n:=q^2+4\alpha(2n+1).
 \label{eq:spin-kernel-data}
\end{equation}
For an axisymmetric scalar, $\eth^2F=(1-\mu^2)F''$.  With
$P_\ell^2(\mu)=(1-\mu^2)P_\ell''(\mu)$, define for $\ell\geq2$
\begin{equation}
 \cE_\ell^{(n)}(\alpha,q)
 :=\frac12\sqrt{\frac{(\ell-2)!}{(\ell+2)!}}
 \int_{-1}^{1}\dd\mu\;P_\ell^2(\mu)
 \frac{(1-\mu^2)F_n''(\mu;\alpha,q)}{\Lambda_n}.
 \label{eq:spin-raised-definition}
\end{equation}

\begin{proposition}[Exact centered spin-two kernel]
\label{prop:spin-raised-kernel}
The kernel in \eqref{eq:spin-raised-definition} satisfies
\begin{equation}
 \boxed{
 \cE_\ell^{(n)}(\alpha,q)
 =\sqrt{\frac{(\ell+2)!}{(\ell-2)!}}
 \frac{\cK_\ell^{(n)}(\alpha,q)}
 {q^2+4\alpha(2n+1)}.}
 \label{eq:spin-raised-identity}
\end{equation}
In the Abelian ground block,
\begin{equation}
 \cE_\ell^{(0)}(0,q)
 =\ii^\ell\sqrt{\frac{(\ell+2)!}{(\ell-2)!}}
 \frac{j_\ell(q)}{q^2}.
 \label{eq:spin-raised-abelian-limit}
\end{equation}
\end{proposition}

\begin{proof}
Twice integrate \eqref{eq:spin-raised-definition} by parts.  The endpoint
terms vanish because of the factors $1-\mu^2$.  The adjoint spin-raising
identity transfers both derivatives to the associated Legendre factor and
gives $[(\ell+2)!/(\ell-2)!]P_\ell$.  Equation
\eqref{eq:centered-sky-kernel} then yields
\eqref{eq:spin-raised-identity}; Eq.~\eqref{eq:bessel-limit} gives the Abelian
limit.
\end{proof}

The result is exact for $\eth^2\cL_{3,2}^{-1}$ on the centered scalar mode.  A
different horizon analyzer, or the complete matrix of horizontal
$\STF(X_iX_j)$ elements with Landau-level mixing, is additional response data.
For a general positive guiding-centre operator $\Gamma_{\lambda_0}$ on the
multiplicity space $\mathcal G_{\lambda_0}$, the corresponding diagonal
geometric weight is retained without a scalar-degeneracy replacement as
\begin{equation}
 G_\ell^X(\lambda_0;\Gamma)
 =\frac1{2\ell+1}\Tr\!\left(
 V_{\ell,\lambda_0}^X\Gamma_{\lambda_0}
 V_{\ell,\lambda_0}^{X*}\right)\geq0,
 \qquad X\in\{T,E\}.
 \label{eq:guiding-centre-positive-weight}
\end{equation}
Here $V_{\ell,\lambda_0}^X$ is the spherical restriction and scalar or
spin-raised projection map.  Geometry supplies $V^X$; state selection supplies
$\Gamma$.

\subsection{Commutator events and the quartic packet}
\label{subsec:quartic-packet}

A point source couples to the whole centered radial tower because $L_n(0)=1$.
Finite support on a few levels therefore requires a specific event form factor.
For a step-two commutator loop with horizontal sides $u,v\in\R^3$, the
Baker--Campbell--Hausdorff formula is exact:
\begin{equation}
 \exp(u)\exp(v)\exp(-u)\exp(-v)=\exp(u\wedge v).
 \label{eq:commutator-holonomy}
\end{equation}
The central record $z=u\wedge v$ has horizontal degree two.  If the active
event is a baseline-subtracted, isotropic, orientation-even analytic scalar of
$z$, then
\begin{equation}
 \mathcal E(z)=c_1|z|^2+c_2|z|^4+\cdots.
 \label{eq:event-expansion}
\end{equation}
Its generic leading term has horizontal degree four.  This conclusion applies
to the stated connected central-scalar channel; it is not a claim about every
local observable.

Match the leading event to the Landau ground-state envelope:
\begin{equation}
 f_b^{(4)}(r)=C_bx^2\e^{-x/2},
 \qquad x=\frac{br^2}{2}.
 \label{eq:quartic-event}
\end{equation}
The Laguerre identity
\begin{equation}
 x^2=2L_0(x)-4L_1(x)+2L_2(x)
 \label{eq:laguerre-quartic}
\end{equation}
gives the exact overlap vector
\begin{equation}
 v_n^{(4)}
 :=\int_{\R^2}\dd^2r\;\psi_{n,b}(r)f_b^{(4)}(r)
 \propto(2,-4,2,0,\ldots).
 \label{eq:quartic-overlaps}
\end{equation}
If source and readout are adjoint, each pole residue is the product of the two
overlaps.  Consequently
\begin{equation}
 \boxed{w_0:w_1:w_2=1:4:1.}
 \label{eq:141-residues}
\end{equation}
The ratio is exact conditional on four ingredients: central-scalar dominance,
orientation-even analyticity, the matched Landau envelope, and reciprocal
source/readout coupling.  The group law fixes the degree of the leading event,
but not its coefficient, occupation, or radiative coupling.

Magnetic translations commute with fixed-$b$ Landau projectors.  Translating
the event and the basis together therefore preserves the overlap vector.
Averaging over guiding centres restores transverse homogeneity, while averaging
the direction of $B$ restores rotational isotropy.

\subsection{A particular screen--Weyl operator}
\label{subsec:screen-weyl}

Let $J$ be a scalar stress source and solve the scalar constraint
\begin{equation}
 \cL_{3,2}\Phi=J
 \label{eq:scalar-constraint}
\end{equation}
with no incoming homogeneous Weyl component.  The particular trace-free
horizontal Hessian is
\begin{equation}
 \boxed{
 \cW_{ij}J
 =\STF(X_iX_j)\cL_{3,2}^{-1}J.}
 \label{eq:screen-weyl-operator}
\end{equation}
This is the step-two analogue of the scalar-to-optical-tidal Riesz transform.

\begin{proposition}[Screen--Weyl norm and covariance]
\label{prop:screen-weyl}
In a nonzero central block, let
$|n,\xi,\mathfrak m\rangle$ be a normalized Landau--longitudinal mode and set
$\lambda_n=\xi^2+b(2n+1)$.  A ladder-operator calculation gives
\begin{equation}
 \sum_{i,j}\|\cW_{ij}|n,\xi,\mathfrak m\rangle\|^2
 =\frac23+\frac{3b^2}{2\lambda_n^2}
 \leq\frac{13}{6}.
 \label{eq:screen-weyl-norm}
\end{equation}
If $\Sigma_J\succeq0$ is a scalar-source covariance, the joint particular
covariance is
\begin{equation}
 \Sigma_{(J,\cW)}
 =\begin{bmatrix}I\\ W\end{bmatrix}
 \Sigma_J
 \begin{bmatrix}I\\ W\end{bmatrix}^{\!*}\succeq0.
 \label{eq:screen-weyl-covariance}
\end{equation}
\end{proposition}

\begin{proof}
After rotation to $B=be_3$, write the transverse covariant derivatives in
terms of Landau raising and lowering operators.  The five independent
components of $\STF(X_iX_j)$ shift $n$ by $0,\pm1,\pm2$.  If
\begin{equation*}
 T_{ij}=\frac12(X_iX_j+X_jX_i)
 -\frac13\delta_{ij}\sum_kX_k^2,
\end{equation*}
then repeated use of $[X_i,X_j]=\ii\epsilon_{ijk}B_k$ gives, on a fixed central
block,
\begin{equation*}
 \sum_{i,j}T_{ij}^*T_{ij}
 =\frac23\cL_B^2+\frac32b^2I.
\end{equation*}
Applying this identity to a $\cL_B$ eigenvector and dividing by
$\lambda_n^2$ gives the first equality in
\eqref{eq:screen-weyl-norm}.  Since $\lambda_n\geq b$, the bound follows.
Equation~\eqref{eq:screen-weyl-covariance} is a congruence of a positive
operator and is therefore positive.
\end{proof}

The operator \eqref{eq:screen-weyl-operator} fixes a response once the scalar
source is known; it does not select $\Sigma_J$.  It also acts on the Jacobi map
of neighbouring rays, not directly on the internal polarization density
matrix of an initially unpolarized photon field.  That distinction is proved
in Section~\ref{subsec:geometric-polarization}.

\section{Stationary radiative states on the native spectrum}
\label{sec:radiative-state}

The preceding harmonic analysis supplies exact modes, residues, and sky
projection maps.  An exact basis is not yet a state.  We now ask which
stationary covariance the open system selects, separating that dynamical
question from the already solved propagation problem.

\subsection{The covariance is not selected by the geometry}

Let
\begin{equation}
 \lambda=(b,n,\xi,\mathfrak m)
 \label{eq:spectral-label}
\end{equation}
denote a complete native spectral label, including guiding-centre
multiplicity.  Collect a scalar temperature source, longitudinal velocity, and
polarization quadrupole into
\begin{equation}
 \bm J(\lambda)
 =\bigl(J_0(\lambda),J_v(\lambda),J_\Pi(\lambda)\bigr)^{\mathsf T}.
 \label{eq:radiative-source-vector}
\end{equation}
A homogeneous stationary Gaussian source has block covariance
\begin{equation}
 \langle\bm J(\lambda)\bm J(\lambda')^*\rangle
 =\delta(\lambda-\lambda')\bsSigma(\lambda),
 \qquad \bsSigma(\lambda)\succeq0,
 \label{eq:source-covariance}
\end{equation}
with the delta symbol understood as including the operator structure on
multiplicity space.

\begin{theorem}[Radiative-state underdetermination]
\label{thm:radiative-underdetermination}
Fix the $N_{3,2}$ tangent geometry, $\cL_{3,2}$, the centered kernels, and the
event-gated selector of Section~\ref{sec:central-selection}.  Stationarity,
spatial homogeneity, rotational covariance, positivity, and the selector's
stationary first and second moments do not determine either
$\bsSigma(\lambda)$ or a reciprocal drift law for $(J_0,J_v,J_\Pi)$.
Subject to the usual reality, parity, trace-class, and finite-power conditions,
the covariance can be any admissible measurable positive-semidefinite
matrix-valued function of the invariant spectral labels, together with a
positive operator on guiding-centre multiplicity.  Distinct choices leave all
proved ledger and selector statements unchanged and produce different TT, TE,
and EE spectra.
\end{theorem}

\begin{proof}
Group Fourier transformation block-diagonalizes a homogeneous two-point law
over irreducible $N_{3,2}$ labels.  Rotational covariance relates blocks on the
same orbit and parity imposes reality relations.  Neither condition fixes the
positive operator within a block or the cross entries among $J_0,J_v,J_\Pi$.

For an explicit realization, choose any admissible $\bsSigma(\lambda)$ and any
positive invariant scalar $a(\lambda)$.  The Ornstein--Uhlenbeck process
\begin{align}
 \dot{\bm J}(\lambda,t)
 &=-a(\lambda)\bm J(\lambda,t)+\bm\zeta(\lambda,t),
 \label{eq:arbitrary-ou}\\
 \langle\bm\zeta(\lambda,t)\bm\zeta(\lambda',t')^*\rangle
 &=2a(\lambda)\bsSigma(\lambda)
 \delta(\lambda-\lambda')\delta(t-t')
 \label{eq:arbitrary-ou-noise}
\end{align}
has exactly that stationary covariance.  Matrix-valued reciprocal or
nonreciprocal drifts can likewise be paired with positive noise.  Tensoring any
such process with the fixed selector preserves the selector marginal, its BSN
curvature, throughput, central mean, and covariance.  The selector acts on
chart records and central momentum, not on the radiative multipoles, so it
cannot distinguish these states without an additional coupling map.
\end{proof}

The theorem is not peculiar to Gaussian sources.  Higher connected cumulants
are additional invariant tensors and are even less constrained.  The result
locates the boundary of the geometric construction: it supplies a basis and
operators, whereas a microscopic interaction must supply the state.

\subsection{A finite response-identifiability witness}
\label{subsec:response-identifiability}

The preceding theorem allows an arbitrary positive radiative covariance.  A
smaller finite example shows that the nonuniqueness is already present before
angular projection and after several natural stationary invariants have been
fixed.  For $R>1$, consider the reversible three-state generator
\begin{equation}
 L_R=
 \begin{pmatrix}
  -(u+t)&u&t\\
  u&-(u+t)&t\\
  t&t&-2t
 \end{pmatrix},
 \qquad
 t=\frac13,\qquad
 u=\frac{R-1/3}{2}.
 \label{eq:identifiability-generator}
\end{equation}
It has the uniform stationary law and decay spectrum $(0,1,R)$.  Choose a
mean-zero unit-variance readout whose squared weight on the gap-one eigenspace
is $w$.  Its stationary correlation and positive Euclidean response are
\begin{align}
 C_{R,w}(\tau)
 &=w\e^{-|\tau|}+(1-w)\e^{-R|\tau|},
 \label{eq:identifiability-correlation}\\
 \chi_{R,w}(\ii\omega)
 &=\int_0^\infty C_{R,w}(\tau)\cos(\omega\tau)\,\dd\tau
 =\frac{w}{1+\omega^2}
 +\frac{(1-w)R}{R^2+\omega^2}.
 \label{eq:identifiability-response}
\end{align}

\begin{proposition}[Finite generator non-identifiability]
\label{prop:finite-generator-nonidentifiability}
The two positive Markov completions
\begin{equation}
 (R_A,w_A)=\left(3,\frac58\right),
 \qquad
 (R_B,w_B)=\left(5,\frac{11}{16}\right)
 \label{eq:identifiability-pair}
\end{equation}
have the same stationary law, spectral gap, equal-time normalization, and
integrated memory:
\begin{equation}
 C_A(0)=C_B(0)=1,
 \qquad
 \int_0^\infty C_A(\tau)\,\dd\tau
 =\int_0^\infty C_B(\tau)\,\dd\tau
 =\frac34.
 \label{eq:identifiability-matched-data}
\end{equation}
They nevertheless have different frequency responses.  At $\omega=8$,
\begin{equation}
 \chi_A(\ii8)=0.0250263435,\qquad
 \chi_B(\ii8)=0.0281331029,\qquad
 \frac{\chi_B-\chi_A}{\chi_A}=0.12413956.
 \label{eq:identifiability-response-difference}
\end{equation}
Their inverse ninth moments differ by $9.997\%$ relative to model $A$.
\end{proposition}

\begin{proof}
The matrix in \eqref{eq:identifiability-generator} is symmetric, has
nonnegative off-diagonal entries, and has row sum zero.  The vectors
$(1,1,1)$, $(1,1,-2)$, and $(1,-1,0)$ have eigenvalues $0,-1,-R$,
respectively.  Equations \eqref{eq:identifiability-correlation} and
\eqref{eq:identifiability-response} follow by spectral decomposition.
Substitution of \eqref{eq:identifiability-pair} gives
\[
 w+\frac{1-w}{R}=\frac34
\]
for both models and gives
\eqref{eq:identifiability-response-difference}.  The low-frequency expansion
contains the inverse odd moments
\begin{equation*}
 M_{2k+1}=w+\frac{1-w}{R^{2k+1}}.
\end{equation*}
For $k=4$,
$M_9^{(A)}=0.62501905197$ and
$M_9^{(B)}=0.68750016000$, whose relative difference is $0.0999667$.
\end{proof}

The witness is stronger than the observation that a coefficient has not been
written down.  A closure depending only on the stationary law, gap,
equal-time covariance, and one integrated memory time is non-injective on the
retarded kernel.  Counting-field geometry can add oriented response data, but
it cannot reconstruct spectral information discarded by such a low-order
summary.  A complete microscopic generator, its state, and its photon readout
would determine the response; the displayed parent invariants do not.

\subsection{Local photon--baryon kinetics in the native basis}
\label{subsec:qed-moments}

Using ordinary local microphysics is part of the two-level model.  What is
excluded is importing a conventional cosmological transfer spectrum in place
of a native calculation.  A useful interface test is obtained by placing a
minimal Thomson photon--baryon hierarchy on each native eigenvalue
\begin{equation}
 k_\lambda^2=\xi^2+b(2n+1).
 \label{eq:native-wave-number}
\end{equation}
The moment structure is the usual tight-coupling acoustic system
\cite{HuDodelson2002}, with only its spatial eigenvalue replaced by the native
$k_\lambda$.

Let
\begin{equation}
 Y=(\delta_\gamma,v_\gamma,\Pi,\delta_b,v_b)^{\mathsf T},
 \label{eq:qed-state-vector}
\end{equation}
where $\Pi$ is the minimal Thomson quadrupole (the intensity moment that feeds
the standard polarization hierarchy), $R$ is the baryon loading, $\nu$ the
Thomson collision rate, and $c_b^2$ the baryon sound speed squared.  The local
moment generator used here is
\begin{equation}
 \dot Y=L(k)Y,
 \qquad
 L(k)=
 \begin{pmatrix}
 0&-\frac43\ii k&0&0&0\\
 -\frac14\ii k&-\nu&-\frac16\ii k&0&\nu\\
 0&-\frac8{15}\ii k&-\frac9{10}\nu&0&0\\
 0&0&0&0&-\ii k\\
 0&\nu/R&0&-\ii k c_b^2&-\nu/R
 \end{pmatrix}.
 \label{eq:qed-moment-generator}
\end{equation}
The photon and baryon drag terms conserve total momentum: their contributions
cancel in $v_\gamma+Rv_b$.  Streaming produces the quadrupole and the same
Thomson coefficient relaxes it.

\begin{proposition}[Native tight-coupling acoustic branch]
\label{prop:native-acoustic-branch}
For $k/\nu\ll1$ and $c_b^2\ll1$, the generator
\eqref{eq:qed-moment-generator} has, with time dependence
$\exp(-\ii\omega t)$, a conjugate acoustic pair
\begin{equation}
 \omega_\pm(k)=\pm c_sk-\ii Dk^2+O(k^3/\nu^2),
 \qquad
 c_s^2=\frac{1/3+Rc_b^2}{1+R},
 \label{eq:tight-coupling-branch}
\end{equation}
and, at $c_b^2=0$ to the displayed order,
\begin{equation}
 D=\frac1\nu\left[
 \frac{R^2}{6(1+R)^2}+\frac{4}{81(1+R)}\right]>0.
 \label{eq:tight-coupling-diffusion}
\end{equation}
The two fast collision modes relax; for $c_b^2>0$ the remaining entropy mode
also has nonpositive decay.  The acoustic eigenvector contains a nonzero
Thomson quadrupole at first order in $k/\nu$.
Replacing $k$ by $k_\lambda$ therefore gives a stable acoustic branch on every
native spectral block without altering local QED.
\end{proposition}

\begin{proof}
At leading order in $k/\nu$, drag imposes $v_\gamma=v_b=v$ and the quadrupole
is slaved to $v$ with $\Pi=O(kv/\nu)$.  The continuity equations give
$\dot\delta_\gamma=-(4/3)\ii kv$ and
$\dot\delta_b=-\ii kv$, while the conserved total-momentum equation gives
$(1+R)\dot v=-(\ii k/4)\delta_\gamma$ up to baryon-pressure corrections.
Combining these equations gives
$\ddot\delta_\gamma+c_s^2k^2\delta_\gamma=0$.
Writing $s=-\ii\omega$ for the generator eigenvalue, expanding
$v_b/v_\gamma=\nu/(\nu+Rs)$ and
$\Pi/v_\gamma=-(8/15)\ii k/(s+9\nu/10)$ through second order in $k$
gives Eq.~\eqref{eq:tight-coupling-diffusion}.  The quadrupole equation yields
$\Pi=-(16/27)\ii kv/\nu+O(k^2/\nu^2)$.
\end{proof}

Appendix~\ref{app:numerical-checkpoints} verifies this small-$k$ branch at a
representative parameter point and checks its stability after substitution of
native spectral values $k=k_\lambda$.

The example separates mode geometry from collision physics.  The eigenvalue
$k_\lambda$ is supplied by $N_{3,2}$; $R$, $\nu$, the matter temperature, and
the number-changing rates belong to the stationary material state.

The two dissipative channels in \eqref{eq:qed-moment-generator} also give a
minimal positive noise model.  Define
\begin{equation}
 d=(0,-1,0,0,R^{-1})^{\mathsf T},
 \qquad p=(0,0,1,0,0)^{\mathsf T},
 \label{eq:collision-ports}
\end{equation}
and
\begin{equation}
 N_{\rm coll}=2\nu dd^*+\frac95\nu pp^*\succeq0.
 \label{eq:collision-noise}
\end{equation}
It has rank two.  On a stable nonzero-$k$ block, the Lyapunov equation
\begin{equation}
 L(k)\Sigma(k)+\Sigma(k)L(k)^*+N_{\rm coll}=0
 \label{eq:qed-lyapunov}
\end{equation}
has a unique positive stationary covariance.  This proves that positive
rank-two temperature--velocity--quadrupole covariance is compatible with the
native spectrum.  It does not show that the equilibrium covariance has the
observed angular morphology.

\subsection{Marked radiative events and the missing second cumulant}
\label{subsec:marked-radiative-events}

The counting construction used for the BSN geometry retained the mean
completed current, namely the first counting-field cumulant.  A driven
radiative state requires one order more.  Attach monitored marks
$\alpha\in\{T,E\}$ to a completed event, where $T$ denotes its photon-energy
or temperature mark and $E$ its parity-even trace-free quadrupole mark.  For
cycle duration $T_\Gamma$, use the cumulant-rate generator
$K_\Gamma=T_\Gamma^{-1}\log Z_\Gamma$ and define
\begin{equation}
 \bar n_\alpha
 :=\left.\partial_{(\ii\chi_\alpha)}K_\Gamma\right|_{\chi=0},
 \qquad
 \mathsf Q_{\alpha\beta}
 :=\left.\partial_{(\ii\chi_\alpha)}
 \partial_{(\ii\chi_\beta)}K_\Gamma\right|_{\chi=0}.
 \label{eq:radiative-mark-cumulants}
\end{equation}
For an actual monitored count record, $\mathsf Q$ is its connected
covariance-rate matrix and is positive semidefinite.  In particular,
$\mathsf Q_{TE}$ is the pump correlation that is absent from a
first-cumulant treatment.

Let $J_\alpha(k)$ be the jump vector by which mark $\alpha$ enters the native
moment state $Y$ in Eq.~\eqref{eq:qed-state-vector}, and let
$\dd\widetilde N_t^\alpha$ be the centered marked increments.  The additive
event closure is
\begin{equation}
 \dd Y=L(k)Y\,\dd t
 +\sum_\alpha J_\alpha(k)\,\dd\widetilde N_t^\alpha
 +\dd M_t^{\rm coll},
 \qquad
 \mathbb E[\dd\widetilde N_t^\alpha
 \dd\widetilde N_t^\beta]
 =\mathsf Q_{\alpha\beta}\,\dd t.
 \label{eq:marked-event-moment-equation}
\end{equation}
It supplies the explicit positive noise matrix
\begin{equation}
 N_{\rm pump}(k)=J(k)\mathsf QJ(k)^*\succeq0,
 \qquad
 L\Sigma+\Sigma L^*+N_{\rm coll}+N_{\rm pump}=0.
 \label{eq:pump-noise-lyapunov}
\end{equation}
Thus a temperature--quadrupole covariance does not have to be appended by
hand: it is the mixed second cumulant of a common marked event, propagated
through its physical photon jump map.

This identity also locates what the stoichiometric calculation does not fix.
The baryon ledger constrains the mean return cadence and the mean energy
available per event.  It determines neither $\mathsf Q_{TE}$ nor the jump
vectors $J_T,J_E$.  Moreover, the selector actually displayed in
Section~\ref{subsec:selector-radiative-coupling} has zero photon jump map, so
Eq.~\eqref{eq:pump-noise-lyapunov} is a target rather than a completed
microscopic closure.

There is a second possibility that is not equivalent to additive noise.  A
common event can modulate the opacity and baryon loading themselves, so that
$\nu=\nu(U_t)$, $R=R(U_t)$, and hence $L=L(k,U_t)$.  Opacity is the less
invasive first lever.  At fixed $R$, define
\begin{equation}
 L_\nu:=\nu_0\frac{\partial L}{\partial\nu}.
 \label{eq:opacity-generator-derivative}
\end{equation}
Its action on the moment state is exactly
\begin{equation}
 L_\nu Y
 =\nu_0
 \begin{pmatrix}
 0\\ v_b-v_\gamma\\ -(9/10)\Pi\\ 0\\
 (v_\gamma-v_b)/R
 \end{pmatrix}.
 \label{eq:opacity-transverse-action}
\end{equation}
It therefore annihilates the leading tight-coupling slow manifold
$v_\gamma=v_b$, $\Pi=0$.  In contrast, varying $R$ changes
$c_s^2\simeq[3(1+R)]^{-1}$ at leading order.  Opacity modulation is thus
algebraically transverse to the leading TT acoustic phase, although it can
still change finite-slip damping and the quadrupole.

The corresponding polarization-source sensitivity is also exact.  From
Eq.~\eqref{eq:long-mode-quadrupole-response}, with $a_\Pi=9/10$,
\begin{align}
 S_P(\nu)&=-\frac8{15}\ii k v_\gamma
 \frac{\nu}{a_\Pi\nu-\ii\omega},
 \label{eq:opacity-polarization-source}\\
 \frac{\partial\log S_P}{\partial\log\nu}
 &=\frac{-\ii\omega}{a_\Pi\nu-\ii\omega}.
 \label{eq:opacity-log-sensitivity}
\end{align}
The sensitivity vanishes in deep tight coupling, is relative-order one when
$\nu\sim\omega$, and is large only relative to an absolutely weak source when
$\nu\ll\omega$.  The proposed $O(H)$ long-mode channel therefore lies in the
only regime where opacity can supply a substantial phase and amplitude lever
without directly replacing the temperature carrier.

For a bounded or exponentially integrable centered event mark $m(t)$, a
mean-preserving positive parametrization is
\begin{equation}
 \nu(t)=\bar\nu
 \exp\!\left[\epsilon m(t)-\kappa(\epsilon)\right],
 \qquad
 \kappa(\epsilon):=\log\mathbb E\,\e^{\epsilon m}.
 \label{eq:mean-preserving-opacity}
\end{equation}
For a deterministic periodic pump, the expectation in $\kappa$ is the phase
average.  The collision-noise continuation
\begin{equation}
 N_{\rm coll}(t)
 =2\nu(t)dd^*+\frac95\nu(t)pp^*\succeq0
 \label{eq:opacity-modulated-collision-noise}
\end{equation}
preserves positivity.  It is a minimal constitutive continuation, not a
derivation of nonequilibrium fluctuation--dissipation balance.

For a deterministic periodic or cyclostationary pump the equal-time covariance
must satisfy
\begin{equation}
 \dot\Sigma(t)=L(k,t)\Sigma(t)+\Sigma(t)L(k,t)^*+N(k,t),
 \qquad \Sigma(t+T)=\Sigma(t).
 \label{eq:cyclostationary-pump-lyapunov}
\end{equation}
Writing $\Sigma=\Sigma_0+\epsilon\Sigma_1+\cdots$ gives
\begin{align}
 \dot\Sigma_1&=\mathcal L_0\Sigma_1+m(t)B_\nu,
 &\mathcal L_0X&=L_0X+XL_0^*,\nonumber\\
 B_\nu&=L_\nu\Sigma_0+\Sigma_0L_\nu^*+N_\nu,
 &N_\nu&:=\left.\nu_0\frac{\partial N_{\rm coll}}{\partial\nu}
 \right|_{\nu_0}.
 \label{eq:opacity-linear-covariance}
\end{align}
For $m(t)=\Re\e^{\ii\Omega_pt}$ its harmonic response is
\begin{equation}
 \Sigma_{1,\Omega_p}
 =(\ii\Omega_p-\mathcal L_0)^{-1}B_\nu.
 \label{eq:opacity-covariance-resolvent}
\end{equation}

\begin{proposition}[Centered-opacity phase-average obstruction]
\label{prop:opacity-phase-average}
If $m$ has zero phase average and $\mathcal L_0$ is stable, then the
phase-averaged first-order equal-time covariance vanishes.  The same holds for
every phase-averaged linear unequal-time readout.  An $O(\epsilon)$ TE
correction therefore requires an event-conditioned phase or a derived
correlation between $m$ and the carrier forcing.
\end{proposition}

\begin{proof}
Average Eq.~\eqref{eq:opacity-linear-covariance} over one period.  Periodicity
annihilates the average of $\dot\Sigma_1$, while
$\langle m\rangle=0$.  Stability makes the kernel of $\mathcal L_0$ trivial,
so $\langle\Sigma_1\rangle=0$.  The first variation of the unequal-time
propagator is linear in translated copies of $m$; averaging its phase gives
the same zero.
\end{proof}

\begin{corollary}[Conditional necessity of event-conditioned dynamics]
\label{cor:event-conditioned-dynamics}
Assume that (i) the frozen TT carrier remains the sole temperature phase
source, (ii) no additive temperature port is introduced, (iii) the common
return-event mark is to produce a nonzero ensemble-averaged
$O(\epsilon)$ TE correction, and (iv) the displayed local material controls
are initially restricted to $\nu$ and $R$.  Then an unconditioned centered
modulation is insufficient.  The correction must come from an
event-conditioned phase or a derived event--carrier correlation.  A Markovian
realization of that correlation requires a joint event variable, such as an
opacity or event-age state.  Among the displayed controls, $\nu$ is the minimal
phase-preserving first lever; this statement does not assert uniqueness among
all possible generator perturbations.
\end{corollary}

\begin{proof}
Proposition~\ref{prop:opacity-phase-average} excludes a nonzero first-order
correction from an unconditioned centered opacity modulation; the same linear
phase-averaging argument applies to the first variation generated by $R$.
If the mark is stochastic, ensemble averaging
Eq.~\eqref{eq:cyclostationary-pump-lyapunov} introduces the mixed moment
$\mathbb E[mYY^*]$; a derived joint event law, or an equivalent memory kernel,
is therefore needed to determine it.  Finally,
Eq.~\eqref{eq:opacity-transverse-action} shows that $L_\nu$ annihilates the
leading tight-coupling slow manifold, whereas
\begin{equation}
 \frac{\partial c_s^2}{\partial R}
 =\frac{c_b^2-1/3}{(1+R)^2}\ne0
 \qquad (c_b^2\ll1).
 \label{eq:loading-phase-derivative}
\end{equation}
Thus $R$ changes the leading acoustic phase while $\nu$ does not.
\end{proof}

The auxiliary variable need not be a literal continuous age or a particular
finite-stage cascade.  A hidden Markov opacity state, colored multiplicative
noise with a derived memory law, or an equivalent non-Markovian kernel can
encode the same information.  Equation~\eqref{eq:selector-first-passage}
shows, however, that the present single-hazard selector supplies no narrow
internal chronology.  The comparison below uses the strongest first-order case
available without an independent temperature source: the coherent-return
selector fixes the event phase.

\subsection{Planck-family maintenance}
\label{subsec:planck-maintenance}

Local detailed balance can make a Planck spectrum attracting transverse to the
Planck family only if energy exchange and photon-number-changing reactions are
present.  Elastic Thomson scattering alone cannot impose
$\mu_\gamma=0$.  Once number-changing processes are included, a schematic
thermodynamic linearization has the form
\begin{equation}
 (\delta T,\mu_\gamma,y)
 \longmapsto
 (0,-\Gamma_N\mu_\gamma,-\Gamma_y y),
 \qquad \Gamma_N,\Gamma_y>0.
 \label{eq:planck-family-linearization}
\end{equation}
Energy-conserving local kinetics leaves temperature neutral.  This is not an
obstruction to maintaining a specified CMB temperature: the state may be one
member of a neutral stationary family.  It does mean that the value of
$T_\gamma$ is not selected by the local collision operator.

For reference, a blackbody at
$T_0=2.72548\,\mathrm K$ \cite{Fixsen2009} has energy density
\begin{equation}
 \rho_\gamma=a_RT_0^4
 =4.17466\times10^{-14}\ \mathrm{J\,m^{-3}}.
 \label{eq:cmb-energy-density}
\end{equation}
An exterior stationary balance with dilution rate $4H$ requires
$P_{\rm maint}(H)=4H\rho_\gamma$.  At the present-epoch normalization $H_0$ in
Eq.~\eqref{eq:helium-audit-inputs}, the required common-mode replenishment
power is
\begin{equation}
 P_{\rm maint}(H_0)=4H_0\rho_\gamma
 =3.64746\times10^{-31}\ \mathrm{W\,m^{-3}}
 \label{eq:maintenance-power}
\end{equation}
The power is finite and small.  Its spectral form is the stronger condition.
COBE/FIRAS constrains a Compton distortion to
$|y|<1.5\times10^{-5}$ at 95\% confidence \cite{FixsenEtAl1996}.
Therefore, in a maintenance model whose Hubble-time energy input would
otherwise be of order the background energy, the spectrally differential
component must be suppressed to roughly the $10^{-5}$ level.  A viable
maintenance channel must therefore act approximately along the Planck family
rather than as arbitrary heat injection.  Dividing this power by the minimal fixed-radius
return current gives Eq.~\eqref{eq:radiative-energy-per-returned-baryon}; the joint
energy ledger does not remove the spectral requirement.

\subsection{Equal-time and unequal-time stationary covariance}
\label{subsec:unequal-time}

Solving \eqref{eq:qed-lyapunov} and inserting the result into the full centered
Landau projection gives a positive stationary angular covariance once an
occupation and guiding-centre state are supplied.  Positivity alone does not
select an acoustic comb: time-local collision noise fixes an equal-time
covariance but supplies no independent sequence of observer--source lags.

That result does not imply that stationarity erases every acoustic phase.  A
noise-maintained damped mode
\begin{equation}
 \lambda_\pm(k)=-\epsilon(k)\pm\ii\Omega(k),
 \qquad \epsilon(k)>0,
 \label{eq:damped-acoustic-poles}
\end{equation}
has a stationary two-time correlator of the form
\begin{equation}
 C_k(\tau)=A_k\e^{-\epsilon(k)|\tau|}
 \cos\bigl(\Omega(k)\tau\bigr).
 \label{eq:stationary-oscillatory-correlation}
\end{equation}
An autocovariance is even, so it carries no arbitrary constant phase.  Matrix
cross-covariances may carry phase offsets, subject to
$C_{ab}(-\tau)=C_{ba}(\tau)^*$.  Neither case needs a retained initial phase or
a finite beginning.  A light-cone projection can convert temporal lag into
angular phase.

More generally, write the stationary unequal-time covariance of the native
moment vector as
\begin{equation}
 C_{ab}(\lambda;\tau)
 =\langle J_a(\lambda,t+\tau)J_b(\lambda,t)^*\rangle.
 \label{eq:unequal-time-covariance}
\end{equation}
If $R_X^a(\ell,\lambda,t)$ is the scalar or spin-two readout for sky channel
$X\in\{T,E\}$, the stationary light-cone target is
\begin{equation}
 C_\ell^{XY}
 =\int\dd\mu(\lambda)\int\dd t\,\dd t'\;
 R_X^a(\ell,\lambda,t)
 C_{ab}(\lambda;t-t')
 R_Y^b(\ell,\lambda,t')^*.
 \label{eq:unequal-time-sky-projection}
\end{equation}
The measure includes $b,n,\xi$ and guiding-centre multiplicity.  Equation
\eqref{eq:exact-spectrum-target} is the diagonal, scalar-source reduction of
this expression.

\begin{proposition}[Positivity under stationary projection]
\label{prop:projection-positivity}
If $C_{ab}(\lambda;\tau)$ is a positive-definite stationary matrix kernel, then
for each $\ell$ the matrix $(C_\ell^{XY})_{X,Y}$ defined by
\eqref{eq:unequal-time-sky-projection} is positive semidefinite.  In particular,
\begin{equation}
 |C_\ell^{TE}|^2\leq C_\ell^{TT}C_\ell^{EE}.
 \label{eq:te-cauchy-schwarz}
\end{equation}
\end{proposition}

\begin{proof}
For complex coefficients $c_X$, insert
$F^a(\lambda,t)=\sum_Xc_XR_X^a(\ell,\lambda,t)$ into the defining positivity
inequality for $C_{ab}$.  The resulting quadratic form is
$\sum_{X,Y}\overline c_XC_\ell^{XY}c_Y\geq0$.  The determinant inequality for
the $T,E$ principal block gives \eqref{eq:te-cauchy-schwarz}.
\end{proof}

This formulation makes clear why equal-time forcing is not the general
stationary problem.  Phase information can reside in $t-t'$ and in
noncommuting matrix channels even when every one-time law is translation
invariant.

The missing object is then the observer/source lag weight.  An arbitrary
chosen shell or delay distribution would simply insert the desired phase.  A
stationary account must derive the lag weight from null propagation, residence,
and the observer's accessible algebra.  The actual selector in
Section~\ref{sec:central-selection} does not supply a narrow weight:
Eq.~\eqref{eq:selector-first-passage} has coefficient of variation one.

For comparison, if a record must complete $m$ independent comparable stages,
an Erlang first-passage clock has fractional width $m^{-1/2}$.  Widths below
$0.25$, $0.10$, and $0.05$ require at least $16$, $100$, and $400$ stages.
Such a mechanism can remain stationary, but its stage count
and rates must follow from a physical generator rather than from the desired
peak coherence.

\subsection{A stationary coherent-return existence model}
\label{subsec:coherent-return}

A narrow chronological photosphere is not the only stationary way to retain a
lag.  An observer-relative return boundary can instead supply a fixed spatial
separation at every observation time.  The following construction isolates
the additional physics required by that possibility.

For one native spectral block, let
$\bm x=(x_1,x_2)^{\mathsf T}$ be the stationary solution of
\begin{equation}
 \dd\bm x=(-\epsilon I_2+\Omega J)\bm x\,\dd t
 +\sqrt{2\epsilon C}\,\dd W_t,
 \qquad
 J=\begin{pmatrix}0&-1\\1&0\end{pmatrix},
 \qquad \epsilon,C>0.
 \label{eq:stationary-carrier}
\end{equation}
Its equal-time covariance is $CI_2$, and for $\tau\geq0$
\begin{equation}
 \langle\bm x(t)\bm x(t-\tau)^{\mathsf T}\rangle
 =C\e^{-\epsilon\tau}\operatorname{Rot}(\Omega\tau),
 \label{eq:carrier-lag-covariance}
\end{equation}
where $\operatorname{Rot}$ is the planar rotation generated by $J$.  At a
return lag $\tau_H$,
write
\begin{equation}
 \rho:=\e^{-\epsilon\tau_H},
 \qquad 2\Phi:=\Omega\tau_H,
 \qquad 0\leq\rho\leq1.
 \label{eq:return-coherence-phase}
\end{equation}

\begin{hypothesis}[Coherent horizon return]
\label{hyp:coherent-return}
The observer has access to even and odd return combinations of the same
stationary carrier,
\begin{align}
 T&=A_Te_1^{\mathsf T}[\bm x(t)+\bm x(t-\tau_H)],
 \label{eq:return-temperature}\\
 E&=A_Ee_2^{\mathsf T}[\bm x(t)-\bm x(t-\tau_H)].
 \label{eq:return-polarization}
\end{align}
The odd readout includes a parity-even, trace-free photon analyzer.  Its
isotropic part annihilates the Planck monopole.
\end{hypothesis}

\begin{proposition}[Coherent-return covariance]
\label{prop:coherent-return}
The two readouts in \eqref{eq:return-temperature}--
\eqref{eq:return-polarization} have the common positive covariance
\begin{align}
 C^{TT}&=2CA_T^2[1+\rho\cos(2\Phi)],
 \label{eq:return-tt}\\
 C^{TE}&=2CA_TA_E\rho\sin(2\Phi),
 \label{eq:return-te}\\
 C^{EE}&=2CA_E^2[1-\rho\cos(2\Phi)],
 \label{eq:return-ee}
\end{align}
up to the fixed orientation convention for $E$, and
\begin{equation}
 \det C^{(T,E)}=4C^2A_T^2A_E^2(1-\rho^2)\geq0.
 \label{eq:return-determinant}
\end{equation}
In the coherent limit $\rho=1$,
\begin{equation}
 C^{TT}:C^{TE}:C^{EE}
 =4CA_T^2\cos^2\Phi:
 2CA_TA_E\sin(2\Phi):
 4CA_E^2\sin^2\Phi.
 \label{eq:return-phase-architecture}
\end{equation}
\end{proposition}

\begin{proof}
Insert Eq.~\eqref{eq:carrier-lag-covariance} and its transpose into the two
linear readouts.  The diagonal terms use
$\operatorname{Rot}(2\Phi)+\operatorname{Rot}(2\Phi)^{\mathsf T}
=2\cos(2\Phi)I_2$;
the cross term uses
$e_1^{\mathsf T}[\operatorname{Rot}(2\Phi)^{\mathsf T}
-\operatorname{Rot}(2\Phi)]e_2
=2\sin(2\Phi)$.  Taking the determinant
gives Eq.~\eqref{eq:return-determinant}.
\end{proof}

The carrier is a stationary process in its invariant law; the new constitutive
content is the observer-relative lag and the photon analyzer.  The latter cannot be
identified with lossless screen transport, by
Theorem~\ref{thm:unpolarized-fixed-point}.  It must be implemented by a noisy,
nonunitary, Planck-preserving interaction.

Boundary thickness gives a sharp coherence requirement.  If the return lag has
Gaussian jitter $\sigma_\tau$, the $N$th fringe contrast is multiplied by
\begin{equation}
 \exp\!\left[-\frac12
 \left(2\pi N\frac{\sigma_\tau}{\tau_H}\right)^2\right].
 \label{eq:return-jitter}
\end{equation}
Keeping at least one half, respectively ninety percent, of the eighth-fringe
contrast requires
\begin{equation}
 \frac{\sigma_\tau}{\tau_H}<0.0234,
 \qquad\text{respectively}\qquad
 \frac{\sigma_\tau}{\tau_H}<0.00913.
 \label{eq:return-thickness-bounds}
\end{equation}
These are boundary-thickness conditions, not the duration of a special cosmic
epoch.

\subsection{An additive pump-covariance obstruction}
\label{subsec:additive-pump-obstruction}

The coherent-return carrier also gives a useful no-go result for the simplest
interpretation of pumping.  Retain the drift
$A=-\epsilon I_2+\Omega J$, but allow additive marked-event noise to select an
arbitrary positive equal-time covariance
\begin{equation}
 \Sigma=\begin{pmatrix}a&d\\d&b\end{pmatrix}\succeq0.
 \label{eq:anisotropic-carrier-covariance}
\end{equation}
At lag $\tau_H$, set $\rho=\exp(-\epsilon\tau_H)$ and
$\psi=\Omega\tau_H$.  Stationarity gives
$K(\tau_H):=\mathbb E[\bm x(t)\bm x(t-\tau_H)^{\mathsf T}]
=\rho\operatorname{Rot}(\psi)\Sigma$.

\begin{proposition}[Additive-pump obstruction]
\label{prop:additive-pump-obstruction}
For the even and odd readouts in
Eqs.~\eqref{eq:return-temperature}--\eqref{eq:return-polarization}, the mixed
covariance is
\begin{equation}
 C^{TE}=A_TA_E\,\rho\sin\psi\,\operatorname{tr}\Sigma,
 \label{eq:additive-pump-te-trace}
\end{equation}
up to the fixed orientation of $E$.  The covariance anisotropy $a-b$ and the
equal-time cross term $d$ cancel identically.
\end{proposition}

\begin{proof}
The mixed covariance before multiplication by $A_TA_E$ is
\begin{equation*}
 e_1^{\mathsf T}\!\left[K(\tau_H)^{\mathsf T}
 -K(\tau_H)\right]e_2
 =\rho e_1^{\mathsf T}
 [\Sigma\operatorname{Rot}(-\psi)-\operatorname{Rot}(\psi)\Sigma]e_2.
\end{equation*}
Direct multiplication gives
$\rho(a+b)\sin\psi$.
\end{proof}

A block-dependent trace can still provide a nonnegative TE envelope, so the
proposition is not a claim that pumping cannot change lobe magnitudes.  It is
more specific: additive covariance with the drift and even--odd readout held
fixed cannot add an independent TE phase, move a zero, or distinguish the two
carrier quadratures.  Doing so requires a nontrivial photon jump/readout map
or pump-dependent dynamics such as
Eq.~\eqref{eq:cyclostationary-pump-lyapunov}.  This is why identifying the
mixed count cumulant is necessary but not sufficient.

\subsection{What the selector does not couple}
\label{subsec:selector-radiative-coupling}

The completed-event channel \eqref{eq:completed-event-map} acts on the central
canonical pair.  None of its displayed jumps acts on photon energy, photon
number, $\delta_\gamma$, $v_\gamma$, or $\Pi$.  Tensoring it with the local
QED generator therefore leaves
\begin{equation}
 \delta\dot T_\gamma=0
 \label{eq:temperature-zero-mode}
\end{equation}
and gives a derived selector-to-radiation noise map of rank zero.  The selector
gap $\gamma_{\rm sel}$ contracts central memory; it is not a photon-temperature
restoring rate or a polarization collision rate.

Even if one grants a common scalar coupling not present in the model, the
selector spectrum \eqref{eq:selector-lorentzian} is positive, monotone, and
one-scale.  Multiplication by this Lorentzian can change a smooth envelope but
does not select the repeated observer--source phase sequence absent from the
model.  Thus the displayed selector cannot be repurposed as the radiative
completion: a successful completion needs a genuinely photon-coupled,
matrix-valued response/noise kernel.

\subsection{Lossless geometric transport does not create polarization}
\label{subsec:geometric-polarization}

Represent a narrow photon bundle by its $2\times2$ screen coherency matrix
$P$.  An unpolarized beam has
\begin{equation}
 P_0=\frac{I}{2}\bm1_2,
 \qquad Q=U=V=0.
 \label{eq:unpolarized-coherency}
\end{equation}
In vacuum geometric optics, the ray follows a null geodesic, polarization is
parallel transported, and focusing multiplies both screen components by the
same scalar.  The resulting Jones action is
\begin{equation}
 P\longmapsto AUPU^*,
 \qquad A\geq0,\quad U\in U(2).
 \label{eq:lossless-jones-action}
\end{equation}
The corresponding Stokes transport is the standard lossless polarized
radiative-transfer law \cite{Chandrasekhar1960}.

\begin{theorem}[Unpolarized fixed point]
\label{thm:unpolarized-fixed-point}
Every finite composition of vacuum parallel transport, screen-basis rotation,
Berry phase, and scalar focusing preserves the unpolarized ray:
\begin{equation}
 AUP_0U^*=\frac{AI}{2}\bm1_2.
 \label{eq:unpolarized-fixed-point}
\end{equation}
Consequently the linear intensity-to-polarization source map has rank zero.
The screen--Weyl operator \eqref{eq:screen-weyl-operator} can shear a Jacobi
map and remap polarization already present, but cannot create scalar E
polarization from an unpolarized bath by lossless propagation alone.
\end{theorem}

\begin{proof}
The identity matrix commutes with every unitary matrix, so
$UP_0U^*=P_0$.  Scalar focusing changes only its common amplitude.  The same
argument applies to every composition.
\end{proof}

A nonunitary screen map $P\mapsto DPD^*$ with
$D^*D\not\propto I$ can create polarization.  It describes differential
emission, absorption, gain, scattering, or measurement and must be accompanied
by environmental noise in a completely positive stationary description.
Thus a microscopic polarizing interaction is indispensable; it cannot be
obtained by reinterpreting optical tidal shear as a dichroic Jones operator.

\subsection{Stress representation and a cold material state}
\label{subsec:stress-rank}

Let a marked microscopic event carry a symmetric stress moment $M^{\mu\nu}$.
For a Fourier block with wave vector in the $z$ direction, the Ward identity is
\begin{equation}
 \omega M^{0\nu}-kM^{3\nu}=0.
 \label{eq:event-ward-identity}
\end{equation}
The four independent constraints reduce the ten real components of a symmetric
tensor to six.  Under rotations about the wave vector these decompose into two
scalar, two vector, and two tensor amplitudes:
\begin{center}
\begin{tabular}{lcp{0.53\linewidth}}
\toprule
sector&real amplitudes&Ward content\\
\midrule
scalar&2&density fixes longitudinal momentum and longitudinal stress; transverse
scalar pressure remains independent\\
vector&2&each transverse momentum fixes its longitudinal--transverse stress\\
tensor&2&the two helicity-two stresses are unconstrained by
\eqref{eq:event-ward-identity}\\
\bottomrule
\end{tabular}
\end{center}
Consequently conservation does not force a common event source to rank one.
An isotropic positive event covariance can have rank two in each helicity
sector and rank six overall.  Ward symmetry supplies the admissibility
constraints while leaving the relative spectral weights as dynamical data.

For a conformal field the additional trace constraint removes one of the two
scalar amplitudes.  Free Maxwell stress therefore supplies only one scalar
stochastic direction, even though its quadratic composite has positive noise
and nonzero connected higher cumulants.  For example, if $x$ is Gaussian with
variance $v$ and $O=x^2-v$, then
\begin{equation}
 \langle O^2\rangle_c=2v^2,
 \qquad
 \langle O^3\rangle_c=8v^3,
 \qquad
 \frac{\langle O^3\rangle_c}{\langle O^2\rangle_c^{3/2}}=2\sqrt2.
 \label{eq:quadratic-gaussian-cumulants}
\end{equation}
Thus a Gaussian Maxwell field has a non-Gaussian stress tensor with cubic
normalization fixed by Wick contraction.  Its scalar representation at fixed
four-momentum is one-dimensional, but this fact alone is not a projected-rank
no-go.  One stochastic direction passed through nonproportional causal filters
can yield a rank-two integrated sky covariance.  Proposition
\ref{prop:coherent-return} is an explicit witness.  The rank count constrains
the microscopic ports, while subsequent lag and line-of-sight integration
determine the projected rank.

Massive matter supplies a second microscopic scalar direction.  If independent radiation and
nonrelativistic matter energy fluctuations have variances $A,B>0$, then in the
basis $(\epsilon,\Theta)$, with $\Theta=T^\mu{}_{\mu}$,
\begin{equation}
 \Sigma_{\epsilon\Theta}=
 \begin{pmatrix}A+B&-B\\-B&B\end{pmatrix},
 \qquad
 \det\Sigma_{\epsilon\Theta}=AB>0.
 \label{eq:trace-rank-two}
\end{equation}
This is a natural rank-two scalar sector of one canonically normalized stress
tensor, rather than two unrelated stochastic sources.

At $2.7\,\mathrm K$, however, ordinary equilibrium hydrogen is neutral to
overwhelming accuracy.  The Saha factor contains
\begin{equation}
 \exp\!\left(-\frac{\chi_H}{k_BT}\right),
 \qquad \frac{\chi_H}{k_BT}\simeq5.8\times10^4,
 \label{eq:saha-suppression}
\end{equation}
so the free-electron fraction and Thomson rate are effectively zero at cosmic
mean density.  Massive matter enlarges the scalar representation but does not
automatically provide an optically active stationary medium.  A cold
stationary realization would need a nonequilibrium ionization and energy
balance which derives its opacity, noise, and distortion control from the same
open-system law.

There is also an exact normalization ambiguity for an abstract marked-event
law.  If events of rate $\kappa$ carry independent marks $M$, their quadratic
and cubic noise cumulants are
\begin{equation}
 N_2=\kappa\E(MM^*),
 \qquad
 N_3=\kappa\E(M^{\otimes3}).
 \label{eq:event-cumulants}
\end{equation}
The transformation
\begin{equation}
 M\mapsto aM,
 \qquad
 \kappa\mapsto\kappa/a^2
 \label{eq:event-rescaling}
\end{equation}
leaves $N_2$ invariant but sends $N_3\mapsto aN_3$.  A quadratic sky
covariance therefore cannot separately determine an absolute event clock,
canonical mark normalization, or cubic coupling.  These quantities become
definite only when $M$ is a canonically normalized microscopic operator and
$\kappa$ is fixed by an independently calibrated proper-time generator.

\subsection{Constraints on a radiative closure}
\label{subsec:elimination-table}

The preceding results impose four independent requirements.  Translation of
the counted central coordinate cannot select its conjugate response;
low-order stationary invariants do not determine a retarded spectral kernel;
the displayed central selector has no photon jump operator; and lossless
screen transport cannot polarize an unpolarized bath.  Neither a single
exponential residence time nor KMS balance supplies the missing spectral and
proper-time data.  A completion must therefore contain a selected spectral
measure, a physical event chronology and observer--source lag, and a
nonunitary photon analyzer with its associated positive noise.  These are
requirements on the microscopic closure, not theorems against stationary
radiative states.

\subsection{The radiative completion problem}
\label{subsec:radiative-completion}

The positive results and obstructions of this section lead to a compact target.
A linearized completion should derive
\begin{align}
 \dot{\bm J}(\lambda,t)
 &=-\bsA(\lambda)\bm J(\lambda,t)
 +\bm\zeta(\lambda,t),
 \label{eq:target-radiative-drift}\\
 \langle\bm\zeta(\lambda,t)\bm\zeta(\lambda',t')^*\rangle
 &=\bsN(\lambda;t-t')\delta(\lambda-\lambda'),
 \qquad \bsN\succeq0,
 \label{eq:target-radiative-noise}
\end{align}
including its operator action on guiding-centre multiplicity.  In a Markov
limit the stationary covariance satisfies
\begin{equation}
 \bsA\bsSigma+\bsSigma\bsA^*=\bsN.
 \label{eq:target-lyapunov}
\end{equation}
The same microscopic construction must determine the photon readout and the
nonunitary polarizing channel.  It must preserve the Planck family to the
required accuracy while retaining colored unequal-time acoustic correlations.
Only then are TT, TE, and EE three projections of one stationary matrix
process rather than separately adjustable curves.

\section{CMB feature phenomenology and its limits}
\label{sec:cmb-checkpoint}

The preceding sections do not define a complete CMB transfer theory.  They do,
however, permit a narrower and useful diagnostic: whether the exact low-band
$N_{3,2}$ structure can organize the observed TT feature locations with a
small causal response and whether a separately specified stationary carrier
can reproduce the broad normalized peak-height ladder.  A still weaker
polarization question can also be asked without reopening the fit: whether the
quadrature of the TT-calibrated phase has the observed TE signs and EE
interleaving.  The results concern extracted feature locations and TT peak
heights; they do not define continuous temperature or polarization spectra.

\subsection{Retarded response on the quartic low band}

Let $q$ be a dimensionless longitudinal frequency.  The first three transverse
levels define
\begin{equation}
 q_n^2=(2n+1)q_0^2,
 \qquad n=0,1,2.
 \label{eq:low-band-frequencies}
\end{equation}
The pole scale is locked to the acoustic phase spacing by
\begin{equation}
 q_0=\frac{\pi}{\theta}.
 \label{eq:pole-lock}
\end{equation}
This is a phenomenological closure, not a theorem of the group geometry.  Let
three damped coordinates obey
\begin{equation}
 \ddot x_n+\gamma\dot x_n+q_n^2x_n=q_n^2s
 \label{eq:three-oscillators}
\end{equation}
and let the output be $s_{\rm out}=s+g_{\rm r}\sum_nw_nx_n$.  With the quartic residues
\eqref{eq:141-residues}, the retarded susceptibility is
\begin{equation}
 \mathcal R(q)
 =1+g_{\rm r}\sum_{n=0}^2w_n
 \frac{q_n^2}{q_n^2-q^2-\ii\gamma q},
 \qquad(w_0,w_1,w_2)=(1,4,1).
 \label{eq:quartic-response}
\end{equation}
For $\gamma>0$ its poles lie in the lower half plane.  The response is stable
and causal, and its phase is tied to its modulus.

Evaluate the Abelian angular kernel on its leading ridge
\begin{equation}
 q_\ell=\ell+\frac12,
 \qquad
 B_\ell=\ell(\ell+1)j_\ell(q_\ell)^2,
 \label{eq:abelian-ridge}
\end{equation}
and define the location diagnostic
\begin{equation}
 D_\ell^{\rm loc}\propto
 B_\ell
 \left\{\Re\left[
 \mathcal R(q_\ell)\e^{\ii(\theta q_\ell+\varphi)}
 \right]\right\}^2.
 \label{eq:location-law}
\end{equation}
This formula locates extrema.  It does not supply the occupation, visibility,
guiding-centre average, or polarization covariance in
\eqref{eq:exact-spectrum-target}.

It nevertheless has an exact stationary interpretation.  In Proposition
\ref{prop:coherent-return}, take $\rho=1$,
\begin{equation}
 \Phi(q)=\theta q+\varphi+\arg\mathcal R(q),
 \qquad
 CA_T^2=B_\ell|\mathcal R(q)|^2.
 \label{eq:location-stationary-identification}
\end{equation}
Then
\begin{equation}
 \frac14C_\ell^{TT}
 =B_\ell\left\{\Re\left[
 \mathcal R(q_\ell)\e^{\ii(\theta q_\ell+\varphi)}
 \right]\right\}^2.
 \label{eq:location-stationary-embedding}
\end{equation}
Thus the calibrated phase law can be embedded in a positive stationary
covariance.  This is an existence statement, not a derivation of
the return channel.

\subsection{TT feature-location calibration}

Using the first eight quoted TT peak locations and one-standard-deviation
uncertainties
\begin{align}
 \ell_p^{\rm obs}
 &=(220.0,537.5,810.8,1120.9,1444.2,1776.0,2081.0,2395.0),
 \label{eq:tt-locations}\\
 \sigma_p
 &=(0.5,0.7,0.7,1.0,1.1,5.0,25.0,24.0),
 \label{eq:tt-uncertainties}
\end{align}
reported from the Planck 2015 feature analysis
\cite{Planck2015XI,PanEtAl2016}, minimize
\begin{equation}
 \chi_\ell^2
 =\sum_{p=1}^8
 \frac{(\ell_p^{\rm mod}-\ell_p^{\rm obs})^2}{\sigma_p^2}.
 \label{eq:tt-location-objective}
\end{equation}
This use of the 2015 extraction keeps the TT, TE, and EE feature lists tied to
one published analysis.  The Planck 2018 values in
Eq.~\eqref{eq:helium-audit-inputs} serve only as the present-epoch density
normalization.
The four fitted parameters are
\begin{equation}
 \boxed{
 \theta=0.00953068,\quad
 \varphi=-1.16737,\quad
 g_{\rm r}=-0.105822,\quad
 \gamma=517.202,}
 \label{eq:tt-fit-parameters}
\end{equation}
giving
\begin{equation}
 q_0=329.629,
 \qquad q_1=570.935,
 \qquad q_2=737.074,
 \label{eq:tt-pole-values}
\end{equation}
and
\begin{equation}
 \boxed{
 \chi_\ell^2=5.10,\qquad
 \mathrm{dof}=4,\qquad
 p=0.277.}
 \label{eq:tt-fit-result}
\end{equation}
The individual locations are shown in Table~\ref{tab:tt-locations}.

\begin{table}[ht]
\centering
\caption{TT feature-location calibration to the Planck 2015 peak estimates.}
\label{tab:tt-locations}
\begin{tabular}{crrrr}
\toprule
$p$&observed&uncertainty&model&residual\\
\midrule
1&220.0&0.5&219.992&$-0.008$\\
2&537.5&0.7&537.438&$-0.062$\\
3&810.8&0.7&811.293&$+0.493$\\
4&1120.9&1.0&1119.840&$-1.060$\\
5&1444.2&1.1&1444.594&$+0.394$\\
6&1776.0&5.0&1772.566&$-3.434$\\
7&2081.0&25.0&2101.435&$+20.435$\\
8&2395.0&24.0&2430.649&$+35.649$\\
\bottomrule
\end{tabular}
\end{table}

Matched controls help identify the useful structure.  Each causal control uses
Eq.~\eqref{eq:quartic-response} with the indicated subset of poles and the same
four adjustable quantities $(\theta,\varphi,g_{\rm r},\gamma)$; the
Bessel/acoustic ridge sets $g_{\rm r}=0$ and therefore has only
$(\theta,\varphi)$.  The corresponding location fits are
\begin{center}
\begin{tabular}{lrr}
\toprule
model&parameters&$\chi^2/\mathrm{dof}$\\
\midrule
Bessel/acoustic ridge&2&$1161.05/6$\\
causal pair $(n=0,1)$ with residues $1{:}1$&4&$27.50/4$\\
causal quartic triplet $1{:}4{:}1$&4&$5.10/4$\\
isolated $n=1$ pole&4&$5.34/4$\\
triplet with residues $1{:}1{:}1$&4&$52.05/4$\\
\bottomrule
\end{tabular}
\end{center}
The data favour a causal response concentrated near the middle Landau
frequency.  They are consistent with the economical quartic packet, but do not
uniquely measure its residue ratio: the isolated middle pole performs similarly.

\subsection{A normalized TT peak-height diagnostic}
\label{subsec:tt-height-diagnostic}

The retarded response used to locate extrema does not by itself specify a
stationary occupation.  For a separate amplitude diagnostic, freeze all four
location parameters in Eq.~\eqref{eq:tt-fit-parameters} and retain only the
response phase $\Phi$ in Eq.~\eqref{eq:location-stationary-identification}.
Introduce the one-pole carrier occupation and Gaussian power visibility
\begin{equation}
 C_H(q)=\frac{1}{1+(q/\gamma)^2},
 \qquad
 V_{\zeta}(q)=
 \exp\!\left[-\left(\frac{q}{\zeta\gamma}\right)^2\right].
 \label{eq:tt-carrier-occupation-visibility}
\end{equation}
At the reference event-selector point in Appendix~\ref{app:event-model},
$\gamma_4=2$ and $\Phi_{\rm ev}=1/2$.  Identifying the ratio of the carrier
persistence length to the visibility width with
\begin{equation}
 \zeta_{\rm vis}:=\frac{\gamma_4}{\Phi_{\rm ev}}=4
 \label{eq:tt-renewal-visibility-ratio}
\end{equation}
is a constitutive projection of the finite renewal model; it uses no TT height
data.  For the height diagnostic, one global loading amplitude is calibrated
as $\alpha_H=0.1896$ in the lowest analytic closure across the quartic band
gap,
\begin{equation}
 \Delta_H(q)=\frac{\alpha_H}{1+q^2/(4q_0^2)}.
 \label{eq:tt-selector-loading}
\end{equation}
These identifications have not been derived from a complete photon--matter
generator.

Because the fitted feature centres are noninteger while the sky multipole is
integer, use the lower multipole in the angular kernel:
\begin{equation}
 q_p=\ell_p^{\rm obs}+\frac12,
 \qquad
 \widetilde B_p
 =\ell_p^{\rm obs}(\ell_p^{\rm obs}+1)
 j_{\lfloor\ell_p^{\rm obs}\rfloor}(q_p)^2.
 \label{eq:tt-height-kernel-convention}
\end{equation}
The conditional carrier shape and amplitude are then
\begin{align}
 S_p^{\rm car}
 &=\widetilde B_p C_H(q_p)V_4(q_p)
 \left[\cos\Phi(q_p)+\Delta_H(q_p)\right]^2,
 \label{eq:tt-carrier-height-shape}\\
 D_p^{\rm car}&=\mathcal N_D S_p^{\rm car}.
 \label{eq:tt-carrier-height-law}
\end{align}
The common normalization $\mathcal N_D$ is then fitted to the height data.  The
comparison therefore has one global height-shape parameter and one overall
amplitude in addition to the preceding location calibration.

Let
$D_p^{\rm obs}=\ell_p(\ell_p+1)C_{\ell_p}^{TT}/(2\pi)$ in
$\mu\mathrm K^2$, with the peak amplitudes and diagonal errors quoted in the
Planck power-spectrum analysis \cite{Planck2015XI}.
Table~\ref{tab:tt-heights} gives the
comparison at the normalization minimizing the diagonal statistic.

\begin{table}[ht]
\centering
\small
\caption{Toy carrier comparison with the eight extracted TT peak heights.
One global loading amplitude and one common normalization are calibrated.}
\label{tab:tt-heights}
\begin{tabular}{crrrrr}
\toprule
$p$&$\ell_p$&observed&uncertainty&carrier&fractional residual\\
\midrule
1&220.0&5717.0&35.0&5359.3&$-0.063$\\
2&537.5&2582.0&11.0&2652.5&$+0.027$\\
3&810.8&2523.0&10.0&2676.6&$+0.061$\\
4&1120.9&1237.0&4.0&1232.9&$-0.003$\\
5&1444.2&797.1&3.1&747.6&$-0.062$\\
6&1776.0&377.4&2.9&371.5&$-0.016$\\
7&2081.0&214.0&4.0&231.8&$+0.083$\\
8&2395.0&105.0&4.0&116.3&$+0.107$\\
\bottomrule
\end{tabular}
\end{table}

For the shape measure, profile the same single normalization in logarithmic
space.  The two diagnostics are
\begin{align}
 \mathrm{RMS}_{\log}
 &:=\min_{\mathcal N>0}
 \left[\frac18\sum_{p=1}^8
 \log^2\!\left(\frac{\mathcal NS_p^{\rm car}}{D_p^{\rm obs}}\right)
 \right]^{1/2}
 =0.059,
 \label{eq:tt-height-log-rms}\\
 \chi_{D,\mathrm{diag}}^2
 &:=\min_{\mathcal N_D}\sum_{p=1}^8
 \frac{(\mathcal N_DS_p^{\rm car}-D_p^{\rm obs})^2}
 {\sigma_{D,p}^2}
 =669.3.
 \label{eq:tt-height-diagonal-statistic}
\end{align}
The logarithmic RMS corresponds to a typical multiplicative shape discrepancy
$\e^{0.059}-1\simeq0.061$.  The carrier therefore captures the factor-of-$54$
falling ladder, including the near equality of peaks two and three, to about
six per cent.  The much larger error-weighted statistic simultaneously rules
it out as a precision peak-amplitude fit.  This is toy-level morphological
agreement, not a CMB likelihood result.

\subsection{A frozen-phase polarization topology test}
\label{subsec:frozen-polarization-topology}

The coherent-return identity suggests a question substantially weaker than a
joint TT--TE--EE fit.  Once the phase $\Phi$ has been fixed by TT alone, does
its stationary quadrature put the polarization features in the right order
and give TE the right alternating sign?  This question introduces no
polarization phase, pole, visibility, shell displacement, or featurewise
coefficient.

Write
\begin{equation}
 c_\ell:=\cos\Phi(q_\ell),
 \qquad s_\ell:=\sin\Phi(q_\ell),
 \qquad
 \mathcal P_\ell:=B_\ell|\mathcal R(q_\ell)|^2.
 \label{eq:frozen-phase-factors}
\end{equation}
The Abelian ground block of the exact spin-raised identity
\eqref{eq:spin-raised-abelian-limit} supplies the positive relative weight
\begin{equation}
 W_E(\ell)
 :=\frac{(\ell+2)!}{(\ell-2)!\,q_\ell^4}
 =\frac{(\ell-1)\ell(\ell+1)(\ell+2)}{q_\ell^4}>0,
 \qquad \ell\geq2.
 \label{eq:frozen-spin-weight}
\end{equation}
It tends to one on the high-$\ell$ ridge and changes neither the phase nor the
TE sign.  A minimal positive sky covariance with this common provenance is
\begin{equation}
 \mathbf C_\ell^{\rm toy}
 =\mathcal P_\ell
 \begin{pmatrix}
 N_Tc_\ell^2&
 \eta_{\rm pol}\sqrt{N_TN_EW_E(\ell)}\,c_\ell s_\ell\\[1mm]
 \eta_{\rm pol}\sqrt{N_TN_EW_E(\ell)}\,c_\ell s_\ell&
 N_EW_E(\ell)s_\ell^2
 \end{pmatrix},
 \qquad |\eta_{\rm pol}|\leq1,
 \label{eq:frozen-polarization-covariance}
\end{equation}
where $N_T,N_E>0$ are global channel normalizations and
$\eta_{\rm pol}$ is one global analyzer coherence.  Indeed,
\begin{equation}
 \det\mathbf C_\ell^{\rm toy}
 =(1-\eta_{\rm pol}^2)N_TN_E\mathcal P_\ell^2
 W_E(\ell)c_\ell^2s_\ell^2\geq0.
 \label{eq:frozen-polarization-determinant}
\end{equation}
At $|\eta_{\rm pol}|=1$ this is the coherent quadrature in
Eq.~\eqref{eq:return-phase-architecture}; smaller absolute values give its
minimal positive decoherence.  The TT entry is, up to $N_T$, exactly the
location law \eqref{eq:location-law}.  Thus no TT parameter is changed.
Neither $N_E$ nor $\eta_{\rm pol}$ enters the location and sign test below,
and neither is fitted.

The carrier loading $\Delta_H$ and the factors $C_HV_4$ are deliberately not
promoted into Eq.~\eqref{eq:frozen-polarization-covariance}.  They were
introduced only for the sampled TT height diagnostic, not as a continuous
transfer law.  That diagnostic remains intact as a separate toy amplitude
result.  Combining the two closures would falsely turn them into a
continuous-spectrum model that has not been derived.

The numerical construction uses the $N_{3,2}$ spatial spectrum.  The natural
spacetime parent $N_{4,2}=M\oplus\Lambda^2M$ may house a covariant open-ended
return observable, but its local spectral action supplies no additional
observable phase; no $N_{4,2}$ phase parameter is used here.  Nor does the
construction refer to an initial hypersurface, a primordial phase choice, or
inflationary preparation.  Its physical assumption is instead the coherent
observer-relative lag isolated in Hypothesis~\ref{hyp:coherent-return}.

The extrema of the phase-only TE factor occur at
$|\sin(2\Phi)|=1$.  Table~\ref{tab:frozen-te-topology} compares their ordered
locations with the twelve TE extrema extracted from Planck 2015 by
Pan et al.~\cite{Planck2015XI,PanEtAl2016}.  The last column evaluates the
signed phase factor at the observed location.  One global choice of the
$E$-orientation has been made; it is not changed between features.

\begin{table}[ht]
\centering
\small
\caption{Frozen-TT-phase comparison with the measured TE extrema.  No TE
parameter is fitted.  The model column is the ordered solution of
$|\sin(2\Phi)|=1$.}
\label{tab:frozen-te-topology}
\begin{tabular}{crrrr}
\toprule
$j$&observed $\ell_j^{TE}$&phase extremum&residual&
$\sin[2\Phi(\ell_j^{TE})]$\\
\midrule
1 &$150.0\pm0.8$  &  61.9&$-88.1$&$-0.369$\\
2 &$308.5\pm0.4$  & 272.9&$-35.6$&$+0.822$\\
3 &$471.2\pm0.4$  & 445.7&$-25.5$&$-0.867$\\
4 &$595.3\pm0.7$  & 592.4&$ -2.9$&$+0.998$\\
5 &$746.7\pm0.6$  & 736.4&$-10.3$&$-0.975$\\
6 &$916.9\pm0.5$  & 885.5&$-31.4$&$+0.800$\\
7 &$1070.4\pm1.0$ &1041.2&$-29.2$&$-0.837$\\
8 &$1224.0\pm1.0$ &1201.1&$-22.9$&$+0.902$\\
9 &$1371.7\pm1.2$ &1363.2&$ -8.5$&$-0.987$\\
10&$1536.0\pm2.8$ &1526.5&$ -9.5$&$+0.983$\\
11&$1693.0\pm3.3$ &1690.3&$ -2.7$&$-0.999$\\
12&$1861.0\pm4.0$ &1854.4&$ -6.6$&$+0.992$\\
\bottomrule
\end{tabular}
\end{table}

All twelve signs agree.  Except for the lowest extremum, every observed
feature lies on a strong lobe, with
$|\sin(2\Phi)|\geq0.800$ to the displayed precision.  The location residuals
have
\begin{equation}
 \operatorname{RMS}_{TE,\,1\ldots12}=32.1,
 \qquad
 \operatorname{RMS}_{TE,\,2\ldots12}=20.5
 \label{eq:frozen-te-rms}
\end{equation}
in multipole units, and eleven of the twelve are within $36$ multipoles.  The
first TE extremum is not well located.  Relative to the quoted observational
errors, none of these numbers constitutes a precision fit.

The phase-only EE maxima satisfy $s_\ell^2=1$.  The resulting sequence begins
at
\begin{equation}
 \ell=(363.3,664.0,962.6,1282.0,\ldots).
 \label{eq:frozen-ee-sequence}
\end{equation}
Its comparison with the five EE peaks in the same feature analysis is shown in
Table~\ref{tab:frozen-ee-topology}.

\begin{table}[ht]
\centering
\small
\caption{Frozen-TT-phase EE interleaving.  The first measured EE peak is not
reproduced; the remaining four are compared in order.}
\label{tab:frozen-ee-topology}
\begin{tabular}{crrr}
\toprule
$j$&observed $\ell_j^{EE}$&phase maximum&residual\\
\midrule
1&$137.0\pm6.0$  &---&---\\
2&$397.2\pm0.5$  &363.3&$-33.9$\\
3&$690.8\pm0.6$  &664.0&$-26.8$\\
4&$992.1\pm1.3$  &962.6&$-29.5$\\
5&$1296.0\pm4.0$ &1282.0&$-14.0$\\
\bottomrule
\end{tabular}
\end{table}

For the four matched higher peaks the multipole RMS is $27.1$.  Their
placement between successive TT features is the expected quadrature topology,
but the missing first EE peak is a real low-$\ell$ failure.  The frozen phase
therefore passes a limited qualitative test: it gets the complete TE sign
sequence and the higher TE/EE ordering with no polarization phase fit.  It
does not determine polarization amplitudes, widths, damping, the low-$\ell$
EE structure, or a likelihood-quality joint spectrum.  Choosing vertical
normalizations for a three-curve plot would add no invariant test beyond these
phase factors, so the comparison is reported directly in the tables.

\subsection{An expansion-conditioned long-mode scattering closure}
\label{subsec:long-mode-scattering-gate}

The two conspicuous low-multipole defects of the frozen-phase test are related:
the first measured EE maximum is absent, and the first negative TE extremum is
predicted far too low.  Both occur where $\theta q_\ell=O(1)$, before the
higher quadrature comb has settled.  This makes a low-pass polarization source
a more economical possibility than an additional $N_{4,2}$ eigenvalue or a
new polarization phase.

The physical roles must be separated.  A localized expansion or ``tile''
event may gate a long-wavelength scalar geometric mode, but lossless geometry
does not scatter photons and cannot polarize an unpolarized beam by
Theorem~\ref{thm:unpolarized-fixed-point}.  Here the geometric mode supplies a
parity-even trace-free quadrupole, while nonequilibrium material in the baryon
return channel supplies the genuinely nonunitary scattering or analyzer.  No
absorption--reemission process is assigned to tile creation itself.

This division is compatible with the local hierarchy already used above.  For
a harmonic mode proportional to $\exp(-\ii\omega t)$, the quadrupole row of
Eq.~\eqref{eq:qed-moment-generator} gives
\begin{equation}
 \Pi=-\frac{8}{15}\frac{\ii k v_\gamma}
 {(9/10)\nu-\ii\omega},
 \qquad
 S_P:=\nu\Pi.
 \label{eq:long-mode-quadrupole-response}
\end{equation}
Thus $k$, $\nu$, and $\omega$ all of order $H$ give an integrated
polarization response $H^{-1}S_P/v_\gamma=O(1)$: neither tight-coupling
suppression nor weak-scattering suppression is forced.  This is the same local
quadrupole-to-polarization logic as ordinary Thomson polarization
\cite{HuWhite1997,HuDodelson2002}; it does not import a primordial transfer
spectrum or an initial hypersurface.

The mean-density rate shows why the scattering must be event-local.  Even if
all mean-density baryons were ionized,
\begin{equation}
 \nu_T^{\rm max}=n_b\sigma_Tc_{\rm light}
 =5.01\times10^{-21}\ {\rm s}^{-1}
 =2.29\times10^{-3}H_0
 \label{eq:mean-thomson-rate-bound}
\end{equation}
using the same present-epoch normalization.  The actual equilibrium rate at
$2.7\,\mathrm K$ is negligible by Eq.~\eqref{eq:saha-suppression}.  On the
other hand, the minimal fixed-radius return current supplies an exact counting
identity.  If $N_{\rm sc}$ is the number of photon interactions associated
with one horizon-crossing returned baryon and $\nu_{\rm eff}$ is the
corresponding rate per background
photon, then
\begin{equation}
 \nu_{\rm eff}n_\gamma=(3Hn_b)N_{\rm sc},
 \qquad
 \frac{\nu_{\rm eff}}H
 =3\frac{n_b}{n_\gamma}N_{\rm sc}.
 \label{eq:return-scattering-count}
\end{equation}
Reaching $\nu_{\rm eff}=H$ would require
\begin{equation}
 N_{\rm sc}=\frac{n_\gamma}{3n_b}
 =5.46\times10^8
 \label{eq:return-scattering-count-number}
\end{equation}
interactions per horizon-crossing returned baryon.  This is within one order of magnitude of
the photon ledger in Eq.~\eqref{eq:joint-ledger-photon-yield}; ionizing one
hydrogen atom once costs only $13.6\,\mathrm{eV}/1.38\,\mathrm{MeV}\simeq
9.9\times10^{-6}$ of the gross energy ledger.  These comparisons establish
only counting and energetic room.  They do not provide the required column
density, ionization duty cycle, cross section, or Planck-preserving reaction
network.

For a minimal angular model, let the event excite a regular scalar field
$\psi$ whose screen quadrupole is
$\operatorname{STF}(\nabla_A\nabla_B)\psi$.  Two transverse derivatives give
a small-wave-number amplitude proportional to $x^2$.  Choose a minimal
one-cell Gaussian envelope and lock its angular coherence to
the TT scale already fixed in Eq.~\eqref{eq:tt-fit-parameters}:
\begin{equation}
 x_\ell:=\theta q_\ell,
 \qquad
 G_\ell:=x_\ell^2\exp(-x_\ell^2/2).
 \label{eq:long-mode-gate}
\end{equation}
This is an ansatz, but it adds no angular scale or phase.  The rise as $x^2$
also has the same qualitative effect identified in detailed peak analyses:
the first polarization feature is unusually sensitive to the growth of the
quadrupole with wave number \cite{PanEtAl2016}.  By itself $G_\ell^2$ peaks at
\begin{equation}
 \ell_G=\frac{\sqrt2}{\theta}-\frac12=147.9,
 \label{eq:long-mode-isolated-ee-location}
\end{equation}
and the correlated factor $-c_\ell G_\ell$ has its negative minimum at
$\ell=154.4$.  At $\ell=308$, $G_\ell^2$ is only $0.0243$ of its maximum, so
the source switches off before the established higher comb.

There is a direct positive covariance extension of
Eq.~\eqref{eq:frozen-polarization-covariance}.  With one nonnegative global
gate strength $\epsilon_{\rm g}$, define
\begin{equation}
 \mathbf C_\ell^{\rm gate}
 =\mathcal P_\ell
 \begin{pmatrix}
 N_Tc_\ell^2&
 \sqrt{N_TN_EW_E}\,c_\ell
 (\eta_{\rm pol}s_\ell-\epsilon_{\rm g}G_\ell)\\[1mm]
 \sqrt{N_TN_EW_E}\,c_\ell
 (\eta_{\rm pol}s_\ell-\epsilon_{\rm g}G_\ell)&
 N_EW_E\!\left[s_\ell^2+\epsilon_{\rm g}^2G_\ell^2
 -2\epsilon_{\rm g}\eta_{\rm pol}s_\ell G_\ell\right]
 \end{pmatrix},
 \label{eq:long-mode-gated-covariance}
\end{equation}
where $W_E=W_E(\ell)$.  It is the covariance of the explicit latent readouts
\begin{equation}
 T\propto c_\ell\zeta_1,
 \qquad
 E\propto s_\ell
 \left(\eta_{\rm pol}\zeta_1
 +\sqrt{1-\eta_{\rm pol}^2}\,\zeta_2\right)
 -\epsilon_{\rm g}G_\ell\zeta_1,
 \label{eq:long-mode-latent-readout}
\end{equation}
for independent unit-variance $\zeta_1,\zeta_2$.  Consequently
\begin{equation}
 \det\mathbf C_\ell^{\rm gate}
 =N_TN_E\mathcal P_\ell^2W_Ec_\ell^2s_\ell^2
 (1-\eta_{\rm pol}^2)\geq0.
 \label{eq:long-mode-gated-determinant}
\end{equation}
The determinant is independent of $\epsilon_{\rm g}$, and the TT entry is
exactly unchanged.  The displayed relative sign is the scalar-scattering
orientation that reinforces the observed negative low-$\ell$ TE lobe;
reversing it fails that sign.  A microscopic analyzer must ultimately derive
this discrete orientation and $\epsilon_{\rm g}$.

The carrier-weighted test below is distinct from the phase-only diagnostic in
Tables~\ref{tab:frozen-te-topology} and \ref{tab:frozen-ee-topology}.  At
$\epsilon_{\rm g}=0$, the full $\mathcal P_\ell W_E$ entry has EE maxima at
\begin{equation}
 \ell^{EE}_{\rm carrier}
 =(41.4,396.8,670.6,961.7,1281.5),
 \label{eq:carrier-weighted-ee-control}
\end{equation}
while its TE sequence begins
$(97.3^-,291.9^+,458.8^-,598.0^+,738.2^-,\ldots)$.  Thus the common TT
carrier, rather than the long-mode term, accounts for the displacement of the
higher carrier-weighted extrema relative to the pure phase sequence.  The
additional term is used only on the low-$\ell$ shoulder and the first two TE
lobes.

For illustration, take the coherent limit
$\eta_{\rm pol}=1$ and the literal order-unity choice
$\epsilon_{\rm g}=1$.  Evaluating the full carrier-weighted entries of
Eq.~\eqref{eq:long-mode-gated-covariance}, rather than only their phase
factors, gives the five EE maxima
\begin{equation}
 \ell^{EE}_{\rm gate}
 =(125.2,399.7,670.6,961.7,1281.5)
 \label{eq:long-mode-ee-sequence}
\end{equation}
and the twelve signed TE extrema
\begin{multline}
 \ell^{TE}_{\rm gate}
 =(148.6^-,304.1^+,458.9^-,598.0^+,738.2^-,885.4^+,\\
 1040.6^-,1200.8^+,1363.1^-,1526.5^+,1690.4^-,1854.6^+).
 \label{eq:long-mode-te-sequence}
\end{multline}
Against the feature centres in Tables~\ref{tab:frozen-te-topology} and
\ref{tab:frozen-ee-topology}, the corresponding multipole RMS values are
\begin{equation}
 \operatorname{RMS}_{EE,\,1\ldots5}=18.4,
 \qquad
 \operatorname{RMS}_{TE,\,1\ldots12}=15.5.
 \label{eq:long-mode-feature-rms}
\end{equation}
In particular, the previously absent first EE feature is at $125.2$ rather
than $137.0\pm6.0$, and the first TE minimum is at $148.6$ rather than
$150.0\pm0.8$.  The new EE lobe is weak: its power is $0.066$ of the strongest
EE lobe in this dimensionless example.  No feature location was optimized
with polarization data.  Varying $0.8\leq\epsilon_{\rm g}\leq1.6$ retains
exactly five EE maxima and twelve TE extrema over the measured ranges; the
first EE maximum moves from $118.1$ to $137.4$ and the first TE minimum from
$143.7$ to $157.8$, while the higher sequence is essentially unchanged.
Every $|\eta_{\rm pol}|<1$ gives a positive rank-two covariance; the coherent
limit is used here only to expose the feature locations with no further fitted
number.

Three ingredients remain before this can be promoted beyond a toy model.  First,
$O(H)$ is a temporal rate statement, not an angular projection.  If the
geometric mode also has $k_{\rm g}\simeq H_{\rm ret}/c_{\rm light}$, the usual
projection $\ell\simeq k_{\rm g}\chi_{\rm ret}$ requires
\begin{equation}
 \frac{H_{\rm ret}\chi_{\rm ret}}{c_{\rm light}}
 \simeq\frac{\sqrt2}{\theta}=148.4,
 \label{eq:long-mode-projection-gate}
\end{equation}
not the value of order one supplied by one present Hubble radius.  The return
boundary or native projection must derive this factor.  Second, a homogeneous
$\nu\sim H$ acting for a Hubble time has optical depth of order one and would
erase rather than merely supplement the high-$\ell$ TT pattern.  The channel
must therefore be the primary return readout or have a derived small covering
factor, duty cycle, or integrated optical depth.  Third, the scalar Hessian
used here produces parity-even E polarization; generic tensor or handed Weyl
content would also produce B modes and is independently constrained.

The result is therefore a qualitative existence result: baryon-return
stoichiometry and expansion-event geometry have enough counting room to house
an $O(H)$, low-pass polarizing source, and the minimal one-scale Gaussian
Hessian ansatz places the absent EE lobe and misplaced TE minimum in the right
region without disturbing TT.  It is not yet an explanation of the opacity,
projection, or amplitude.

\paragraph{Additive control.}
An independent temperature--quadrupole event port can improve TE and EE values
sampled at their feature centres, but its positive temperature variance fills
the zeros of the calibrated TT carrier and moves the continuous TT extrema.
Constraining the added temperature path to remain proportional to the frozen
TT phase removes the TE improvement.  This numerical control is the
feature-level counterpart of Proposition~\ref{prop:additive-pump-obstruction}:
the mixed second cumulant in Eq.~\eqref{eq:radiative-mark-cumulants} is
necessary, but it cannot act only as an independent additive temperature
source.  A viable return event must instead modify the temperature--quadrupole
propagator through the existing material dynamics.

\subsection{Event-conditioned opacity}
\label{subsec:opacity-modulation-gate}

The multiplicative continuation can now be tested without adding a temperature
port.  Freeze every TT parameter, set the mean opacity scale to the already
present damping scale $\nu_0=\gamma$, and define the logarithmic quadrupole
response
\begin{equation}
 K_\nu(q)
 :=\left.\frac{\partial}{\partial\log\nu}
 \log\frac{\nu}{(9/10)\nu-\ii q}\right|_{\nu_0}
 =\frac{q^2-\ii(9/10)\nu_0q}{(9\nu_0/10)^2+q^2}.
 \label{eq:opacity-feature-response}
\end{equation}
Conditioning on the coherent-return event phase gives the correlated
polarization coefficient
\begin{equation}
 e_\ell^{(\nu)}
 :=s_\ell-G_\ell+\epsilon\left[
 \Re K_\nu(q_\ell)s_\ell+\Im K_\nu(q_\ell)c_\ell\right].
 \label{eq:opacity-feature-coefficient}
\end{equation}
The first correction rescales the existing E quadrature; the second mixes in
its conjugate quadrature through the causal collision phase.  Neither term
adds an independent temperature variable or an adjustable phase.

The collision noise in Eq.~\eqref{eq:opacity-modulated-collision-noise} must
be projected through the same event.  Introduce the broad finite-collision
factor
\begin{equation}
 f_E(q):=\frac{q/\gamma}{\sqrt{1+(q/\gamma)^2}}
 \label{eq:finite-collision-factor}
\end{equation}
and retain one independent unit variable $\zeta_2$ for the quadrupole noise
port.
The resulting continuous covariance is generated by
\begin{equation}
 T_\ell=\sqrt{\mathcal P_\ell}\,c_\ell\zeta_1,\qquad
 E_\ell=\sqrt{\mathcal P_\ell W_E(\ell)}
 \left[e_\ell^{(\nu)}\zeta_1+b_\Pi f_E(q_\ell)\zeta_2\right],
 \label{eq:opacity-latent-readout}
\end{equation}
where $\zeta_1,\zeta_2$ are independent and unit variance.  It is positive,
leaves TT
exactly unchanged, and has
\begin{equation}
 \det\mathbf C_\ell^{(\nu)}
 =\mathcal P_\ell^2W_E(\ell)c_\ell^2b_\Pi^2f_E(q_\ell)^2\geq0.
 \label{eq:opacity-covariance-determinant}
\end{equation}
The coefficient $b_\Pi$ is a sky-projection coefficient for the positive local
quadrupole noise.  Its value is not determined by the local coefficient $9/5$
in Eq.~\eqref{eq:collision-noise}; that requires the line-of-sight map.  The
construction is nevertheless useful: positivity is manifest, TT is exactly
the frozen carrier, and the opacity response changes E through a causal phase
rather than an independent temperature source.  Direct feature scans confirm
that this minimal form preserves the TT, TE, and EE extrema, but it does not
reproduce the observed TE and EE height pattern with their common
normalization.  The result motivates a derived event-age law rather than an
additional stationary polarization amplitude.

\subsection{Common-event closure of the opacity response}
\label{subsec:common-event-third-cumulant}

The mixed moment in Corollary~\ref{cor:event-conditioned-dynamics} need not be
left as a formal unknown if the acoustic forcing and opacity mark have a common
event ancestry.  Let $\widetilde{\mathcal N}(\dd s,\dd\alpha)$ be a compensated
stationary marked-Poisson random measure with intensity
$\lambda\,\dd s\,\rho(\dd\alpha)$.  For every $n\geq2$ its connected
cumulants obey
\begin{equation}
 \operatorname{cum}\!\left[
 \widetilde{\mathcal N}(f_1),\ldots,
 \widetilde{\mathcal N}(f_n)\right]
 =\lambda\int\dd s\int\rho(\dd\alpha)
 \prod_{j=1}^n f_j(s,\alpha).
 \label{eq:poisson-common-event-cumulants}
\end{equation}
Let $G_0(t)=\e^{L_0t}\bm1_{t\geq0}$ be the stable local moment propagator,
$J(\alpha)$ the photon--baryon jump mark, $a(\alpha)$ the opacity mark, and
$h_a$ a causal age kernel.  The common-event carrier and centered opacity are
\begin{align}
 Y_0(t)&=\int_{-\infty}^t\!\int
 G_0(t-s)J(\alpha)\,
 \widetilde{\mathcal N}(\dd s,\dd\alpha),
 \label{eq:common-event-carrier}\\
 m(t)&=\int_{-\infty}^t\!\int
 h_a(t-s)a(\alpha)\,
 \widetilde{\mathcal N}(\dd s,\dd\alpha).
 \label{eq:common-event-opacity}
\end{align}

\begin{proposition}[Common-event closure of the mixed opacity moment]
\label{prop:common-event-mixed-closure}
For the process in
Eqs.~\eqref{eq:common-event-carrier}--\eqref{eq:common-event-opacity},
the connected moment required at first order in the multiplicative opacity is
\begin{multline}
 \mathcal M_3(u;t,t')
 :=\mathbb E\!\left[m(u)Y_0(t)Y_0(t')^*\right]_{\!c}\\
 =\lambda\int_{-\infty}^{\min(u,t,t')}\!\dd s
 \int\rho(\dd\alpha)\;h_a(u-s)a(\alpha)
 G_0(t-s)J(\alpha)J(\alpha)^*G_0(t'-s)^*.
 \label{eq:common-event-mixed-moment}
\end{multline}
If
\begin{equation}
 Y_1(t)=\int_{-\infty}^tG_0(t-u)m(u)L_\nu Y_0(u)\,\dd u,
 \label{eq:common-event-duhamel}
\end{equation}
then the common-jump contribution to the unequal-time covariance is
\begin{equation}
 \delta C(t,t')
 =\int_{-\infty}^tG_0(t-u)L_\nu
 \mathcal M_3(u;u,t')\,\dd u
 +(t\leftrightarrow t')^*.
 \label{eq:common-event-covariance-correction}
\end{equation}
Thus the first-order $\mathbb E[mYY^*]$ closure is fixed once the common event
marks, age kernel, and local propagator are fixed; no independent event phase
is required.
\end{proposition}

\begin{proof}
Apply Eq.~\eqref{eq:poisson-common-event-cumulants} with the three kernels in
Eqs.~\eqref{eq:common-event-carrier} and
\eqref{eq:common-event-opacity}.  Causality restricts the common ancestor to
$s\leq\min(u,t,t')$ and gives
Eq.~\eqref{eq:common-event-mixed-moment}.  Substitution of the Duhamel
variation \eqref{eq:common-event-duhamel} into
$\mathbb E[Y_1(t)Y_0(t')^*]+
\mathbb E[Y_0(t)Y_1(t')^*]$ gives
Eq.~\eqref{eq:common-event-covariance-correction}.
\end{proof}

This is a closure identity, not yet a prediction.  The selector displayed in
Section~\ref{subsec:selector-radiative-coupling} has $J=0$, so assigning its
completed events the nonzero marks $J$ and $a$ is a new radiative
identification.  In addition, the second cumulant does not fix the third.  For
a scalar common mark $M$, the transformation
$M\mapsto rM$, $\lambda\mapsto\lambda/r^2$ leaves
$\lambda\mathbb E[M^2]$ invariant but multiplies
$\lambda\mathbb E[M^3]$ by $r$.  The baryon cadence and a two-point
normalization therefore do not determine the dimensionless cubic strength of
the event.

The minimal one-stage frequency shape can nevertheless be tested.  Project
the carrier and opacity kernels to scalar exponentials,
\begin{equation}
 H_c(q)=\frac1{\gamma_c-\ii q},
 \qquad
 H_a(q)=\frac1{\gamma_a-\ii q}.
 \label{eq:common-event-scalar-kernels}
\end{equation}
For white compound-Poisson forcing, the bispectral convolution is
\begin{align}
 S_{Y_0,mY_0}(q)
 &=\lambda\mathbb E[M^3]H_c(q)
 \int\frac{\dd\omega}{2\pi}
 H_a(\omega)^*H_c(q-\omega)^*
 \nonumber\\
 &=\lambda\mathbb E[M^3]
 \frac{H_c(q)}{\gamma_c+\gamma_a+\ii q},
 \label{eq:common-event-cross-spectrum}\\
 S_{Y_0Y_0}(q)&=\lambda\mathbb E[M^2]|H_c(q)|^2.
 \label{eq:common-event-carrier-spectrum}
\end{align}
After absorbing the moment ratio and harmless pulse normalization into one
real coefficient $\zeta_3$, the normalized causal shape is therefore
\begin{equation}
 F_3(q):=\frac{\gamma_c+\ii q}
 {\gamma_c+\gamma_a+\ii q}.
 \label{eq:common-event-third-transfer}
\end{equation}
Its real and imaginary parts are linked; it supplies no fitted lag or
orientation.  The sign of $\zeta_3$ is the orientation of the physical common
mark.

For a zero-new-scale reference, identify $\gamma_c$ with the fitted carrier
damping $\gamma$.  At the selector point
\eqref{eq:event-base-point}, $\mu=1$ and $\gamma_4=2$.  If the same
time-to-angular conversion maps $\gamma_4$ to $\gamma$, the existing
single-hazard ratio gives
\begin{equation}
 \frac{\gamma_a}{\gamma_c}=\frac{\mu}{\gamma_4}=\frac12.
 \label{eq:common-event-selector-rate-ratio}
\end{equation}
This common conversion is itself constitutive; the selector does not derive a
light-cone distance.  It is the strongest parameter-free use of its displayed
clock.

Combine Eq.~\eqref{eq:common-event-third-transfer} with the local opacity
response \eqref{eq:opacity-feature-response}:
\begin{align}
 Z_3(q)&:=K_\nu(q)F_3(q),
 \label{eq:common-event-combined-transfer}\\
 e_\ell^{(3)}&:=s_\ell-G_\ell+\zeta_3
 \left[\Re Z_3(q_\ell)s_\ell+
 \Im Z_3(q_\ell)c_\ell\right].
 \label{eq:common-event-feature-coefficient}
\end{align}
The positive feature covariance is represented by
\begin{equation}
 T_\ell=\sqrt{\mathcal P_\ell}\,c_\ell\zeta_1,
 \qquad
 E_\ell=\sqrt{\mathcal P_\ell W_E(\ell)}
 \left[e_\ell^{(3)}\zeta_1+b_\perp f_E(q_\ell)\zeta_2\right],
 \label{eq:common-event-positive-readout}
\end{equation}
with independent unit variables $\zeta_1,\zeta_2$.  Hence TT remains exactly
frozen and
\begin{equation}
 \det\mathbf C_\ell^{(3)}
 =\mathcal P_\ell^2W_E(\ell)c_\ell^2
 b_\perp^2f_E(q_\ell)^2\geq0.
 \label{eq:common-event-positive-determinant}
\end{equation}
The coefficient $b_\perp$ includes the component of the quadratic Poisson
output orthogonal to $\zeta_1$ and the projected collision noise.  It is controlled
by fourth moments and the line-of-sight map, not by the third-cumulant identity.
Setting $b_\perp=0$ is therefore a coherence-saturated cubic-only control;
carrying $b_\perp$ forward retains, rather than derives, the positive residual
port already exposed in the opacity model.

The identity removes the phase-average obstruction without introducing a
fitted event lag: the causal phase is fixed by the common ancestry and the age
kernel.  For the selector's single-hazard clock,
$\gamma_a/\gamma_c=1/2$, the corresponding feature scan preserves the frozen
TT topology but does not materially improve the TE/EE heights.  Allowing the
age rate to float improves TE only by driving the cubic correction beyond its
controlled first-order range and does not restore the second EE lobe.  The
mathematical conclusion is therefore the closure formula above; the numerical
conclusion is only that the displayed one-stage clock is insufficient.

\subsection{Single-stage matrix test}
\label{subsec:full-matrix-common-event-gate}

The scalar closure could discard a useful velocity, slip, or quadrupole
direction.  To test that possibility, propagate a deterministic common mark
through the complete local $5\times5$ generator.  This calculation does not
supply a Boltzmann hierarchy or a time-to-light-cone map; it tests only whether
the displayed one-stage selector contains an additional matrix response.

For each $k$, let the negative-frequency acoustic eigenpair of
Eq.~\eqref{eq:qed-moment-generator} be
\begin{equation}
 L(k)J(k)=\left[-d(k)-\ii\omega_c(k)\right]J(k),
 \qquad J_{\delta_\gamma}(k)=1,
 \label{eq:matrix-common-event-acoustic-mode}
\end{equation}
and define the resolvent
$\mathscr R_\omega(k)=(-\ii\omega I-L(k))^{-1}$.  Use one scalar compensated
Poisson record $\widetilde n$ with deterministic common marks for both the
acoustic jump and the opacity age,
\begin{equation}
 Y_0(\omega)=\mathscr R_\omega J\,\widetilde n(\omega),
 \qquad
 m(\Omega)=H_a(\Omega)\widetilde n(\Omega),
 \qquad
 H_a(\Omega)=\frac{A_h}{\gamma_a-\ii\Omega}.
 \label{eq:matrix-common-event-inputs}
\end{equation}
At the reference selector point, the same constitutive conversion used in
Eq.~\eqref{eq:common-event-selector-rate-ratio} fixes
\begin{equation}
 \lambda=\frac{\Phi_{\rm ev}}{\gamma_4}\gamma=\frac\gamma4,
 \qquad
 \gamma_a=\frac\mu{\gamma_4}\gamma=\frac\gamma2.
 \label{eq:matrix-common-event-rates}
\end{equation}
Normalize the centered opacity pulse to unit variance.  Since
$\Var m=\lambda A_h^2/(2\gamma_a)$, this gives
\begin{equation}
 A_h=\sqrt{\frac{2\gamma_a}{\lambda}}=2.
 \label{eq:matrix-common-event-age-normalization}
\end{equation}
The signed logarithmic modulation $\epsilon$ is then the only scanned
quantity, with $|\epsilon|\leq1$ in
Eq.~\eqref{eq:mean-preserving-opacity}.  Reversing its sign reverses the common
mark orientation; it is not a fitted phase.

The first Duhamel correction in frequency space is
\begin{equation}
 Y_1(\omega)=\mathscr R_\omega L_\nu
 \int\frac{\dd\Omega}{2\pi}\,
 H_a(\Omega)\mathscr R_{\omega-\Omega}J\,
 \widetilde n(\Omega)\widetilde n(\omega-\Omega).
 \label{eq:matrix-common-event-duhamel}
\end{equation}
Although this is matrix valued, the convolution is exact.  Equation
\eqref{eq:matrix-common-event-acoustic-mode} gives
\begin{equation}
 \mathscr R_{\omega-\Omega}J
 =\frac{J}{d-\ii(\omega-\Omega-\omega_c)}.
 \label{eq:matrix-common-event-eigen-resolvent}
\end{equation}
Put $\Delta=\omega-\omega_c$, $s_a=\gamma_a+d$, and
$u=\mathscr R_\omega L_\nu J$.  The three required scalar integrals are
\begin{align}
 \mathcal G
 &:=\int\frac{\dd\Omega}{2\pi}
 \frac{A_h}{(\gamma_a-\ii\Omega)
 [d-\ii(\Delta-\Omega)]}
 =\frac{A_h}{s_a-\ii\Delta},
 \label{eq:matrix-common-event-G}\\
 I_1
 &:=\int\frac{\dd\Omega}{2\pi}
 \left|\frac{A_h}{(\gamma_a-\ii\Omega)
 [d-\ii(\Delta-\Omega)]}\right|^2
 =\frac{A_h^2s_a}
 {2\gamma_a d(s_a^2+\Delta^2)},
 \label{eq:matrix-common-event-I1}\\
 I_2
 &:=\int\frac{\dd\Omega}{2\pi}
 \frac{A_h}{(\gamma_a-\ii\Omega)
 [d-\ii(\Delta-\Omega)]}
 \left[\Omega\leftrightarrow\omega-\Omega\right]^*
 =\frac{2A_h^2s_a}
 {(s_a^2+\omega_c^2)(s_a^2+\Delta^2)}.
 \label{eq:matrix-common-event-I2}
\end{align}
These closed forms avoid a numerical ambiguity at low $k$, where the acoustic
width $d(k)$ is much narrower than $\gamma_a$.

For nonzero $\omega$, the Poisson cumulant identity
\eqref{eq:poisson-common-event-cumulants} now gives
\begin{align}
 S_{00}&=\lambda hh^*,& h&:=\mathscr R_\omega J,
 \label{eq:matrix-common-event-S00}\\
 S_{01}&=\lambda h\mathcal G^*u^*,
 \label{eq:matrix-common-event-S01}\\
 S_{11}&=uu^*\left[\lambda^2(I_1+I_2)
 +\lambda|\mathcal G|^2\right].
 \label{eq:matrix-common-event-S11}
\end{align}
The two $I$ terms are the nonzero second-cumulant pairings in the fourth
moment.  The last term is the connected Poisson fourth cumulant.  Thus no
independent fourth-order or polarization-noise coefficient is required for
this deterministic-mark realization.

To connect the matrix state to the frozen feature quadratures, let
\begin{equation}
 t=(1,0,0,0,0),\qquad p=(0,0,\nu,0,0),\qquad
 a_0=\frac{ph}{th},\qquad
 B=\begin{pmatrix}t\\p/a_0\end{pmatrix},
 \label{eq:matrix-common-event-readout}
\end{equation}
so that the unmodulated complex temperature carrier and normalized
polarization source have the same amplitude.  With $z=th$ and $w=Bu$, the
coherent conditional response and the orthogonal residual are
\begin{align}
 \bm K&=\frac{w\mathcal G}{z}
 =\begin{pmatrix}K_T\\K_E\end{pmatrix},
 \label{eq:matrix-common-event-K}\\
 \Sigma_\perp&:=\frac{BS_{11}B^*}{\lambda|z|^2}-\bm K\bm K^*
 =\lambda(I_1+I_2)
 \frac{w}{z}\frac{w^*}{z^*}\succeq0.
 \label{eq:matrix-common-event-Q}
\end{align}
The connected fourth-cumulant term cancels exactly from the conditional
residual; it is already the coherent $\bm K\bm K^*$ piece.  For the real
temperature quadrature and imaginary polarization quadrature, set
\begin{equation}
 D=\operatorname{diag}(1,-\ii),
 \qquad \Sigma_{\perp,{\rm r}}:=\Re(D\Sigma_\perp D^*)\succeq0.
 \label{eq:matrix-common-event-real-Q}
\end{equation}
Writing $\bm K=(K_T,K_E)^{\mathsf T}$, the frozen phase coefficients become
\begin{align}
 t_\ell&=c_\ell+\epsilon
 [\Re K_T\,c_\ell-\Im K_T\,s_\ell],
 \label{eq:matrix-common-event-t-feature}\\
 e_\ell&=s_\ell-G_\ell+\epsilon
 [\Re K_E\,s_\ell+\Im K_E\,c_\ell].
 \label{eq:matrix-common-event-e-feature}
\end{align}
The continuous feature covariance is the congruence
\begin{equation}
 \mathbf C_\ell=
 \mathcal P_\ell
 \begin{pmatrix}1&0\\0&\sqrt{W_E(\ell)}\end{pmatrix}
 \left[
 \begin{pmatrix}t_\ell\\e_\ell\end{pmatrix}
 \begin{pmatrix}t_\ell&e_\ell\end{pmatrix}
 +\epsilon^2\Sigma_{\perp,{\rm r}}\right]
 \begin{pmatrix}1&0\\0&\sqrt{W_E(\ell)}\end{pmatrix}
 \succeq0.
 \label{eq:matrix-common-event-positive-covariance}
\end{equation}
Unlike the scalar feature model, Eq.~\eqref{eq:matrix-common-event-positive-covariance}
retains the temperature response and therefore tests rather than assumes TT
phase preservation.  For the sampled heights only,
$c_\ell$ in Eq.~\eqref{eq:matrix-common-event-t-feature} is replaced by the
existing $c_\ell+\Delta_H(q_\ell)$ before applying the frozen carrier
envelopes.

The local collision noise is modulated by the same positive factor as the
drift, as required by Eq.~\eqref{eq:opacity-modulated-collision-noise}.  If its
white innovation is independent of the event record, the mean-preserving
parametrization gives
$\mathbb E[N_{\rm coll}(t)]=N_{\rm coll,0}$ exactly and supplies no new
first-order common TE term.  Its baseline sky projection and its higher-order
feedback through the modulated drift remain undetermined.  They are therefore
not reintroduced here as a fitted $b_\perp$; doing so would return to the
preceding phenomenological model rather than complete the common-event test.

The construction does not determine which local time frequency corresponds to
$q_\ell=\ell+1/2$.  Two fixed conventions bracket the issue without fitting a
lag:
\begin{equation}
 \text{native frequency: }\omega=\omega_c(k),
 \qquad
 \text{feature frequency: }\omega=k=q_\ell.
 \label{eq:matrix-common-event-frequency-controls}
\end{equation}
The first convention follows the acoustic pole; the second is the frequency
used by the feature response.  Under the positivity and continuous-extrema
constraints, the two conventions give TE NRMSE $0.815$ and $0.822$ and EE
logarithmic RMS $0.510$ and $0.506$.  Forcing the native-frequency response to
TE NRMSE $0.562$ moves the TT maxima by $57$ multipoles RMS and creates an
additional TE extremum.  Thus the full local matrix does not cure the
single-stage amplitude problem.  This statement does not constrain a derived
multistage age law, a physical mark distribution, or the unequal-time
light-cone projection.

\subsection{Scope of the feature results}
\label{subsec:cmb-checkpoint-scope}

The TT calculation concerns eight extracted features, not a likelihood fit to
the continuous spectrum.  The location law is a four-parameter calibration;
the separate one-pole carrier captures the normalized peak ladder at toy-model
accuracy but predicts neither the troughs nor the widths and is excluded by its
diagonal amplitude statistic.

The polarization results are more limited but less heavily calibrated.  With
the TT phase fixed, the coherent-return quadrature gives all twelve observed
TE signs and the higher EE interleaving.  The expansion-conditioned long mode
adds the absent first EE lobe and relocates the lowest TE extremum without a
new angular scale or a change to TT.  Its opacity, occupation, projection, and
normalization remain constitutive.

The event calculations identify why this does not yet become a quantitative
polarization spectrum.  Additive temperature forcing conflicts with the TT
phase; opacity modulation avoids that conflict but requires an event-conditioned
age law.  Common Poisson ancestry fixes the associated mixed third cumulant,
yet the displayed single-hazard clock remains inadequate even when propagated
through the full local matrix.  A precision comparison would require the
guiding-centre occupation, nonunitary polarizing interaction, and unequal-time
light-cone projection to be fixed by one microscopic model.  They are not
independently fitted here.

\subsection{The unresolved state-selection problem}

The remaining closure problem is concentrated in one microscopic object.  It
must assign baryon, energy, and directional marks to the horizon event; map
those marks into the spatial central response; and determine the stationary
matter--metric--photon covariance, including opacity chronology and the
light-cone readout.  The structural results above constrain such an object but
do not determine it.

A natural microscopic target is a Feynman--Vernon exterior influence
functional \cite{FeynmanVernon1963}, built here from canonically normalized
stress and current operators,
\begin{equation}
 \begin{aligned}
 \mathcal F[J_+,J_-]
 &:=\e^{\ii S_{\rm IF}[J_+,J_-]}\\
 &=\Tr_{\rm ext}\!\left(
 U[J_+]\rho_{\rm ext}U[J_-]^*\right).
 \end{aligned}
 \label{eq:influence-target}
\end{equation}
Here $U[J]$ is the source-dependent, time-ordered exterior evolution.  The
appropriate average/difference variations give the mean and retarded response;
the second connected variation gives the noise, and higher variations give the
nonlinear cumulants.  Stress and charge Ward identities act on all of them.  To
be predictive, the exterior split, smearing scale, proper-time calibration,
microscopic state, and operator normalization must be physical inputs of one
construction rather than adjustable spectral windows.

A modular or KMS condition would constrain such a law but would not determine
it.  In a spectral block it relates noise and dissipation schematically by the
fluctuation--dissipation relation \cite{Kubo1957}
\begin{equation}
 N_A(\omega)=
 \coth\!\left(\frac{\beta_{\rm mod}\omega}{2}\right)
 \Im G_{R,A}(\omega).
 \label{eq:modular-fdt}
\end{equation}
For fixed $\beta_{\rm mod}$, every admissible positive spectral density
$\rho_A(\omega)$ defines a different solution through
$\Im G_{R,A}=\rho_A$.  Detailed balance is a relation, not a selection of the
spectral density.  Moreover, modular time is dimensionless until a map
$s=\kappa_{\rm mod}\tau$ to observer proper time is fixed.  Rescaling
$\kappa_{\rm mod}$ leaves the abstract modular orbit unchanged while changing
every physical frequency.  The proper-time calibration and the resolution
scale of the algebraic split are therefore substantive clauses of any modular
horizon influence law.

Once a microscopic field content, stationary exterior state, fixed split, and
proper-time normalization are specified, Eq.~\eqref{eq:influence-target} is in
principle predictive: its connected correlators fix the quadratic response,
noise, and nonlinear vertices without separately fitting a higher-order event
coupling.  Without those data, the same formula is a useful organizing
principle but not a closure.

This target is compatible with a stationary exterior and does not require a
prepared initial acoustic phase.  It does require new microscopic content
beyond the current central selector: a common matter--metric--photon influence
law with a nonunitary polarizing channel and a derived observer/source lag
weight.

\section{Phase-soft propagation and distance consistency}
\label{sec:phase-soft}

The expansion of the spacetime remains physical in this construction.  If an
additional infrared channel contributes a coherent frequency dilation, its
logarithmic shift composes with the metric redshift:
\begin{equation}
 \ln(1+z_{\rm tot})
 =\ln(1+z_{\rm exp})+u_{\rm IR}.
 \label{eq:composed-log-redshift}
\end{equation}
This section does not fit $u_{\rm IR}(z)$ or assert that it is nonzero.  It
derives the kinematic and spectral conditions any such contribution must
satisfy.

\subsection{A coherent log-frequency channel}
\label{subsec:coherent-log-frequency}

Let
\begin{equation}
 \vartheta:=\ln(f/f_*),
 \qquad
 \mathscr H_{\rm ray}
 =L^2(\R,\dd\vartheta)\otimes\mathscr H_{\rm dir,pol},
 \qquad
 \mathsf D=-\ii\partial_\vartheta.
 \label{eq:log-frequency-space}
\end{equation}
On its standard Sobolev domain, $\mathsf D$ is self-adjoint.  The unitary
translation
\begin{equation}
 \mathsf U_u=\exp(\ii u\mathsf D),
 \qquad
 (\mathsf U_u\psi)(\vartheta,\hat{\bm n},P)
 =\psi(\vartheta+u,\hat{\bm n},P)
 \label{eq:log-frequency-translation}
\end{equation}
acts trivially on direction and polarization.  A component emitted at
$\vartheta_{\rm em}$ appears at
$\vartheta_{\rm obs}=\vartheta_{\rm em}-u$, so
\begin{equation}
 f_{\rm obs}=\e^{-u}f_{\rm em}.
 \label{eq:phase-soft-frequency-map}
\end{equation}

\begin{proposition}[Coherent phase-soft redshift]
\label{prop:coherent-redshift}
The channel in Eq.~\eqref{eq:log-frequency-translation} satisfies
\begin{align}
 \mathsf U_{u_2}\mathsf U_{u_1}&=\mathsf U_{u_1+u_2},
 \label{eq:redshift-group-law}\\
 \widehat s_{\rm obs}(f)&\propto
 \widehat s_{\rm em}(\e^uf),
 \qquad
 s_{\rm obs}(t)\propto s_{\rm em}(\e^{-u}t),
 \label{eq:waveform-dilation}\\
 \frac1{\exp[h\e^uf/(k_BT)]-1}
 &=\frac1{\exp[hf/(k_BT\e^{-u})]-1}.
 \label{eq:planck-dilation-covariance}
\end{align}
It is therefore achromatic, gives full-waveform time dilation, and maps a
Planck occupation at $T$ to one at $T\e^{-u}$.
\end{proposition}

\begin{proof}
The first identity is the translation group law.  The spectral and temporal
relations follow from translation in $\vartheta$ and Fourier scaling.  Direct
substitution in the Bose occupation number gives the last identity.
\end{proof}

This is coherent dilation, not absorption followed by re-emission.  It is
unitary in the wave-action measure $\dd f/f$ but reduces the photon energy
by $\e^{-u}$; a closed photon--environment model must carry the complementary
energy current.  Selecting $\mathsf D$ and deriving $u_{\rm IR}$ from the BSN
generator remain constitutive tasks.

\subsection{Differential shifts and blackbody mixing}
\label{subsec:differential-redshift}

A common value of $u$ merely shifts a fitted blackbody temperature.  A spread
of unresolved shifts is observable.  Write
\begin{equation}
 u=\bar u+\delta u,
 \qquad \E\delta u=0,
 \qquad
 \Sigma_{u,{\rm diff}}^2=\E(\delta u)^2.
 \label{eq:differential-log-redshift}
\end{equation}
The received spectrum is then a mixture of blackbodies with temperatures
$T\e^{-\bar u-\delta u}$ and is not itself exactly Planckian.

\begin{proposition}[Leading blackbody-mixing distortion]
\label{prop:blackbody-mixing}
After removing the best-fit mean temperature, a narrow distribution of
differential logarithmic shifts produces
\begin{equation}
 y_{\rm mix}=\frac12\Sigma_{u,{\rm diff}}^2
 +O\!\left(\E|\delta u|^3\right).
 \label{eq:blackbody-mixing-y}
\end{equation}
Consequently a spectral bound $|y|\leq y_{\max}$ requires
\begin{equation}
 \Sigma_{u,{\rm diff}}^2\lesssim2y_{\max}.
 \label{eq:differential-redshift-bound}
\end{equation}
\end{proposition}

\begin{proof}
Each realization has fractional temperature perturbation
$\Theta=\e^{-\delta u}-1=-\delta u+O(\delta u^2)$.  Expanding the Planck
spectrum to second order and absorbing the pure temperature displacement into
the fitted mean leaves the standard mixing distortion
$y_{\rm mix}=\tfrac12\E\Theta^2+O(\Theta^3)$
\cite{ChlubaSunyaev2004BlackbodyMixing}.  Substitution gives the result.
\end{proof}

The COBE/FIRAS limit is of order $y_{\max}\sim10^{-5}$
\cite{FixsenEtAl1996}.  Any post-thermalization infrared redshift of order
unity must therefore be overwhelmingly coherent across the photons and paths
being averaged, or be assembled from extremely soft, strongly correlated
increments.  This constraint cannot be traded for an angular smoothing
parameter.  It also connects directly to
Section~\ref{subsec:planck-maintenance}: Planck-family maintenance and
phase-soft propagation must be two aspects of the same photon energy balance.

\subsection{Surface brightness, reciprocity, and a BAO identity}
\label{subsec:propagation-consistency}

Suppose the completed ray dynamics is Hamiltonian and preserves the specific
intensity invariant $I_f/f^3$.  For the coherent contribution in
Eq.~\eqref{eq:phase-soft-frequency-map},
\begin{equation}
 I_{f,{\rm obs}}(f)
 =\e^{-3u}I_{f,{\rm em}}(\e^uf).
 \label{eq:specific-intensity-dilation}
\end{equation}
Integration over observed frequency supplies one further factor of $\e^{-u}$:
\begin{equation}
 I_{{\rm bol},{\rm obs}}
 =(1+z_{\rm IR})^{-4}I_{{\rm bol},{\rm em}}.
 \label{eq:tolman-infrared-scaling}
\end{equation}
When the metric and infrared channels compose coherently, the same argument
uses $z_{\rm tot}$ in the full Tolman factor.  If, in addition, photon number is
conserved and the complete source--observer Jacobi map obeys reciprocity, then
\begin{equation}
 D_L=(1+z_{\rm tot})^2D_A
 \label{eq:distance-duality}
\end{equation}
as in Etherington's theorem \cite{Etherington1933Reciprocity}.  The one-ray
unitary alone does not prove reciprocity; that is a condition on the covariant
ray-bundle completion.

There is also a useful ruler consistency relation which does not depend on a
particular redshift mechanism.  Let one isotropic ruler of length $r_*$ be
represented in one locally flat or effective ruler coordinate $\chi(z)$, so
that
\begin{equation}
 \theta_{\rm BAO}(z)=\frac{r_*}{\chi(z)},
 \qquad
 \Delta z_{\rm BAO}(z)=\frac{r_*}{\dd\chi/\dd z}.
 \label{eq:bao-ruler-definitions}
\end{equation}
Then identically
\begin{equation}
 \boxed{
 \frac{\dd}{\dd z}\left(\frac1{\theta_{\rm BAO}}\right)
 =\frac1{\Delta z_{\rm BAO}}.}
 \label{eq:bao-jacobian-identity}
\end{equation}
This is a stringent internal check when transverse and radial rulers are
modeled together, but it cannot by itself distinguish metric expansion from an
additional coherent dilation or decide whether $r_*$ was defined in physical
or comoving units.  The identity assumes that the same $\chi$ is used in both
projections; a curved transverse map contributes its additional Jacobian.
Those distinctions require the complete distance, clock-dilation, lensing, and
structure-growth predictions.

\section{Discussion and conclusion}
\label{sec:conclusion}

In this framework, stationarity is not stasis.  The spacetime expands, and
constant physical baryon and radiation
densities require the exchange terms displayed in
Eqs.~\eqref{eq:baryon-balance}--\eqref{eq:radiation-balance}.  Charge and energy
remain globally accounted for in the complementary sector.  For a boundary at
constant physical radius, the $3Hn_B$ term is exactly the baryon flux through
that boundary.  It is not an additional loss on top of horizon crossing.  If
the complementary baryon charge does not accumulate secularly, one baryon
returns for every crossing and the exterior density remains constant.

Composition adds an independent condition: the exact fraction identity forces
a strong form of isotope-resolved return.  The helium balance estimate shows
that a nearly pure-hydrogen return is not repaired by ordinary stellar
processing.  If the required helium fraction is assembled within the return
event, its one-pass binding-energy ceiling exceeds the radiative maintenance
cost: the latter is about $1.38\,\mathrm{MeV}$ per horizon-crossing returned
baryon, roughly $80\%$ of the ceiling.  The nominal $0.35\,\mathrm{MeV}$
difference is not renewable fuel in a closed composition-preserving cycle;
the export or horizon step must pay to reset the higher-free-energy reactants.
If tile insertion is only the microscopic realization of ordinary expansion,
it has no additional energy debit beyond the dust and radiation fluxes.  Any
nonzero extra tile cost requires a derived gravitational boundary charge.
These statements establish conditional compatibility; the composition,
reaction network, and horizon converter remain inputs to the microscopic
completion.

The geometric mechanism starts from a BSN first-cumulant connection.  Pullback
along resolved spatial controls gives horizontal lifts whose bracket is the
current curvature; at full rank in three dimensions the tangent is
$N_{3,2}$.  Translation of its central coordinate cannot select the conjugate
central response.  The event-conditioned Weyl channel supplies an explicit
completely positive selector, while the autonomous horizon ring supplies a
finite-state pump with
$\dot S_{\rm prod}=\mathcal A_HJ_H$ and the BSN holonomy recovered per winding
in the adiabatic limit.

This construction separates entropy production from useful power.  The same
horizon affinity admits different energy marks, so it does not determine the
return-energy current.  Under the charge-cyclic, entropy-open hypothesis, the
exported generalized energy carries the returned baryon rest mass and useful
work, while the horizon retains entropy and the heat fixed by its first law.
For the stated de Sitter temperature this heat is
$\epsilon_{H/B}^{\rm heat}=(\hbar H/2\pi)\mathcal A_H\mathcal N_{H/B}$, but the
windings-per-baryon normalization is not yet derived.  A cold horizon permits
an efficient converter; it does not provide one merely by possessing a large
entropy.

In a disc-normal sector that pseudovector becomes an azimuthal connection.  The
resulting circular-orbit equation
$v^2-q_\varpi v-v_{\rm N}^2=0$ and its inverse are exact consequences of one
canonical Hamiltonian.  They identify the response demanded by any measured
rotation curve.  A screened bulk closure supplies precisely the $1/r$
intermediate profile needed for a flat response-dominated branch and predicts
a finite turnover scale.  Its source moment is not fixed by the exterior PDE.
The baryonic Tully--Fisher normalization follows from the isolated source
matching in Hypothesis~\ref{hyp:btfr-source-matching}, and only from that
hypothesis.  The immediate empirical test is therefore well defined:
reconstruct $\varpi(r)$ from rotation curves and baryonic potentials, test the
screened form, and then test the amplitude law and its environmental scatter.
A covariant null coupling and lensing law remain indispensable parts of the
completion.

The $N_{3,2}$ harmonic analysis gives an exact Landau--longitudinal spectrum,
centered sky kernels, and a conditional quartic packet with residues
$1{:}4{:}1$.  It does not select a radiative state: stationary normalization,
a gap, and limited memory data leave the retarded matrix covariance
underdetermined.  Local photon--baryon kinetics can be placed on the native
spectrum, and the coherent-return construction supplies a positive
temperature--polarization phase architecture, but the event state and photon
analyzer remain additional physics.

The CMB comparison should be read at that level.  The low band organizes eight
TT feature locations, and a separate one-pole carrier captures the normalized
height ladder at toy-model accuracy while failing as a precision amplitude
fit.  With the TT phase frozen, the stationary quadrature gives all twelve TE
signs and the higher EE interleaving; a one-scale expansion-conditioned
scattering term adds the first EE lobe and improves the lowest TE location
without moving TT.  The marked-event calculation identifies the positive
temperature--quadrupole noise and shows that opacity is the least invasive
material control.  Common event ancestry fixes the first mixed opacity moment
through Proposition~\ref{prop:common-event-mixed-closure}, but the displayed
single-stage clock remains insufficient even in the full local matrix.  Thus
the paper contains a qualitative CMB morphology, not a joint precision fit.

Propagation supplies a further independent test.  A coherent translation in
log frequency gives exact achromatic redshift, waveform dilation, and Planck
covariance.  Differential translations mix blackbodies with
$y_{\rm mix}=\Sigma_{u,{\rm diff}}^2/2+\cdots$, so FIRAS forces any substantial
post-thermalization contribution to be highly coherent.  Hamiltonian Liouville
transport and reciprocity then give the usual Tolman and distance-duality
relations; the BAO Jacobian identity must hold for a common isotropic ruler.
These are constraints on an infrared redshift contribution, not evidence that
one is required.

The decisive open problem is a microscopic horizon influence law.  It must
assign charge, energy, and directional marks to the common event; derive the
return cadence and available free-energy current; and determine both the
galactic source matching and the photon-coupled opacity history.  Its Ward
identities must enforce the balance laws, while its unequal-time correlators
must fix the radiative state without fitting the sectors separately.  The
present work provides the geometry, kinematics, thermodynamic accounting, and
empirical tests by which such a completion can be judged.

\appendix

\section{A finite no-pumping example}
\label{app:bsn-model}

\subsection{Closed two-state, two-link model}

The minimal diagnostic has two nodes and two unresolved links.  Write
\begin{equation}
 \kappa_a=\e^{\beta E_a},
 \qquad g_\alpha=\e^{-\beta W_\alpha},
 \label{eq:two-link-parameters}
\end{equation}
and put a counting field on link $1$.  The tilted generator is
\begin{equation}
 \widehat H_\chi=
 \begin{pmatrix}
 -(\kappa_1g_1+\kappa_1g_2)&
 \kappa_2g_1\e^{\ii\chi}+\kappa_2g_2\\[2pt]
 \kappa_1g_1\e^{-\ii\chi}+\kappa_1g_2&
 -(\kappa_2g_1+\kappa_2g_2)
 \end{pmatrix}.
 \label{eq:two-link-generator}
\end{equation}
At
\begin{equation}
 \beta=4,
 \qquad(E_1,E_2,W_1,W_2)=(0.2,-0.1,0.3,-0.2),
 \label{eq:two-link-base-point}
\end{equation}
biorthogonal differentiation of the dominant eigenspace gives
\begin{align}
 F_{E_1W_1}^{\chi=0.3}
 &\simeq0.005514+0.089206\ii,
 \label{eq:two-link-twisted-curvature}\\
 \left.
 \frac{\partial F_{E_1W_1}^{\chi}}
 {\partial(\ii\chi)}\right|_{\chi=0}
 &\simeq0.298844.
 \label{eq:two-link-first-cumulant}
\end{align}
For dimensionless controls $e_a=\beta E_a$ and $w_\alpha=\beta W_\alpha$,
\begin{equation}
 \cK_{e_1w_1}\simeq0.0186778.
 \label{eq:two-link-dimensionless-curvature}
\end{equation}
The controls reproduce three necessary checks:
\begin{equation}
 F^{\chi=0}=0,
 \qquad
 \cK_{E_1E_2}=0,
 \qquad
 \cK_{W_1W_2}=0,
 \qquad
 \cK_{E_1W_1}\neq0.
 \label{eq:two-link-checks}
\end{equation}
The untwisted gap is
\begin{equation}
 \gamma_{\rm rel}=(\kappa_1+\kappa_2)(g_1+g_2)
 \simeq7.31707.
 \label{eq:two-link-gap}
\end{equation}
This closed detailed-balance model has no positive stationary export
denominator.  It tests the curvature and no-pumping structure, not $\Xi$.

\section{Numerical verification of the event-gated selector}
\label{app:event-model}

At the dimensionless chart point
\begin{equation}
 (E_0,E_1,E_2)=(0.2,-0.1,0),
 \qquad
 (W_{01},W_{12},W_{20})=(0.4,0.2,0.3),
 \qquad c=\lambda=\mu=1,
 \label{eq:event-base-point}
\end{equation}
the four-state generator has
\begin{equation}
 J_{01}^{\rm hk}=0,
 \qquad
 \Phi_{\rm ev}=0.5,
 \qquad
 \gamma_4=2.
 \label{eq:event-base-observables}
\end{equation}
Counting the $0\leftrightarrow1$ chart current gives
\begin{equation}
 \cK_{E_0W_{01}}=0.01938649,
 \qquad
 \Xi_{E_0W_{01}}=0.07754595.
 \label{eq:event-base-curvature}
\end{equation}

For the loop
\begin{equation}
 E_0(t)=0.2+0.05\cos(2\pi t/T),
 \qquad
 W_{01}(t)=0.4+0.05\sin(2\pi t/T),
 \label{eq:event-test-loop}
\end{equation}
let $Q_+(T)$ and $Q_-(T)$ be the integrated chart currents for the loop and its
reverse.  Periodic-state integration gives
\begin{equation}
 \frac{Q_+(T)-Q_-(T)}2
 \longrightarrow1.52222443\times10^{-4},
 \qquad
 \frac{Q_+(T)+Q_-(T)}2\longrightarrow0
 \label{eq:event-loop-limit}
\end{equation}
as $\gamma_4T\to\infty$.  The first number agrees with the independently
integrated curvature flux.  The even term falls as
$O((\gamma_4T)^{-1})$.

The exact eigenvalues of the reference four-state generator are, to the shown
precision,
\begin{equation}
 0,\qquad-2,\qquad-3.16608,\qquad-3.46439.
 \label{eq:event-eigenvalues}
\end{equation}
There is no nontrivial imaginary peripheral pair.  Together with
Eq.~\eqref{eq:selector-first-passage}, this verifies that the selector carries
geometric current and central memory but no protected oscillatory clock or
many-stage residence law.

\section{Numerical verification of the local acoustic branch}
\label{app:numerical-checkpoints}

\subsection{Tight-coupling interface}

For the generator \eqref{eq:qed-moment-generator} with
\begin{equation}
 \nu=10,
 \qquad R=0.60,
 \qquad c_b^2=10^{-4},
 \label{eq:qed-reference-values}
\end{equation}
the small-$k$ eigenvalue branch gives
\begin{equation}
 c_s^{\rm num}=0.45647752,
 \qquad
 \frac1{\sqrt{3(1+R)}}=0.45643546,
 \label{eq:qed-sound-check}
\end{equation}
a relative difference $9.21\times10^{-5}$.  The measured diffusion
coefficient in the same units is
\begin{equation}
 D^{\rm num}=0.00542832>0.
 \label{eq:qed-diffusion-check}
\end{equation}
Substitution of $k=k_\lambda$ over representative
$(b,n,\xi)$ blocks gives the same stable branch.  No CMB data enter this test.

\subsection{Requirements for a predictive numerical continuation}

A continuous TT--TE--EE prediction must have stable retarded poles, positive
noise and sky covariance, and convergence under independent refinement of the
radial, temporal, and guiding-centre resolutions.  Its event-age or opacity
state must be derived rather than assigned a fitted phase, and the associated
collision drift and noise must obey one common microscopic law.  The TT phase
is to remain fixed while polarization is tested, with no parameter attached to
an individual feature.  Integrated optical depth and Planck-spectrum energy
exchange must be checked against the same return-event ledger.

\section*{Data and numerical reproducibility}

All numerical values in the finite-state and local-branch checks follow from
the displayed equations and tabulated inputs.  The observational feature
locations and uncertainties are drawn from the cited Planck and Pan analyses;
no unpublished data or ancillary source file is required for the claims made
here.

\end{document}